\documentclass{jfm}
\usepackage{graphicx}
\usepackage{epstopdf, epsfig}
\usepackage{hyperref}
\usepackage{amsmath}
\usepackage{amssymb}
\usepackage{placeins}

\usepackage{bm}
\usepackage{natbib}

\usepackage{color,soul}

\sethlcolor{blue}
\definecolor{lightblue}{rgb}{0.8,0.8,1}

\title{Large Eddy Simulations of bubbly flows and breaking waves with Smoothed Particle Hydrodynamics}

\author{J. R. C. King\aff{1}
  \corresp{\email{jack.king@manchester.ac.uk}},
  S. J. Lind\aff{1}
  B. D. Rogers\aff{1}
  P. K. Stansby\aff{1}    
 \and R. Vacondio\aff{2}}

\affiliation{\aff{1}School of Engineering, The University of Manchester, Manchester, UK
\aff{2}Department of Engineering and Architecture, Università di Parma, Parma, Italy}

\begin{document}
\newcommand{\rra}[1]{\sethlcolor{white}\hl{#1}}    
\newcommand{\rrb}[1]{\sethlcolor{white}\hl{#1}}   
\newcommand{\rrc}[1]{\sethlcolor{white}\hl{#1}}  

\newcommand{\rrha}[1]{\sethlcolor{white}\hl{#1}}   
\newcommand{\rrhb}[1]{\sethlcolor{white}\hl{#1}}   
\newcommand{\rrhc}[1]{\sethlcolor{white}\hl{#1}}  

\maketitle

\begin{abstract}
For turbulent bubbly flows, multi-phase simulations resolving both the liquid and bubbles are prohibitively expensive in the context of different natural phenomena. One example is breaking waves, where bubbles strongly influence wave impact loads, acoustic emissions, and atmospheric-ocean transfer, but detailed simulations in all but the simplest settings are infeasible. An alternative approach is to resolve only large scales, and model small scale bubbles adopting sub-resolution closures. Here we introduce a large eddy simulation (LES) Smoothed Particle Hydrodynamics (SPH) scheme for simulations of bubbly flows. The continuous liquid phase is resolved with a semi-implicit isothermally compressible SPH framework. This is coupled with a discrete Lagrangian bubble model. Bubbles and liquid interact via exchanges of volume and momentum, through turbulent closures, bubble breakup and entrainment, and free-surface interaction models. By representing bubbles as individual particles, they can be tracked over their lifetimes, allowing closure models for sub-resolution fluctuations, bubble deformation, breakup and free-surface interaction in integral form, accounting for the finite timescales over which these events occur. We investigate two flows: bubble plumes, and breaking waves, and find close quantitative agreement with published experimental and numerical data. In particular, for plunging breaking waves, our framework accurately predicts the Hinze scale, bubble size distribution, and growth rate of the entrained bubble population. This is the first coupling of an SPH framework with a discrete bubble model, with potential for cost effective simulations of wave-structure interactions and more accurate predictions of wave impact loads.
\end{abstract}

\section{Introduction\label{intro}}

Flows involving complex, dynamic free-surface motion are found widely in industry and nature, with fuel sloshing in aircraft wings and wave impacts on coastal and offshore structures being prime examples. For waves in particular, the violent motion of the free surface often results in the entrainment of bubbles at the free surface, which can have significant effects on the overall dynamics and peak loads, and plays a major role in the exchange of gas between the ocean and atmosphere.

With a greater understanding of the effect of bubbles in breaking waves as our motiviation, we seek improved approaches to their numerical simulation. The topic of bubble entrainment in breaking waves has been the subject of considerable experimental (e.g.~\citet{rapp_1990,deane_2002}) and numerical~\citep{chen_1999wave,deike_2015,deike_2016,ma_2011,kirby_2014} research, and with advances in computational resources and mesh-adaptivity, in recent years, researchers have begun to conduct multi-phase simulations with the aim of resolving even the smallest bubble and droplet scales~\citep{mostert_2022,deike_2015,deike_2016}. This work has elucidated fundamental aspects of the process of wave breaking, including the degree of three-dimensionality in the flow~\citep{mostert_2022}, non-locality in the bubble break-up cascade~\citep{chan_2020,chan_2021b}, and the underlying physical mechanisms controlling bubble break-up~\citep{riviere_2021,ruth_2022}. Whilst high-fidelity simulations are desirable for obtaining fundamental insight, their applicability in more industrially relevent settings is limited due to computational costs. Even with adaptive mesh refinement, the computational cost of multi-phase simulations as in~\citet{mostert_2022} is significant: of the order of a month\rrb{, and half a million CPU hours across several hundred cores.}

An alternative numerical approach is to model the presence of bubbles (rather than resolving individual bubbles), with some form of population balance equation and, typically, an assumption that the bubbles are spherical. With this approach, there are two options for the treatment of the dispersed phase. Firstly, it may be modelled as a continuum, through a bubble volume fraction, or number density field, which is subject to an evolution equation. Here the evolution equations for the dispersed phase are partial differential equations (PDEs), which are integrated in the same numerical framework as the equations of motion of the continuum liquid phase. In the mesh-based literature, such schemes are referred to as \emph{Eulerian-Eulerian}, and have been developed for the simulation of bubble plume dynamics~\citep{sokolichin_1994,becker_1994,pfleger_2001}, air entrainment in breaking waves~\citep{ma_2011,kirby_2014}, and liquid jet breakup~\citep{edelbauer_2017}. Since the dynamics of bubbles dispersed in a liquid have a strong dependence on the bubble size, the treatment of polydisperse bubble distributions here is problematic. Typically, a population of bubbles is segregated into groups of similar sizes, each group requiring two additional evolution equations. Bubble breakup and coalescence may then be incorporated by source and sink terms exchanging mass (or number density) between bubble groups. An additional limitation is that this approach does not allow the tracking of individual bubbles over their lifetimes, but only of statistical averages, and this constrains potential additional physical models, such as those for bubble break up. Although numerically this approach allows bubbles larger than the discretisation lengthscale of the continuous liquid phase (but with small number density), the assumptions used in the derivation of such models limit bubble sizes to smaller than the resolution of the continuous phase~\citep{lakehal_2002}. 

The second option for treating the dispersed phase is to model\rrb{ the bubbles individually, representing each bubble as a Lagrangian particle.} Again, such approaches have been widely developed for mesh-based schemes, for example the work by~\citet{fraga_2016} focussing on bubble plumes, or the studies on turbulence-bubble interactions~\citep{mazzitelli_2003,mazzitelli_2003b,pozorski_2009,breuer_2017,olsen_2017}, and cavitation bubble clouds~\citet{fuster_2011,maeda_2018}. In the mesh-based community, these methods are described as \emph{Eulerian-Lagrangian} schemes. \rrb{Each individual bubble interacts} with the continuous phase through exchanges of momentum (and sometimes volume, as in~\citet{finn_2011}, although in many cases where the concentration of the dispersed phase is small, volume exchanges are neglected). Schemes of this type allow for the tracking of individual bubbles, and a continuous polydisperse bubble distribution poses no additional challenge\rrb{. However, the resolution of the liquid phase imposes an upper limit on the maximum bubble size which can be represented~\mbox{\citep{fraga_2016}}, and for very small bubbles, the computational costs increase with a) the increasing number of bubbles, and b) the increasing stiffness of the equation of motion of small bubbles due to the closure model for the drag force.}

Temporarily setting aside the presence of bubbles, mesh-free methods have shown significant promise for simulations of breaking waves in recent decades. Smoothed Particle Hydrodynamics (SPH) is one mesh-free method, originally developed for astrophysical simulations~\citep{gingold_1977,lucy_1977}, and since applied with considerable success to a range of terrestrial flows, including those with dynamically evolving free surfaces~\citep{monaghan_2012}. The fluid is discretised by a set of Lagrangian particles, and spatial derivatives are approximated by weighted sums of fluid properties at neighbouring particles. Whilst tracking a deforming surface undergoing topological changes is a complex task in mesh-based methods, for SPH, little additional effort is required. There are now a wide variety of SPH schemes and related methods capable of simulating breaking waves, including weakly compressible SPH~\citep{dualsphysics}, incompressible schemes~\citep{lind_2012,chow_2018,guo_2018}, and the Moving Particle Semi-Implicit method~\citep{khayyer_2009}. Although multi-phase SPH schemes are well established (e.g.~\citet{hammani_2020}), and capable of simulating multiple bubbles~\citep{zhang_2015}, and bubble-free-surface interactions~\citep{sun_2021}, we seek to avoid the cost of explicitly resolving both phases. We observe that the terms ``Eulerian-Eulerian'' and ``Eulerian-Lagrangian'' used above are misnomers (and somewhat ambiguous) in the context of SPH-based methods, while appropriate for Eulerian mesh-based numerical methods. In this work we refer to the two approaches as ``continuous-continuous'' and ``continuous-discrete'', descriptions which remain clear even when used to describe schemes with non-Eulerian methods for the continuous phase. Of the bubble modelling approaches described above, developments in SPH lag behind mesh-based methods. A model based on the continuous-continuous approach has only recently been introduced to SPH~\citep{fonty_2019}, but with very promising results for air entrainment in flow over a spillway~\citep{fonty_2020}, though the method is currently limited to simplified closure models for interphase momentum exchange. We are not aware of any continuous-discrete SPH models for bubbly flows, although the SPH implementations closest in philosophy to this approach are perhaps the multi-phase dusty gas \rrb{formulations, original developed by~\mbox{\citet{monaghan_1995,maddison_1996}}, and more recently extended by~\mbox{\citet{laibe_2012}}.}

Herein, we present an SPH implementation of the continuous-discrete approach for bubbly free-surface flows. The liquid is resolved via large eddy simulations (LES) using a semi-implicit isothermally compressible SPH framework, whilst \rra{each bubble is} represented as \rra{a} discrete Lagrangian particle which interacts with the liquid via exchanges of momentum, volume, and sub-resolution turbulence closures. We particularly focus on integral models for bubble entrainment, break-up and free-surface interaction, with application to breaking waves.

The remainder of the paper is set out as follows. In Section~\ref{sec:ge} we introduce the governing equations of our model. Section~\ref{sec:ni} presents details of the numerical implementation. In Section~\ref{sec:plume} we test our simulation framework against numerical and experimental data for bubble plumes, and in Section~\ref{sec:bw} we use our model to simulate the air entrainment in breaking waves. Section~\ref{sec:conc} is a summary of our conclusions. Further validation of our LES model is provided in Appendix~\ref{les}. A table of all symbols used in the work is given in Table~\ref{los}.

\section{Governing equations}\label{sec:ge}

\begin{figure}
\centerline{\includegraphics[width=0.6\textwidth]{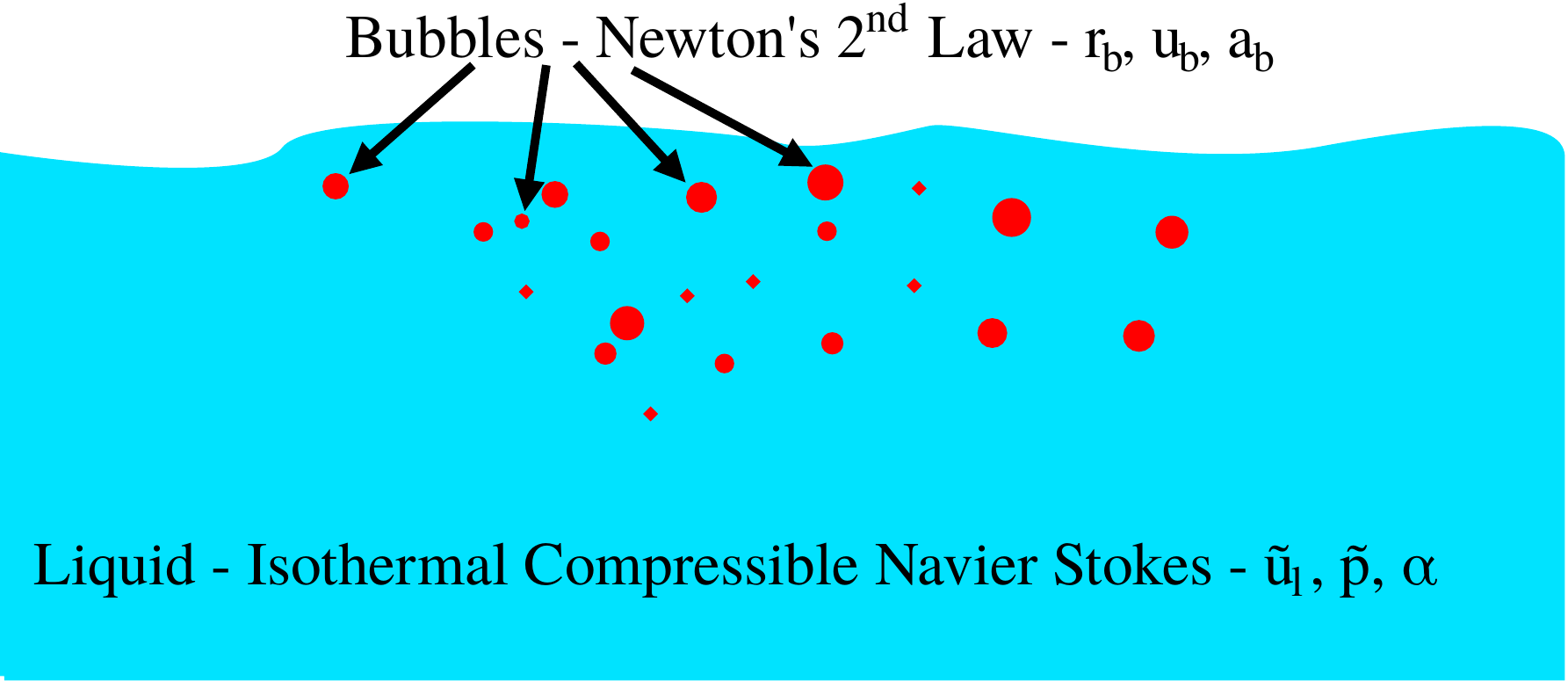}}
\caption{An illustration of the configuration considered herein. The liquid is treated as a continuum (blue), and with dispersed bubbles (red) treated as discrete particles.\label{fig:elschem}}
\end{figure}

The system we consider is a continuous liquid phase, containing a dispersed bubble phase, as illustrated in Figure~\ref{fig:elschem}. The liquid phase is governed by the isothermal compressible Navier-Stokes equations, whilst \rra{each bubble is modelled as a} discrete particle which obeys Newton's second law. In the following, the subscripts $l$ and $b$ indicate properties in the liquid (continuous) and dispersed bubble phases respectively. Where the subscript $l,b$ appears, it denotes a liquid property evaluated at a bubble. We consider the liquid to be isothermally compressible, with sound speed $c$, density $\rho_{l}$ and viscosity $\mu_{l}$. The bubbles are assumed to be spherical, comprised of gas with density $\rho_{b}$, and with a constant liquid-bubble surface tension of $\gamma$. The entire system is subject to a gravitational acceleration of $g\bm{e_{g}}$, where $\bm{e_{g}}$ is a unit vector. Following non-dimensionalisation by suitable integral length- and velocity-scales, the problem is parameterised by the Reynolds number $Re$, the Mach number $Ma$, the Froude number $Fr$, the (integral scale) Weber number $We$, and the density ratio $\beta=\rho_{l}/\rho_{b}$. Although our numerical framework is able to capture acoustic signals, in the present work these are not of interest, and are damped out\rrb{ by our implicit treatment of the pressure}, and hence $Ma$ is treated as a numerical, rather than physical, parameter. \rrb{In the limit $Ma=0$ ($c\to\infty$), our framework collapses to an incompressible framework. This approach of permitting weak compressibility, with an artificial sound speed is common in SPH (fully explicit weakly-compressible SPH being the most widely used variant in engineering simulations), but a difference in our approach is to use an implicit treatment, as in~\mbox{\citet{khayyer_2009}}, permitting larger time-steps, and resulting in more accurate pressure fields.}

\subsection{Liquid phase}

\rrb{The liquid phase is modelled as an isothermally compressible continuum, with} a Large Eddy Simulation (LES) scheme. Due to the assumption of isothermal \rrb{flow}, we can write
\begin{equation}\frac{d\tilde{p}}{d\rho_{l}}=c^{2},\end{equation}
where $\tilde{p}$ is the (implicitly) filtered pressure\rrb{, and $c$ is an artificial speed of sound.}. The filtered continuity equation may be then expressed, in \rra{a Lagrangian} frame of reference, as an evolution equation for the pressure
\begin{equation}\frac{d}{d{t}}\left(\alpha\tilde{p}\right)=-\frac{\alpha}{Ma^{2}}\nabla\cdot\tilde{\bm{u}}_{l},\label{eq:mass_le}\end{equation}
where $\alpha$ is the liquid volume fraction, and $\tilde{\bm{u}}_{l}$ is the \rra{implicitly} filtered liquid velocity. \rrb{Again we note that in the limit of $Ma=0$,~\mbox{\eqref{eq:mass_le}} becomes $\nabla\cdot\tilde{\bm{u}}_{l}=0$, and incompressibility is recovered.} We next assume that the liquid is weakly compressible; the compressibility $1/\rho_{l}c^{2}\ll1$, allowing us to neglect terms involving the density variation in the (filtered) momentum equation, which is written as
\begin{equation}\frac{d}{d{t}}\left(\alpha\tilde{\bm{u}}_{l}\right)=-\nabla\left(\alpha\tilde{p}\right)+\frac{\alpha}{Fr^{2}}\bm{e_{g}}+\frac{1}{Re}\nabla\cdot\left(\alpha\left(1+\nu_{srs}\right)\nabla\tilde{\bm{u}}_{l}\right)+\bm{M},\label{eq:mom_le}\end{equation}
where $\nu_{srs}$ is a dimensionless sub-resolution viscosity, determined by the LES closure model and the term $\bm{M}$ represents the momentum exchange between the liquid and bubble phases. 
In practice we solve~\eqref{eq:mass_le} and~\eqref{eq:mom_le} in a frame which deviates from perfectly Lagrangian by a small velocity $\bm{u}_{ps}$, referred to in the SPH literature as a shifting velocity, and introduced to add stability to the numerical solution. This results in a small error \rrha{which scales with the resolution of the discretisation scheme,} associated with the advection terms $\bm{u}_{ps}\cdot\nabla$ which are omitted from~\eqref{eq:mass_le} and~\eqref{eq:mom_le}. This approach is widely used in the SPH literature~\citep{lind_2012}.

\subsubsection{LES model}

\rrb{We denote the implicit filter width of the SPH framework as $\widetilde{\Delta}$.} The filtered equations are closed with a model for the sub-resolution viscosity, which is comprised of a shear-induced and bubble-induced eddy viscosity component: $\nu_{srs}=\nu_{S}+\nu_{B}$. The bubble-induced viscosity $\nu_{B}$ accounts of the production of sub-resolution turbulence by bubbles following the model of~\citet{sato_1975}, and is described in Section~\ref{sec:bit}. For the shear-induced turbulence, we use the mixed-scale model of~\citet{lubin_2006}, with
\begin{equation}\nu_{S}=Re{C}_{M}\widetilde{\Delta}^{1+\xi}\lvert\widetilde{S}\rvert^{\xi/2}\left(q_{c}^{2}\right)^{\left(1-\xi\right)/2},\label{eq:msm}\end{equation}
where $\widetilde{\mathcal{S}}=\left(\nabla\tilde{\bm{u}}_{l}+\left(\nabla\tilde{\bm{u}}_{l}\right)^{T}\right)/2$ is the resolved strain rate tensor, $q_{c}^{2}$ is the test-filtered kinetic energy, $\xi=0.5$ and ${C}_{M}=0.06$. We evaluate $q_{c}$ by explicitly filtering $\tilde{\bm{u}_{l}}$ with a Shepard filter (see Section~\ref{sec:sph}). \rrb{We note that multi-phase LES closure models are still an open area of research. The separation of the effective viscosity into shear- and bubble-induced components, following~\mbox{\citet{kirby_2014}}, is a modelling simplification which presumes that the effects of bubbles and free surfaces on~\mbox{\eqref{eq:msm}} may be neglected.} \rrb{As in~\mbox{\citet{kirby_2014}}, the} turbulent dissipation rate \rrb{is proportional to the shear induced viscosity and the square of the strain rate tensor norm, and} is calculated as
\begin{equation}\varepsilon=\frac{1+\nu_{S}}{Re}\left\lvert\tilde{\mathcal{S}}\right\rvert^{2}.\end{equation}

In the development of our method, we explored several additional closure models, including a standard Smagorinsky model, the dynamic Germano model~\citep{germano_1991,lilly_1992}, with both local (via Shepard filtering) and Lagrangian (along streamlines) averaging. We choose the mixed-scale model for our simulations because it is known to yield good results in flows with deforming free surfaces~\citep{lubin_2006}, and in tests of the decay of single phase isotropic turbulence (included in Appendix~\ref{les}), it yielded improved results \emph{in our numerical framework} compared with the other closure models.

\subsection{Dispersed bubble phase}

The dispersed phase is represented by a discrete set of $N_{b}$ Lagrangian bubbles $\mathcal{B}$, each with a velocity $\bm{u}_{b}$, position $\bm{r}_{b}$, radius $a_{b}$, and volume $V_{b}=4\pi{a}_{b}^{3}/3$. The system of bubbles is governed by
\begin{subequations}
\begin{align}
\frac{d\bm{r}_{b}}{dt}&=\bm{u}_{b}\label{eq:u_b}\\
V_{b}\frac{d\bm{u}_{b}}{dt}&=\bm{F}_{d}+\bm{F}_{l}+\bm{F}_{vm}+\bm{F}_{g}\label{eq:mom_b}
\end{align}\end{subequations}
for each bubble in $\mathcal{B}$. Here $\bm{F}_{d}$, $\bm{F}_{l}$, $\bm{F}_{vm}$ and $\bm{F}_{g}$ are the drag, lift, virtual mass and buoyancy forces acting on the bubble due to the surrounding liquid. These forces are evaluated for each bubble $i\in\mathcal{B}$ through closure models as in (e.g.)~\citet{fraga_2016} and~\citet{kirby_2014}, with
\begin{subequations}
\label{eq:bforces}\begin{align}
\bm{F}_{d,i}&=\frac{1}{2}C_{d}\beta\pi{a}_{b,i}^{2}\left\lvert\bm{u}_{rel,i}\right\rvert\bm{u}_{rel,i},\label{eq:fd}\\
\bm{F}_{l,i}&=C_{l}\beta{V}_{b,i}\bm{u}_{rel,i}\times\nabla\times\bm{u}_{l,i},\label{eq:fl}\\
\bm{F}_{vm,i}&=C_{vm}\beta{V}_{b,i}\left(\frac{d\bm{u}_{l}}{dt}-\frac{d\bm{u}_{b,i}}{dt}\right),\label{eq:fvm}\\
\bm{F}_{g,i}&=\frac{\left(1-\beta\right)V_{b,i}}{Fr^{2}}\bm{e_{g}},\end{align}
\end{subequations}
where \rrb{$\beta$ is the density ratio,} the relative velocity between the bubble and liquid phase at bubble $i$ is given by $\bm{u}_{rel,i}=\bm{u}_{l}\left(\bm{r}_{b,i}\right)-\bm{u}_{b,i}$, and $C_{d}$, $C_{l}$ and $C_{vm}$ are drag, lift and virtual mass coefficients respectively. Note the absence of the tilde in the liquid velocity appearing in the definition of relative velocity, and in~\eqref{eq:fl} and~\eqref{eq:fvm}. Following~\citet{breuer_2017}, the sub-resolution fluctuating part of the liquid velocity is modelled stochastically, with
\begin{equation}\bm{u}_{l}\left(\bm{r}_{b,i}\right)=\tilde{\bm{u}}_{l}\left(\bm{r}_{b,i}\right)+\bm{u}^{\prime}_{l}\left(\bm{r}_{b,i}\right)\end{equation}
comprising the LES filtered velocity and a sub-resolution fluctuating part. The calculation of $\bm{u}^{\prime}_{l}\left(\bm{r}_{b,i}\right)$ is described in Section~\ref{sec:langevin}. \rrb{Following~\mbox{\citet{kirby_2014}}, t}he drag coefficient in~\eqref{eq:fd} is \rrb{modelled using the standard drag curve of~\mbox{\citet{clift_1978}} as}
\begin{equation}{C}_{d}=\begin{cases}0.44&{Re}_{b,i}>1000;\\\frac{24}{Re_{b,i}}\left(1+0.15{Re}_{b,i}^{0.687}\right)&Re_{b,i}\le1000,\end{cases}\end{equation}
where the relative bubble Reynolds number is related to integral-scale Reynolds number by
\begin{equation}Re_{b,i}=2a_{b,i}\left\lvert\bm{u}_{rel,i}\right\rvert{Re}.\end{equation}
Following~\citet{kirby_2014} the coefficient of virtual mass and the lift coefficient are set to $C_{vm}=C_{l}=0.5$. The momentum exchange from bubble $i$ back to the liquid phase is
\begin{equation}\bm{M}_{b,i}=\left(-\bm{F}_{d,i}-\bm{F}_{l,i}-\bm{F}_{vm,i}\right)/\beta,\end{equation}
\rrb{with $\beta$ the density ratio as defined in Section~\mbox{\ref{sec:ge}}.}

\subsubsection{Langevin model}\label{sec:langevin}

The fluctuating velocity component ``felt'' by the bubbles is calculated through the integration of a stochastic model following work by~\citet{pozorski_2009} and~\citet{breuer_2017}. The fluctuation velocity obeys the Langevin equation
\begin{equation}d\bm{u}_{l}^{\prime}=-\bm{\mathsf{G}}\bm{u}_{l}^{\prime}dt+\sqrt{2\sigma_{srs}^{2}}\bm{\mathsf{B}}\bm{dW},\label{eq:langevin1}\end{equation}
where $\bm{dW}$ is a Wiener process, and $\bm{\mathsf{G}}$ and $\bm{\mathsf{B}}$ are drift and diffusion matrices respectively. The quantity $\sigma_{srs}$ is the standard deviation of the fluctuating velocity, and related to the sub-resolution turbulent kinetic energy $k_{srs}$ by
\begin{equation}\sigma_{srs}=\sqrt{\frac{2}{3}k_{srs}}.\end{equation} 
The sub-resolution turbulent kinetic energy is estimated from the double filtered velocity, as
\begin{equation}k_{srs}=\frac{1}{2}\left\lvert\bm{\tilde{u}}_{l}-\bm{\hat{\tilde{u}}}_{l}\right\rvert^{2},\end{equation}
where the $\hat{}$ indicates explicit filtering with a test filter, descibed in Section~\ref{sec:sph}.

Following~\citet{breuer_2017}, by taking advantage of the fact that a Langevin equation may be analytically integrated, we transform~\eqref{eq:langevin1} into the recursion equation
\begin{equation}\bm{u}_{l}^{\prime}\left(\bm{r}_{b,i},t+\delta{t}\right)=\bm{\mathsf{E}}\bm{u}_{l}^{\prime}\left(\bm{r}_{b,i},t\right)+\bm{\mathsf{W}}\bm{\zeta},\label{eq:langevin2}\end{equation}
where $\delta{t}$ is a time increment, $\bm{\mathsf{E}}$ is the exponential of the drift matrix $\bm{\mathsf{G}}$, $\bm{\mathsf{W}}$ is the square root of the velocity fluctuation covariance matrix, and $\bm{\zeta}$ is a random vector whose components are normally distributed. Denoting the filtered relative velocity (excluding the fluctuations) as $\bm{\tilde{u}}_{rel}$, we define the matrix
\begin{equation}\bm{\mathsf{R}}=\frac{1}{\left\lvert\bm{\tilde{u}}_{rel}\right\rvert^{2}}\bm{\tilde{u}}_{rel}\otimes\bm{\tilde{u}}_{rel},\end{equation}
and the exponential of the drift matrix is then given by
\begin{equation}\bm{\mathsf{E}}=\left(E_{\parallel}-E_{\perp}\right)\bm{\mathsf{R}}+E_{\perp}\bm{\mathsf{I}},\end{equation}
where $\bm{\mathsf{I}}$ is the identity matrix, and
\begin{subequations}
\begin{align}E_{\parallel}=&\exp\left(\frac{-\delta{t}}{\tau_{\parallel}}\right)\\
E_{\perp}=&\exp\left(\frac{-\delta{t}}{\tau_{\perp}}\right).\end{align}\end{subequations}
\begin{subequations}
Here
\begin{align}\tau_{\parallel}=&\tau_{srs}\left(1+\frac{\left\lvert\bm{\tilde{u}}_{rel}\right\rvert}{\sigma_{srs}^{2}}\right)^{-\frac{1}{2}}\\
\tau_{\perp}=&\tau_{srs}\left(1+4\frac{\left\lvert\bm{\tilde{u}}_{rel}\right\rvert}{\sigma_{srs}^{2}}\right)^{-\frac{1}{2}}\end{align}\end{subequations}
are the sub-resolution timescales associated with fluctuations parallel and perpendicular to the filtered velocity, and the sub-resolution timescale $\tau_{srs}$ is related to the velocity fluctuations by
\begin{equation}\tau_{srs}=\frac{C\tilde{\Delta}}{\sigma_{srs}},\end{equation}
with the constant $C=1$. The factors relating $\tau_{\parallel}$ and $\tau_{\perp}$ to the sub-resolution timescale $\tau_{srs}$ account for the crossing trajectory and continuity effects~\citep{pozorski_2009}. The square root of the covariance matrix is given by
\begin{equation}\bm{\mathsf{W}}=\left(W_{\parallel}-W_{\perp}\right)\bm{\mathsf{R}}+W_{\perp}\bm{\mathsf{I}},\end{equation}
where
\begin{subequations}
\begin{align}W_{\parallel}=&\sigma_{srs}\sqrt{1-\exp\left(-\frac{2\delta{t}}{\tau_{\parallel}}\right)}\\
W_{\perp}=&\sigma_{srs}\sqrt{1-\exp\left(-\frac{2\delta{t}}{\tau_{\perp}}\right)}.\end{align}\end{subequations}

\subsubsection{Bubble-induced turbulence model}\label{sec:bit}

As mentioned above, we evaluate the bubble-induced turbulence, following~\citet{kirby_2014}, based on the model of~\citet{sato_1975}, where the contribution of an individual bubble to the turbulent viscosity is proportional to the product of the bubble diameter and the relative velocity. In our discrete bubble framework, the contribution of an individual bubble $j$ is
\begin{equation}\nu_{B,b,j}=C_{\nu,B}2a_{b,j}\left\lvert\bm{u}_{rel,j}\right\rvert,\end{equation}
where the constant $C_{\nu,B}=0.6$ as in~\citet{kirby_2014}. To obtain the bubble-induced turbulent viscosity in the liquid $\nu_{B}$, we interpolate $\nu_{B,b}$ from the bubbles to the liquid phase, as described in Section~\ref{sec:interp}.

\subsubsection{Bubble entrainment and breakup}

Our intention is to simulate flows where bubbles are entrained at the free surface. We use an entrainment model, similar in principle to that of~\citet{ma_2011} and~\citet{kirby_2014}, in which a fraction of the turbulent kinetic energy of the liquid is assumed to be converted into surface energy as bubbles are created (or entrained) at the free surface. We further include a model for bubble breakup, which is based on the imbalance between the restoring pressure on a bubble due to surface tension, and the deforming stress due to the turbulent motion of the liquid, based on the models of~\citet{martinez_1999a,martinez_1999b} and~\citet{martinez_2010}. These models cannot be clearly explained without reference to our discretisation scheme, and we defer detailed description of them to later, in Section~\ref{sec:be}.

\section{Numerical implementation}\label{sec:ni}

\subsection{SPH discretisation}

The liquid phase is represented by set of discrete particles $\mathcal{P}$, each of which we label $i\in\left[1,N\right]$, where $N$ is the total number of particles. \rrb{In the present work we treat the liquid as weakly compressible, but this weak compressibility is treated implicitly, by solving an elliptic equation (Helmholtz, as opposed to Poisson) as in incompressible SPH frameworks. Hence, i}n the present work the core of the SPH scheme closely follows~\citet{king_2021} and~\citet{lind_2012}. All particles carry a (constant) liquid mass $m$, and with the assumption of a weakly compressible liquid, the associated liquid volume $V_{l}$ is assumed constant. The position of particle $i$ is denoted $\bm{r}_{i}$, its smoothing length is $h_{i}$, and the total volume associated with the particle is $V_{i}$. We denote the difference in the property $\left(\cdot\right)$ of two particles $i$ and $j$ as $\left(\cdot\right)_{ij}=\left(\cdot\right)_{i}-\left(\cdot\right)_{j}=-\left(\cdot\right)_{ji}$. In SPH, values and derivatives of field variables at the location of particle $i$ are calculated using a weighted sum of the values of the field variables at the neighbouring particles $j\in\mathcal{P}_{i}$, where the weights are obtained from a kernel function $W\left(\left\lvert\bm{r}_{ij},h_{i}\right\rvert\right)=W_{ij}$ and its derivatives. Here, $\mathcal{P}_{i}$ is the set of neighbours of particle $i$, and contains all particles $j$ with $\left\lvert\bm{r}_{ij}\right\rvert\le{r}_{s,i}$, where $r_{s,i}$ is the support radius of the kernel of particle $i$. Throughout this work we use the Wendland C2 kernel~\citep{wendland_1995} for which the support radius is $r_{s,i}=2h_{i}$. We use an initial particle spacing of $\delta{r}=h_{0}/1.3$, where $h_{0}$ is the smoothing length when $\alpha=1$. \rra{In all cases, we set the implicit filter scale to the local smoothing length: $\widetilde{\Delta}_{i}=h_{i}$. }For a derivation and analysis of SPH fundamentals, we refer the reader to~\citet{price_2012,fatehi_2011} and~\citet{monaghan_2012}. In a perfectly Lagrangian framework, particles follow streamlines, which can result in highly anisotropic particle distributions, particularly around stagnation points, degrading the accuracy of the simulation. To regularise the particle distribution, we use the particle shifting technique of~\citet{lind_2012} to set $\bm{u}_{ps}$ as in~\citet{king_2021}. \rrc{The ability of the underlying SPH methodology to accurately simulate wave propagation has been demonstrated previously, for example by~\mbox{\citet{lind_2012}}, and~\mbox{\citet{skillen_2013}}.} \rra{Note, we do not include surface tension effects in the single phase SPH simulation - rather they only appear in the closure terms governing bubble dynamics.}

\subsubsection{Interpolation between phases}\label{sec:interp}

\begin{figure}
\centerline{\includegraphics[width=0.9\textwidth]{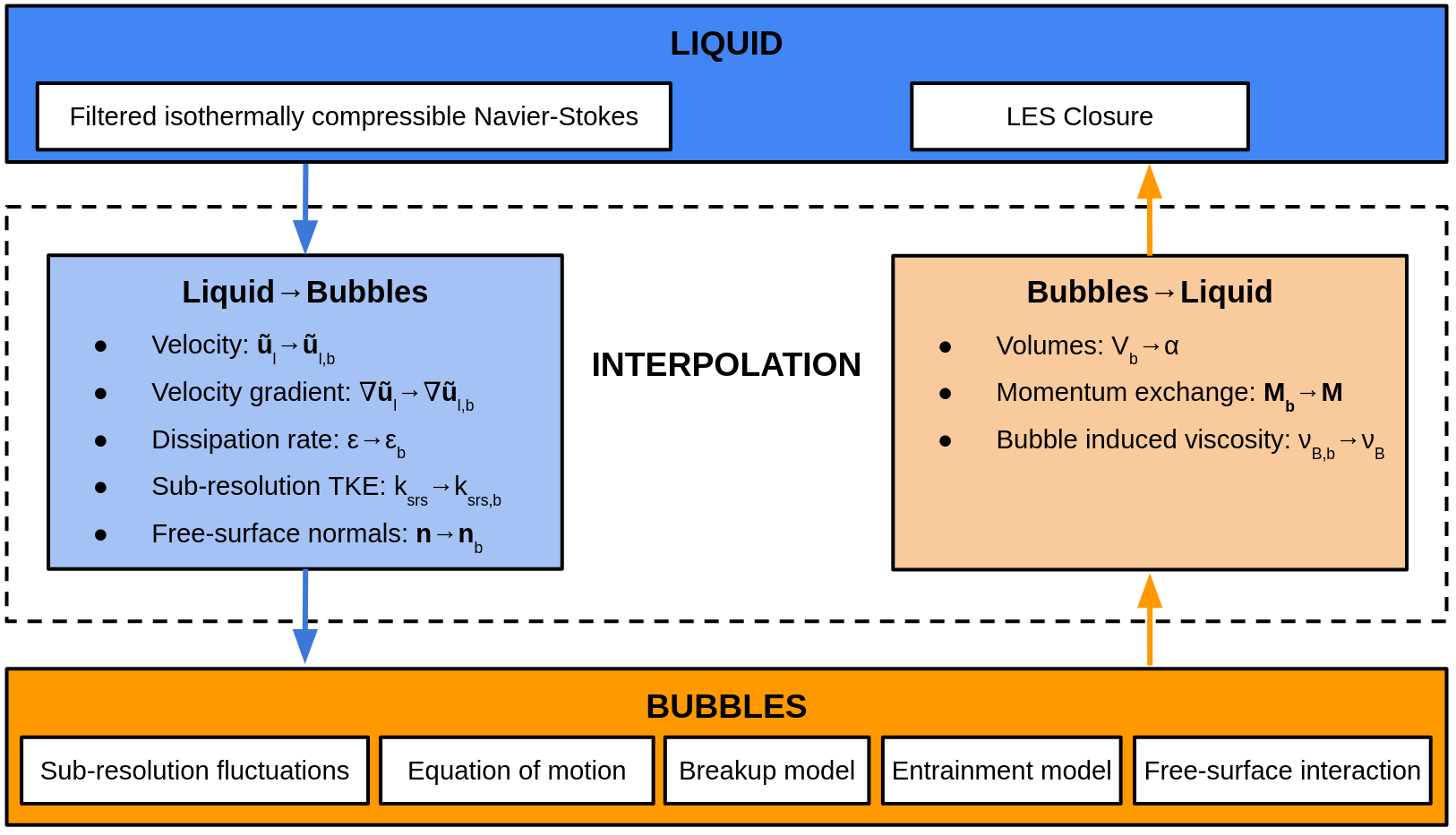}}
\caption{A schematic diagram of the numerical framework, showing the liquid and bubble properties which are interpolated between phases.\label{fig:schem}}
\end{figure}

The interaction between the liquid and bubbles occurs both through the modification of the liquid volume fraction $\alpha$ due to the presence of the bubbles, the momentum exchange $\bm{M}$ between the phases, and the bubble-induced turbulent viscosity $\nu_{B}$. To achieve this, it is necessary to interpolate between the phases: that is, to calculate the value of a liquid property at a bubble location, and vice versa. Figure~\ref{fig:schem} provides a diagrammatic overview of the framework, including those properties which are exchanged between phases. Numerically, the interphase interpolation is as follows. The total volume associated with SPH particle $i$ is defined as
\begin{equation}V_{i}=V_{l,i}\left(1+\displaystyle\sum_{j\in\mathcal{B}_{i}}W\left(\bm{r}_{b,j}-\bm{r}_{i},h_{i}\right)V_{b,j}\right),\label{eq:bvols}\end{equation}
where $\mathcal{B}_{i}$ is the set of all bubbles within the support radius of particle $i$. \rra{The summation in the right hand side of~\mbox{\eqref{eq:bvols}} represents the contribution to the volume of SPH particle $i$ of all individual bubbles in $\mathcal{B}_{i}$.} The liquid volume fraction of particle $i$ is then 
\begin{equation}\alpha_{i}=\frac{V_{l,i}}{V_{i}}\label{eq:alp}.\end{equation}
As the volume of the bubbles is accounted for in the liquid phase, the SPH particles effectively expand in the proximity of gas bubbles. To retain an accurate SPH approximation, the smoothing length of each SPH particle must be adjusted accordingly, to maintain
\begin{equation}\displaystyle\sum_{j\in\mathcal{P}_{i}}W\left(\bm{r}_{j}-\bm{r}_{i},h_{i}\right)V_{j}\approx1\qquad\forall{i}\in\mathcal{P}\label{eq:pou}.\end{equation}
With both $V_{i}$ (through~\eqref{eq:bvols}) and the partition of unity~\eqref{eq:pou} having a non-linear dependence on $h_{i}$, we cannot choose $h_{i}$ explicitly to satisfy~\eqref{eq:pou}. However, by setting 
\begin{equation}h_{i}=h_{0}\left(\frac{V_{i}}{V_{l,i}}\right)^{\frac{1}{d}},\label{eq:seth}\end{equation}
in which $d=3$ is the number of spatial dimensions, we obtain a system in which the SPH particle distribution expands and contracts in response to the bubble volumes. The response is not instantaneous, but occurs over a finite time, as the volume effects propagate through the particle distribution. However, when averaged over time, this system results in a discretisation for which~\eqref{eq:pou} is satisfied. \rra{This effect is discussed further in Section~\mbox{\ref{sec:tempevol}}.}

The momentum exchange between the bubble and liquid phases is evaluated at each bubble (denoted $\bm{M}_{b}$), and then interpolated back to the liquid phase through
\begin{equation}\bm{M}_{i}=V_{i}\displaystyle\sum_{j\in\mathcal{B}_{i}}\bm{M}_{b,j}W\left(\bm{r}_{b,j}-\bm{r}_{i},h_{i}\right)\label{eq:b2l}.\end{equation}
The bubble-induced turbulence at each bubble $\nu_{B,b}$ is interpolated back to each SPH particle in the same manner to obtain $\nu_{B}$.
Evaluation of the lift, drag and virtual mass forces for each bubble require the knowledge of the filtered liquid velocity and liquid velocity gradients, and the turbulent kinetic energy, at each bubble location. Additionally, the bubble entrainment model described in Section~\ref{sec:be} requires the turbulent dissipation rate to be interpolated from the liquid to each bubble location. These properties are interpolated from SPH particles to bubble locations through 
\begin{equation}\phi_{b,j}=\displaystyle\sum_{i\in{P}}\phi_{i}W\left(\bm{r}_{b,j}-\bm{r}_{i},h_{i}\right)V_{i}\qquad\forall{j\in\mathcal{B}},\label{eq:l2b}\end{equation}
where $\phi_{i}$ is the value at particle $i$ of the property to be interpolated, and $\phi_{b,j}$ is the interpolated property at bubble $j$. In our code, we construct an array for each particle $i$ containing the indices $\mathcal{P}_{i}$, and a global array containing the indices of the bubble neighbours of all particles: $\left[\mathcal{B}_{1}\dots\mathcal{B}_{i}\dots\mathcal{B}_{N}\right]$. 

\subsubsection{SPH operators}\label{sec:sph}

For the test filter used to evaluate $k_{srs}$, and $q_{c}$ in the LES closure model we use a normalised Shepard filter
\begin{equation}\hat{\tilde{\phi}}_{i}=\frac{\displaystyle\sum_{j\in\mathcal{P}_{i}}\tilde{\phi}_{j}W_{ij}V_{j}}{\displaystyle\sum_{j\in\mathcal{P}_{i}}W_{ij}V_{j}}.\label{eq:shep}\end{equation}
First derivatives are discretized according to
\begin{equation}\langle\nabla\tilde{\phi}\rangle_{i}=\displaystyle\sum_{j\in\mathcal{P}_{i}}\left(\tilde{\phi}_{j}-\tilde{\phi}_{i}\right)\nabla{W}^{\star}_{ij}{V}_{j}\qquad\langle\nabla\cdot\tilde{\bm{u}}\rangle_{i}=\displaystyle\sum_{j\in\mathcal{P}_{i}}\left(\tilde{\bm{u}}_{j}-\tilde{\bm{u}}_{i}\right)\cdot\nabla{W}^{\star}_{ij}{V}_{j}\label{eq:grad},\end{equation}
where the corrected kernel gradient $\nabla{W}^{\star}_{ij}$ due to~\citet{bonet_lok} is used, as detailed in~\citet{king_2021}. This provides first order consistency for first derivatives. \rra{The angled brackets indicate the quantity is an SPH approximation to the gradient}. The Laplacian is approximated using the formulation of~\citet{morris_1997} as
\begin{equation}\langle\nabla^{2}\tilde{\phi}\rangle_{i}=\displaystyle\sum_{j\in\mathcal{P}_{i}}\frac{2\tilde{\phi}_{ij}}{\lvert\bm{r_{ij}}\rvert^{2}}\bm{r_{ij}}\cdot\nabla{W}_{ij}{V}_{j},\label{eq:lap}\end{equation}
and for the inhomogeneous ``div-grad'' operator with spatially varying coefficient $\kappa$, we use
\begin{equation}\langle\nabla\cdot\left(\kappa\nabla\phi\right)\rangle_{i}=\displaystyle\sum_{j\in\mathcal{P}_{i}}\frac{2\bar{\kappa}_{ji}\tilde{\phi}_{ij}}{\lvert\bm{r_{ij}}\rvert^{2}}\bm{r_{ij}}\cdot\nabla{W}_{ij}{V}_{j},\end{equation}
with $\bar{\kappa}_{ji}={2\kappa_{i}\kappa_{j}}/\left(\kappa_{i}+\kappa_{j}\right)$ the harmonic mean.
To evaluate the \rrb{shifting} velocity $\bm{u}_{ps}$, we calculate the gradient of the particle number density as
\begin{equation}\nabla{\rho}_{N,i}=\displaystyle\sum_{j\in\mathcal{P}_{i}}\left[1+\frac{1}{4}\left(\frac{W_{ij}}{W_{ii}}\right)^{4}\right]\nabla{W}_{ij}V_{j},\end{equation}
and then set the \rrb{shifting} velocity as
\begin{equation}\bm{u}_{ps,i}=\frac{h_{i}^{2}}{4\delta{t}}\begin{cases}\nabla{\rho}_{N,i}&\forall{i}\in\mathcal{P}_{I}\\
\left(\bm{n}_{i}\cdot\nabla{\rho}_{N,i}\right)\bm{n}_{i}&\forall{i}\in\mathcal{P}_{FS},\label{eq:ups}\end{cases}\end{equation}
where $\bm{n}_{i}$ is the unit vector normal to the surface at particle $i$, $\mathcal{P}_{I}$ is the set of internal particles, and $\mathcal{P}_{FS}$ is the set of free-surface particles\rrb{. This treatment of the shifting velocity at the free surface, and the identification of free-surface particles, follow~\mbox{\citet{lind_2012}} and~\mbox{\citet{king_2021}}}. \rra{The surface normal vectors are evaluated as}
\begin{equation}\bm{n}^{\star}_{i}=\frac{1}{h_{i}}\displaystyle\sum_{j\in\mathcal{P}_{i}}V_{j}\nabla{W}_{ij},\label{eq:surfnorm}\end{equation}
\rra{where the term $1/h_{i}$ normalises the magnitude of the vector relative to the resolution. Following~\mbox{\cite{chow_2018}}, the surface-normal vectors are then smoothed using the normalised Shepard filter as in~\mbox{\eqref{eq:shep}}, with $\bm{n}_{i}=\hat{\bm{n}_{i}^{\star}}$.}

\subsection{Fractional step approach for isothermally compressible liquid\label{sec:isoT}}

Our approach to the time integration of~\eqref{eq:mass_le} and~\eqref{eq:mom_le} combines the methods used in~\citet{king_2021}, and the approach described in~\citet{khayyer_2009,khayyer_2016}. We use a fractional step algorithm to solve~\eqref{eq:mass_le} and~\eqref{eq:mom_le} in our SPH framework. Following the classic projection method of~\citet{chorin_1968}, initially introduced to SPH in~\citet{cummins_1999}, the right hand side of~\eqref{eq:mom_le} is split, with viscous and advective terms being applied in a predictor step, and the pressure gradient and any divergence-free body forces (e.g. gravity) being used in a projection step to obtain a velocity field which satisfies the continuity equation~\eqref{eq:mass_le}. Splitting~\eqref{eq:mom_le} as described above, we obtain
\begin{subequations}
\begin{align}\alpha^{n+1}\tilde{\bm{u}}^{\star}_{l}=&\alpha^{n}\tilde{\bm{u}}^{n}_{l}+\delta{t}\left[\frac{1}{Re}\nabla\cdot\left(\alpha^{n}\left(1+\nu_{srs}\right)\nabla\tilde{\bm{u}}_{l}^{n}\right)+\bm{M}\right]\label{eq:fs1}\\
\alpha^{n+1}\tilde{\bm{u}}_{l}^{n+1}=&\alpha^{n+1}\tilde{\bm{u}}_{l}^{\star}-\delta{t}\left[\nabla\left(\alpha^{n+1}\tilde{p}^{n+1}\right)-\frac{\alpha^{n+1}}{Fr^{2}}\bm{e_{g}}\right]\label{eq:fs2},\end{align}\end{subequations}
where $\tilde{\bm{u}}^{\star}_{l}$ is an intermediate velocity, which is not required to be compatible with the continuity equation~\eqref{eq:mass_le}. We require the velocity field at the end of the time step $\tilde{\bm{u}}_{l}^{n+1}$ to satisfy~\eqref{eq:mass_le}, and we take the divergence of~\eqref{eq:fs2}, to get
\begin{multline}
\alpha^{n+1}\nabla\cdot\tilde{\bm{u}}_{l}^{n+1}=\\-\tilde{\bm{u}}_{l}^{n+1}\cdot\nabla\alpha^{n+1}+\nabla\cdot\left(\alpha^{n+1}\tilde{\bm{u}}_{l}^{\star}\right)-\delta{t}\nabla^{2}\left(\alpha^{n+1}\tilde{p}^{n+1}\right)+\frac{\delta{t}}{Fr^{2}}\bm{e_{g}}\cdot\nabla\alpha^{n+1}.\label{eq:divfs2}\end{multline}
Substituting the right hand side of~\eqref{eq:divfs2} into the final term of~\eqref{eq:mass_le} evaluated at time step $n+1$, and replacing the time derivative with a backwards Euler difference equation, yields
\begin{multline}
\frac{\left(\alpha^{n+1}\tilde{p}^{n+1}-\alpha^{n}\tilde{p}^{n}\right)}{\delta{t}}=\\\frac{1}{Ma^{2}}\left[\tilde{\bm{u}}_{l}^{n+1}\cdot\nabla\alpha^{n+1}-\nabla\cdot\left(\alpha^{n+1}\tilde{\bm{u}}_{l}^{\star}\right)+\delta{t}\nabla^{2}\left(\alpha^{n+1}\tilde{p}^{n+1}\right)-\frac{\delta{t}}{Fr^{2}}\bm{e_{g}}\cdot\nabla\alpha^{n+1}\right].\label{eq:ppe1}\end{multline}
As~\eqref{eq:ppe1} explicitly contains $\tilde{\bm{u}}_{l}^{n+1}$, we substitute~\eqref{eq:fs2} back into~\eqref{eq:ppe1}, note that $\left(1/\alpha\right)\nabla\alpha=\nabla\ln\alpha$, and collect terms containing $\tilde{p}^{n+1}$, obtaining
\begin{multline}
\nabla^{2}\left(\alpha^{n+1}\tilde{p}^{n+1}\right)-\frac{Ma^{2}\alpha^{n+1}\tilde{p}^{n+1}}{\delta{t}^{2}}-\nabla\left(\ln\alpha^{n+1}\right)\cdot\nabla\left(\alpha^{n+1}\tilde{p}^{n+1}\right)\\=\frac{\alpha^{n+1}}{\delta{t}}\nabla\cdot\left(\tilde{\bm{u}}_{l}^{\star}\right)-\frac{Ma^{2}\alpha^{n}\tilde{p}^{n}}{\delta{t}^{2}}.\label{eq:ppe2}\end{multline}
Note that in the combined limits of single phase ($\alpha=1$) incompressible ($Ma=0$) flow, the standard Poission equation is recovered. 

\subsection{Temporal evolution of both phases}\label{sec:tempevol}

We now describe the complete algorithm for the temporal evolution of the complete system. In the following, each operation is applied to every SPH particle $i\in\mathcal{P}$ or bubble $i\in\mathcal{B}$, and we have dropped the subscripts $i$ for clarity. We introduce the superscripts $n$, $\star$ and $n+1$ to represent properties at the current, intermediate, and next time steps. The algorithm is as follows.
\begin{enumerate}
\item Advect particles to intermediate positions according to $\bm{r}^{\star}=\bm{r}^{n}+\delta{t}\tilde{\bm{u}}_{l}$, and bubbles to new positions according to $\bm{r}_{b}^{n+1}=\bm{r}_{b}^{n}+\delta{t}\bm{u}_{b}^{n}$.
\item Construct boundary conditions via mirror particles, calculate bubble entrainment and breakup if any (as described in Section~\ref{sec:be}), and build neighbour lists $\mathcal{P}_{i}$ and $\mathcal{B}_{i}$.
\item Calculate the \rrb{shifting} velocity $\bm{u}_{ps}$ from~\eqref{eq:ups}, based on a modified form of the Fickian shifting introduced by~\citet{lind_2012}, described in detail in~\citet{king_2021}.
\item Interpolate bubble volumes into SPH particle locations using~\eqref{eq:bvols} and~\eqref{eq:alp} to obtain volume fractions $\alpha^{n+1}$. Adjust smoothing lengths via~\eqref{eq:seth}. Note that this is a first order approximation to $\alpha^{n+1}$, but allows us to retain an explicit scheme for the volume fractions.
\item Evaluate $\nabla\tilde{\bm{u}}_{l}$, $\varepsilon$, and $k_{srs}$, and interpolate to bubble positions through~\eqref{eq:l2b}. 
\item Evolve~\eqref{eq:langevin2} to obtain $\bm{u}_{l}^{\prime}$ at bubble locations, and evaluate forces in bubbles via~\eqref{eq:bforces}.
\item Update bubble velocities by integrating~\eqref{eq:mom_b}, evaluate $\bm{M}_{b}$ and $\nu_{B,b}$, and interpolate back to SPH particle positions through~\eqref{eq:b2l}.
\item Evaluate remaining terms in~\eqref{eq:fs1}, and evaluate $\tilde{\bm{u}}_{l}^{\star}$.
\item Construct and solve equation~\eqref{eq:ppe2} to obtain $\alpha^{n+1}\tilde{p}^{n+1}$, and hence $\tilde{p}^{n+1}$. The system~\eqref{eq:ppe2} is solved using a BiCGStab algorithm with Jacobi preconditioning, with Neumann boundary conditions on solid surfaces, and homogeneous Dirichlet boundary conditions on free surfaces as in~\citet{king_2021}.
\item Evaluate pressure gradient, and project $\tilde{\bm{u}}_{l}^{\star}$ onto $\tilde{\bm{u}}_{l}^{n+1}$ using~\eqref{eq:fs2}.
\item Advect particles to final positions $\bm{r}^{n+1}=\bm{r}^{n}+\delta{t}\left(\tilde{\bm{u}}_{l}^{n}+\tilde{\bm{u}}_{l}^{n+1}+2\bm{u}_{ps}\right)/2$.
\end{enumerate}

The value of $\delta{t}$ is set adaptively according to criteria for the Courant condition and viscous diffusion, as in~\citet{xenakis_2015}.
\begin{equation}\delta{t}=0.2\min\left(\frac{h}{\displaystyle\max_{\mathcal{P}}\left(\lvert\tilde{\bm{u}}_{0}\rvert\right)},Re{h}^{2}\right),\end{equation}
in which $\displaystyle\max_{\mathcal{P}}$ is the maximum value over the set of all particles $\mathcal{P}$. 
Although our numerical scheme is capable of capturing acoustic waves in the liquid, this does not impose an additional (stability related) constraint on the time-step, as the acoustic physics are treated implicitly in the fractional step approach. The acoustic part of the system is unconditionally stable, although for larger time-steps, the acoustic waves are subject to greater numerical dissipation. There is a trade-off here, between computational costs, and the degree to which acoustic information is of interest. Regardless of the application, there is a benefit of the present approach over a perfectly incompressible approach. Whilst both methods result in smooth pressure fields with no spurious noise (as is commonly experienced in explicit, weakly compressible SPH), the additional terms in~\eqref{eq:ppe2} accounting for compressibilty increase the diagonal dominance of the linear system which represents the discrete form of this equation. This increased diagonal dominance renders the system more amenable to solution using iterative methods, which will converge more quickly. In the present application we are not interested in capturing the acoustic waves, and do not impose an additional time-step constraint proportional to $hMa$. Despite this, for $Ma=0.05$, we still obtain a reduction in the necessary number of iterations to solve~\eqref{eq:ppe2} by a factor of typically around $5$. For our SPH framework, the solution of~\eqref{eq:ppe2} is the most expensive aspect of the simulation, so this presents a significant computational saving. With this in mind, we consider $Ma$ in this framework to be a numerical parameter (rather than a physical one), as in weakly compressible SPH schemes. The value of $Ma=0.05$ is consistent with the artificial sound speeds widely used in weakly compressible SPH, ensures the density fluctuations are well below $1\%$~\citep{monaghan_1994}, and is used throughout this work.

\begin{figure}
\centerline{\includegraphics[width=0.49\textwidth]{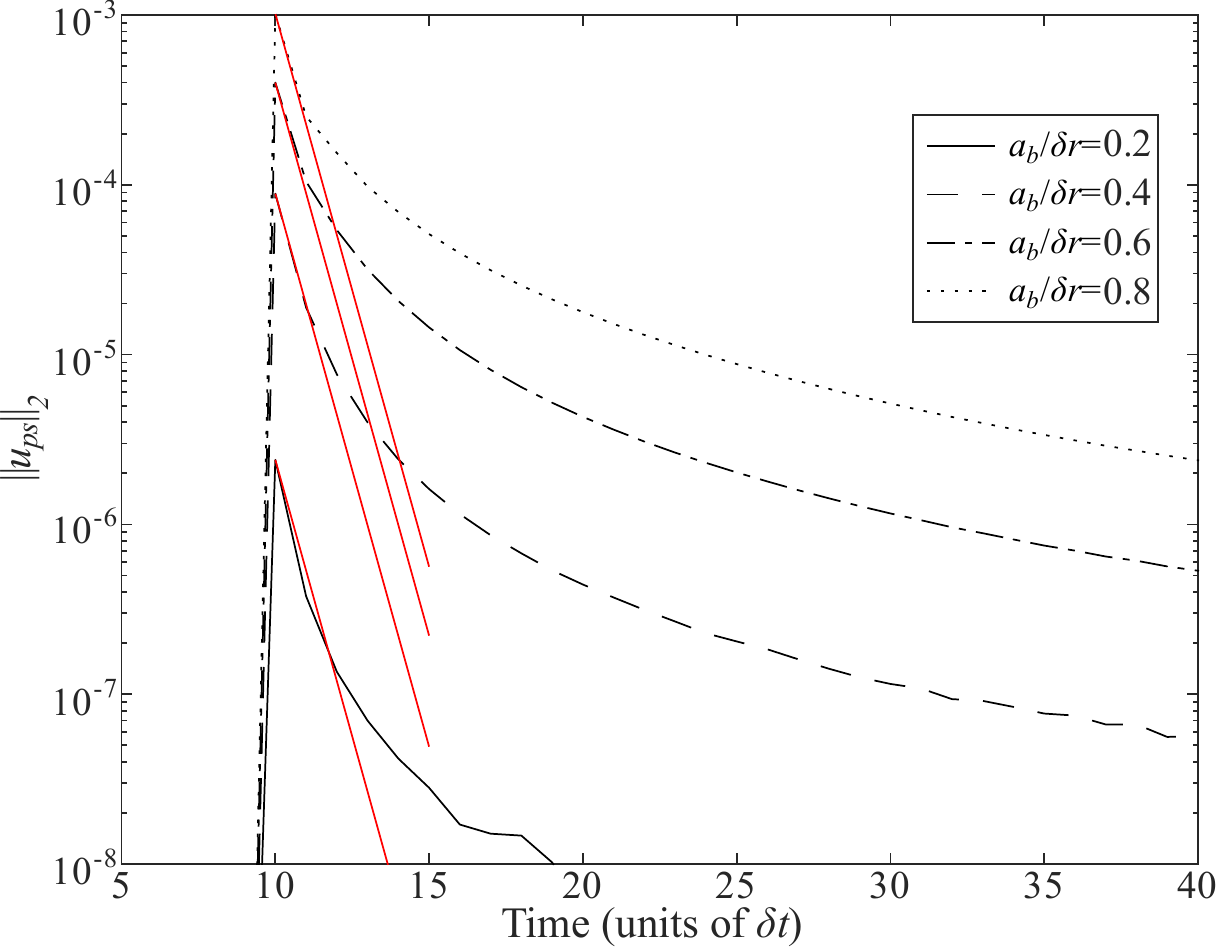}}
\caption{Variation of the $L_{2}$ norm of the shifting velocity magnitude $\left\lvert\bm{u}_{ps}\right\rvert$ with time (in units of $\delta{t}$) for a triply-periodic unit cube of fluid at rest, subject to the instantaneous addition of a single bubble. The different patterned lines indicate different bubble sizes $a_{b}$ relative to the SPH particle spacing $\delta{r}$. The red lines indicate an exponential decay with characteristic time $2\delta{t}/3$.\label{fig:redist}}
\end{figure}

\rra{As mentioned in Section~\mbox{\ref{sec:interp}}, SPH particle volumes are adjusted to account for bubble volumes, resulting in a redistribution of SPH particles over a finite time to retain an approximately uniform particle number density. To verify that this effect is small, we perform a simulation over a triply periodic cubic domain with unit side length, and the liquid initially at rest. We set $\delta{r}=1/20$, and neglect gravity. After $10$ SPH time-steps, we introduce a stationary bubble at the centre of the domain. Firstly, we observe that the resulting pressure and velocity fields remain zero everywhere, as does the velocity of the bubble. The response to the addition of the bubble is purely numerical, in that only the particle distribution is affected. We calculate the $L_{2}$ norm of $\left\lvert\bm{u}_{ps}\right\rvert$ over the domain which, being proportional to the SPH particle concentration gradient, allows us to quantify the effect. Figure~\mbox{\ref{fig:redist}} shows $\lvert\lvert\bm{u}_{ps}\rvert\rvert_{2}$ plotted against time-step number for several values of bubble radius $a_{b}$. For all bubble sizes, there is an initial peak when the bubble is created, where the SPH particle volumes have increased to accommodate the bubble, but no particle redistribution has taken place. The magnitude of this peak is proportional to the cube of the bubble radius, as expected. For bubbles with $a_{b}\le0.1\delta{r}$, the peak in $\lvert\lvert\bm{u}_{ps}\rvert\rvert_{2}$ is below $10^{-6}$. After the peak, the traces all decay. The decay is initially exponential, with characteristic time approximately $2\delta{t}/3$ (slope lines shown in red in Figure~\mbox{\ref{fig:redist}}) for all bubble radii, but at later times the decay rate slows. We believe this reduction in decay rate is a consequence of the finite volume of the bubble which must be accommodated within the particle distribution. For larger bubbles, the particle redistribution must be less localised in space, and consequently in time. We have confirmed this by running this test for a range of different values of $\delta{t}$ spanning an order of magnitude. We perform the same test with a fixed $a_{b}=0.4\delta{r}$, for a range of values of $h_{0}/\delta{r}\in\left[1.1,1.5\right]$ (noting that all other simulations in this work are performed with $h/\delta{r}=1.3$), and found a negligible variation in the peak value of $\lvert\lvert\bm{u}_{ps}\rvert\rvert_{2}$ with $h_{0}/\delta{r}$, whilst the initial decay timescale varied from $0.55\delta{t}$ for $h_{0}/\delta{r}=1.1$, to $0.8\delta{t}$ for $h_{0}/\delta{r}=1.5$. In all these cases, the characteristic time of decay is less than the value of the SPH time-step - in other words: the particle redistribution happens quickly. With the results shown in Figure~\mbox{\ref{fig:redist}} in mind, we observe that the mean value of $\lvert\lvert\bm{u}_{ps}\rvert\rvert_{2}$ for the simulation of a breaking wave studied in Section~\mbox{\ref{sec:bw}} is $1.4\times{10}^{-2}$. Whilst there are physical processes in our model occuring on shorter timescales than the redistribution process (e.g. bubble break-up, sub-resolution fluctuations), the effect of particle redistribution due to the presence of bubbles is more than an order of magnitude smaller than to the particle redistribution due to the fluid motion which is a standard aspect of SPH.}

\subsection{Bubble entrainment, breakup, and free-surface interaction}\label{sec:be}

\subsubsection{Entrainment}


Our intention is to simulate flows where bubbles are entrained at the free surface. We use an entrainment model, similar in ethos to that of~\citet{ma_2011} and~\citet{kirby_2014}, in which a fraction of the turbulent kinetic energy of the liquid is assumed to be converted into surface energy as bubbles are created (or entrained) at the free surface. \rra{Based on this energy balance, t}he energy available for bubble creation at SPH particle $i$ in a given time step is 
\begin{equation}E_{bc,i}=C_{\varepsilon}\beta\alpha_{i}\varepsilon_{i}V_{i}\delta{t}\label{eq:ebc},\end{equation}
where $C_{\varepsilon}=0.01$ is a constant set empirically. The surface energy of a bubble of radius $a_{b}$ is 
\begin{equation}E_{se}=\frac{4\pi{a}_{b}^{2}}{We}\label{eq:bse}.\end{equation}
The closure models used to evaluate the forces on each bubble are based on the assumption of spherical non-interacting bubbles~\citep{fraga_2016}. When the concentration of bubbles is large (and hence $\alpha$ is small), these assumptions cease to be valid. Therefore, we impose an additional constraint on bubble entrainment, such that $1-\alpha\lesssim1/3$, by denoting the volume available for bubble entrainment as
\begin{equation}V_{bc,i}=\frac{1}{W\left(0\right)}\left(\frac{3}{2}-\frac{1}{\alpha_{i}}\right)\label{eq:vbc},\end{equation}
where $W\left(0\right)$ is the maximum value of the SPH kernel. $V_{bc,i}$ is the volume of bubbles which must be entrained at SPH particle $i$ to result in $\alpha_{i}=2/3$. \rra{We note that our algorithm does not impose a hard limit of $\alpha>2/3$, but rather an approximate limit, as it is possible for bubbles to converge on a particular region of the flow subsequent to entrainment. Although this limit may influence the resulting physics of our simulations, it is a necessary limitation of the model of this form: we can either limit $\alpha$, or permit $\alpha$ to stray into territory where our underlying assumptions cease to be valid. Given that spherical bubbles in a cubic-packed lattice would yield a volume fraction of $\alpha\approx0.52$, this limit does not seem overly stringent. Furthermore, we note that the simulations of breaking waves in~\mbox{\citet{kirby_2014}} yielded values of $\alpha>2/3$.} Following~\citet{kirby_2014}, we only entrain bubbles at the free surface, and only when $\varepsilon$ exceeds a certain threshold. Where others~\citep{ma_2011,kirby_2014} treated the bubbles as a continuum, we treat them discretely, and hence our algorithm differs somewhat, despite the principles of energy balancing being the same. Where~\citet{kirby_2014} modelled the bubbles through a set of bubble groups, each with a characteristic size, we are able to model individual bubbles, and hence obtain a continuous bubble size distribution. To achieve this, our entrainment algorithm is as follows.

At every time-step, for every free-surface particle $i\in\mathcal{P}_{FS}$ for which $\varepsilon_{i}>0.2$ and $\alpha>2/3$:
\begin{enumerate}
\item Evaluate the available energy $E_{bc,i}$ and volume $V_{bc,i}$ for bubble creation, according to~\eqref{eq:ebc} and~\eqref{eq:vbc}. Initialise counters for new bubbles, and new potential bubbles: $n_{nb}=0$ and $n_{npb}=0$.
\item Generate a potential bubble radius $a_{b,p}\in\left[We^{-3/5}/40,\delta{r}\right]$ with uniform probability distribution. Increment the new potential bubble counter: $n_{npb}=n_{npb}+1$.
\item Evaluate the surface energy $E_{se,p}$ of the potential bubble via~\eqref{eq:bse}. If $E_{bc,i}\ge{E}_{se,p}$, there is enough energy. If $E_{bc,i}<E_{se,p}$ and $E_{bc,i}/E_{se,p}>\zeta_{E}$, where $\zeta_{E}\in\left[0,1\right]$ is a uniformly distributed random number, then there is not enough energy for this potential bubble. 
\item Evaluate the volume of the potential bubble $V_{b,p}=4\pi{a}_{b,p}^{3}/3$. If $V_{bc,i}\ge{V}_{b,p}$, there is enough volume available to create this bubble. If $V_{bc,i}<V_{b,p}$ and $V_{bc,i}/V_{b,p}>\zeta_{V}$, where $\zeta_{V}\in\left[0,1\right]$ is a uniformly distributed random number, then there is not enough volume available for this potential bubble.
\item If the checks in steps (iii) and (iv) were passed, then make a new bubble $j$ with radius $a_{b,j}=a_{b,p}$, position $\bm{r}_{b,j}=\bm{r}_{i}+\bm{\zeta}_{r}\delta{r}$, where $\bm{\zeta}_{r}$ is a random vector with elements in $\left[0,1\right]$, $\bm{u}_{b,j}=\tilde{\bm{u}}_{l,i}$, and $\delta{r}$ is the initial particle spacing. Denote the time of bubble creation as $T_{b,j}$. Increment the new bubble counter: $n_{nb}=n_{nb}+1$.
\item If the checks in steps (iii) and (iv) were passed, reduce the available energy and volume by
\begin{subequations}
\begin{align}
E_{bc,i}&=E_{bc,i}-E_{se,p}\\
V_{bc,i}&=V_{bc,i}-V_{b,p}.
\end{align}
\end{subequations}
\item Check whether to continue entraining bubbles. If all the inequalities
\begin{subequations}
\begin{align}
E_{bc,i}&>0\\
V_{bc,i}&>0\\
n_{nb}&<10\\
n_{npb}&<n_{nb}+5\end{align}
\end{subequations}
are satisfied, return to step (ii). Otherwise, move on to the next SPH particle.
\end{enumerate}

The addition of the stochastic processes in steps (iii) and (iv) to allow some bubbles for which $E_{se,p}>E_{bc,i}$ and $V_{b,p}>V_{bc,i}$, based on the remainders, prevents the creation of large bubbles from being overly suppressed. 

The maximum potential bubble size $a_{b,p}\le\delta{r}$ is set to ensure that the bubbles are smaller than the implicit LES filter scale $\widetilde{\Delta}$. The minimum potential bubble size $a_{b,p}\ge{We}^{-3/5}/40$ is equal to $0.05a_{H}$, where $a_{H}$ is the Hinze scale bubble radius evaluated a priori. This limit is imposed to prevent the generation of a very large number of very small bubbles, which would impose a significant computational cost on the simulation, but which play little role in the overall dynamics of the flow. In future, we consider that the present framework could be extended to treat bubbles with $a_{b}\ll{a}_{H}$ as a continuum, as is done for all bubble sizes in ``Eulerian-Eulerian'' schemes. \rra{The value of $C_{\varepsilon}$ influences the total entrained volume, but has little influence on the bubble distribution in time, space and size. For all simulations of breaking waves we set $C_{\varepsilon}=0.01$, which provides an approximate match in terms of the total entrained bubble volume with the DNS results of~\mbox{\citet{deike_2016}}.}

\subsubsection{Breakup}

As in work by~\citet{ma_2011,martinez_2010,martinez_1999a,martinez_1999b,chan_2021a} and~\citet{chan_2021b}, we use a stochastic breakup model based on energy balance considerations. Our model is similar to~\citet{martinez_2010}, and is based on the imbalance between the surface restoring pressure and the stress on the bubble surface due to the motion of the liquid. For a bubble of size $a_{b}$, the surface restoring pressure is $6/\left(2a_{b}We\right)$. The stress exerted on the bubble by the turbulent motion of the liquid is $\frac{1}{2}{c}_{def}\left(2\varepsilon{a}_{b}\right)^{2/3}$, where $c_{def}=8.2$ as given in~\citet{batchelor}. The net deforming stress on the bubble is given by the difference, from which we obtain an expression for a characteristic deformation velocity
\begin{equation}u_{def}=\text{sgn}\left(c_{def}\left(2\varepsilon{a}_{b}\right)^{\frac{2}{3}}-\frac{6}{{a}_{b}{We}}\right)\sqrt{\left\lvert{c}_{def}\left(2\varepsilon{a}_{b}\right)^{\frac{2}{3}}-\frac{6}{{a}_{b}{We}}\right\rvert}\end{equation}
In works such as~\citet{ma_2011,martinez_2010,chan_2021a} and~\citet{chan_2021b} which model the dispersed phase through a continuous bubble number density field, a characteristic timescale of breakup is often given by
\begin{equation}T_{bu}=\frac{2a_{b}}{u_{def}}=\frac{2a_{b}}{\sqrt{c_{def}\left(2\varepsilon{a}_{b}\right)^{\frac{2}{3}}-\frac{6}{{a}_{b}{We}}}},\end{equation}
assuming $u_{def}>0$.
In this work, we treat each bubble individually, and hence we can construct a model which attempts to account for the residence time of the bubble, and the fact that deformation and breakup occur over a finite time~\citep{risso_1998}. For every bubble we numerically integrate the deformation velocity to obtain a ``deformation distance''
\begin{equation}L_{def}=\max\left\{\displaystyle\int_{t_{bc,i}}^{t}u_{def}\left(\tau\right)d\tau,0\right\},\label{eq:defdist}\end{equation}
which is a measure of how deformed the bubble is. Here $t_{bc,i}$ is the time of creation of bubble $i$. Since we allow $u_{def}<0$,~\eqref{eq:defdist} accounts for both deformation of the bubble due to turbulence, and relaxation of the bubble back towards a spherical shape. We assume that the bubble breaks once the deformation distance exceeds the bubble diameter: when $L_{def}>2a_{b}$. 

\begin{figure}
\centerline{\includegraphics[width=0.6\textwidth]{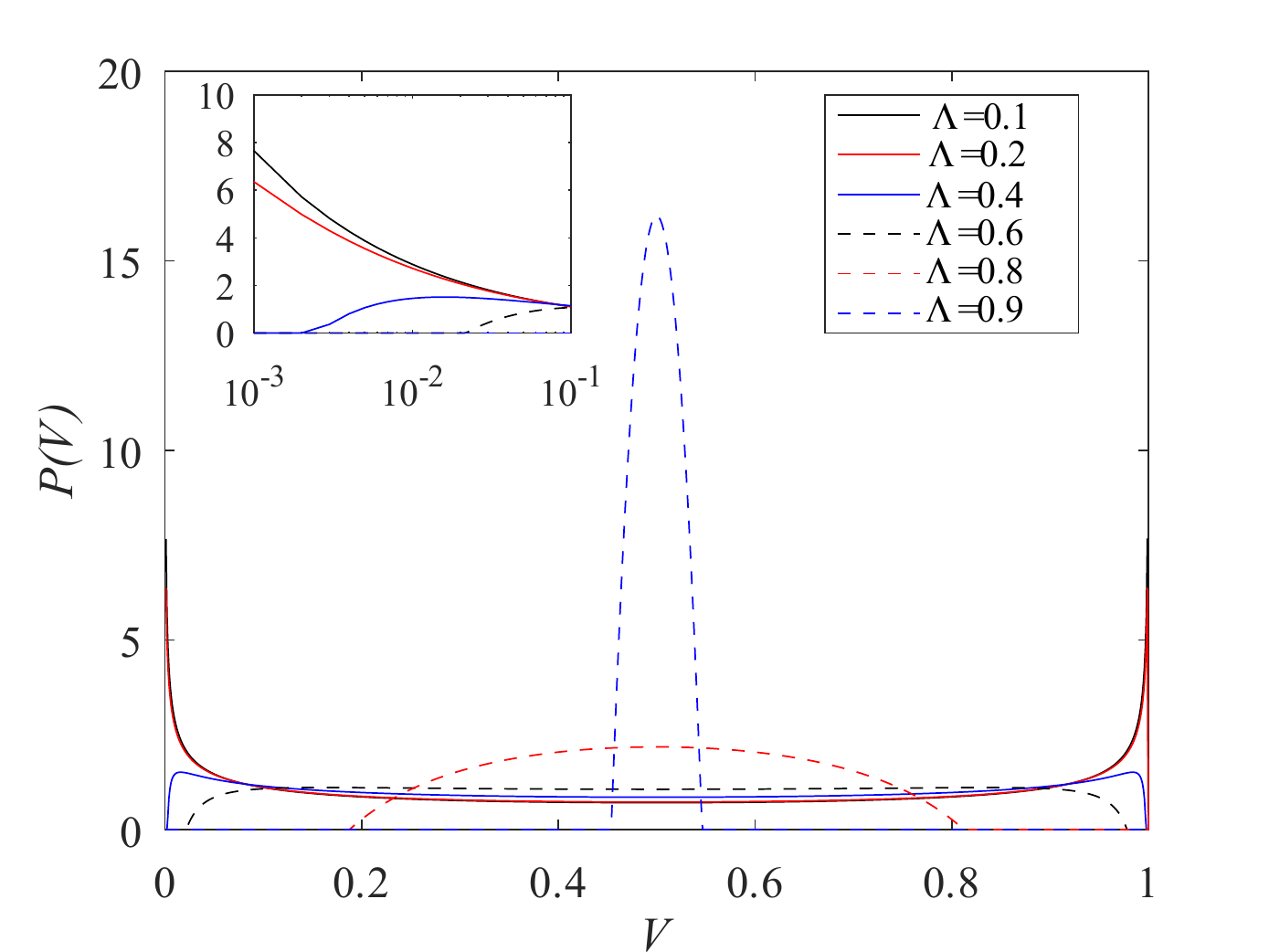}}
\caption{The probability density function of child bubble volumes for different values of $\Lambda$, given by the breakup model of~\citet{martinez_2010}.\label{fig:breakupspecmodels}}
\end{figure}

We assume all breakup events are binary; that is, a parent bubble splits into two child bubbles. Once a bubble has been marked for a breakup event, we set $V=V_{b,child}/V_{b,parent}\in\left[0,1\right]$ randomly with the probability distribution given by~\citet{martinez_2010}, where
\begin{equation}P\left(V\right)\propto\begin{cases}V^{-2/3}\left(1-V\right)^{-2/3}\left(V^{2/9}-\Lambda^{5/3}\right)\left[\left(1-V\right)^{2/9}-\Lambda^{5/3}\right]&V\in\left[V_{min},V_{max}\right]\\0&\text{otherwise}\end{cases}\label{eq:pdf}\end{equation}
Here $V_{min}$ is set to limit the size distributions to regions where the expression for $P\left(V\right)$ is positive, with an additional hard limit of $V_{min}\ge10^{-3}$, to ensure that breakup events do not occur in a capillary-action dominated regime. $P\left(V\right)$ is even about $V=1/2$, and $V_{max}=1-V_{min}$. To normalise $P\left(V\right)$, and evaluate the cumulative density function (required for generating a random number with distribution $P\left(V\right)$), we integrate~\eqref{eq:pdf} numerically. $\Lambda$ represents a critical bubble radius (relative to the parent bubble), below which the confining stresses due to surface tension exceed the turbulent stresses - it is the largest bubble which will not break due to the turbulent flow (at a given instant in time and space). For each breaking event, we evaluate $\Lambda$ as
\begin{equation}\Lambda=\left(\frac{12}{c_{def}We}\right)^{3/5}\bar{\varepsilon}^{-2/5}\left(2a_{b,p}\right)^{-1},\end{equation}
where
\begin{equation}\bar{\varepsilon}=\frac{1}{t-t_{bc,i}}\displaystyle\int_{t_{bc,i}}^{t}\varepsilon\left(\tau\right){d}\tau,\end{equation}
is the mean dissipation rate of the bubble lifetime, with $\varepsilon$ the instantaneous dissipation rate in the liquid at bubble $i$. In the above model, a critical value of $\Lambda_{crit}=2^{-6/45}\approx0.912$ exists, and the model is not valid for $\Lambda>\Lambda_{crit}$ ($P\left(V\right)<0$ $\forall{V}$). This effectively imposes an additional limit on the smallest bubble which can break, which is not necessarily consistent with the breakup criteria obtained by integration of $L_{def}$. Even if the evaluation of $L_{def}$ from~\eqref{eq:defdist} suggests a bubble should break, if it has a corresponding $\Lambda>\Lambda_{crit}$ it does not, as the child bubble sizes are undefined in~\eqref{eq:pdf}. In practice, this situation rarely occurs, and has neglible impact on the overall bubble population dynamics. Figure~\ref{fig:breakupspecmodels} shows $P\left(V\right)$ for various values of $\Lambda$. We see that for small $\Lambda$ the child size distribution is largely flat, but with peaks at very small and very large bubbles (more apparent in the inset). As $\Lambda$ increases, these peaks reduce, and for $\Lambda=0.4$ the distribution is nearly flat. For larger $\Lambda$, the possible range of child sizes decreases, with an increasing dominance of equal-sized breakup events. For $\Lambda=\Lambda_{crit}$, all breakup events result in $V=1/2$.

Whilst binary break-up models have been utilised in a range of works~\citep{ma_2011,kirby_2014,martinez_1999a,martinez_1999b,martinez_2010}, recent work~\citep{riviere_2021,ruth_2022} has suggested that the binary breakup mechanism presents some limitation. DNS of individual bubbles breaking in isotropic turbulence~\citep{riviere_2021} show that even for a binary breakup event, a bubble filament joining breaking bubbles forms, and this filament ruptures into one or many small bubbles due to capillary effects. This mechanism provides a non-local element to the bubble size cascade, and has been suggested as responsible for the majority of sub-Hinze scale bubble formation~\citep{ruth_2022}. A breakup model constructed on the framework above, but which includes the formation of multiple small bubbles by capillary action, is an active area of development for us. However, at this stage we prefer to adopt a simple binary breakup model, and focus on the free-surface-bubble interactions.

\subsubsection{Free-surface interactions}

In the vicinity of free surfaces, the assumption that the fluid around the bubble is uniform and infinite does not hold, and the closure models for the forces in~\eqref{eq:bforces} are not valid. The dynamics of the interactions between bubbles and free surfaces are complex, even for individual bubbles in stationary reservoirs, and can include a direct rise to surface bursting, oscillatory bouncing behaviour~\citep{sanada_2005,sunol_2010}, and long-term persistence of the bubble at the free surface. Despite this, most mesh-based Eulerian-Lagrangian schemes simply represent the free surface via a rigid free slip condition (e.g.~\citet{fraga_2016}), or through a volume-of-fluid (VOF) approach to resolve the gas phase above the free surface (e.g.~\citet{zhang_2005,pan_2021}). In the latter approach, both the deceleration of bubbles on approach to the free surface and the free-surface deformation are captured, although this occurs solely through modification of the relative densities in the closure models used to evaluate the forces on the bubbles. These closure models still assume a uniform liquid field surrounding the bubbles, and the complex and small scale physics involved are not included. Detailed modelling of free surface-bubble interactions are beyond the scope of this work. However, some mechanism to describe the behaviour of the interaction is necessary, to prevent bubbles from simply rising due to buoyancy and leaving the domain. For each SPH particle, a surface normal vector $\bm{n}$ is evaluated (and smoothed), as in~\citet{king_2021}. The surface normal vectors are then interpolated to each bubble position through~\eqref{eq:l2b} to obtain $\bm{n}_{b}$. For each bubble we construct a parameter $\psi_{fs}$ which identifies when a bubble is in proximity to the free surface as
\begin{equation}\psi_{fs}=\frac{\left\lvert\bm{n}_{b}\right\rvert}{\left\lvert\bm{n}_{0}\right\rvert{V}_{lb}},\end{equation}
where $\left\lvert\bm{n}_{0}\right\rvert$ is the magnitude of the surface normal vector at a plane surface, and $V_{lb}$ is the SPH volume interpolated to the bubble location (i.e.~\eqref{eq:l2b} with $\phi=1$). \rra{$\lvert\bm{n}_{0}\rvert$ is dependent only on the choice of SPH kernel and ratio of smoothing length to SPH particle spacing, and is hence constant across all our simulations, taking the precomputed value of $\lvert\bm{n}_{0}\rvert=0.353$.} For bubbles far from the free surface, $\psi_{fs}\approx0$. For bubbles within the support radius of free-surface SPH particles, $\psi_{fs}$ smoothly increases, to approximately $1$ when bubbles are located at the free surface. 
Denoting $\hat{\bm{n}}_{b}=\bm{n}_{b}/\left\lvert\bm{n}_{b}\right\rvert$, the relative normal velocity between a bubble and the free surface is $\bm{u}_{rel}\cdot\hat{\bm{n}}_{b}$. Bubbles with $\psi_{fs}>0.1$ and $t-t_{bc}>T_{c}$ are flagged to interact with the free surface. Here $T_{c}=\delta{r}/\left\lvert\bm{u}_{rel}\cdot\hat{\bm{n}}_{b}\right\rvert$ is a threshold age (which varies as a bubble evolves), designed to prevent bubbles from being destroyed at the moment of entrainment. When a bubble is marked for free-surface interaction, it is given an expected merge time $t_{m}=t+T_{c}$, at which it will be located at the free surface. For bubbles interacting with the free surface,~\eqref{eq:mom_b} is modified as
\begin{equation}V_{b,i}\frac{d\bm{u}_{b,i}}{dt}=\bm{F}_{d}+\bm{F}_{l}+\bm{F}_{vm}+\bm{F}_{g}+\bm{F}_{fs},\label{eq:mom_b_fs}\end{equation}
where the free-surface interaction force $\bm{F}_{fs}$ is
\begin{multline}
\bm{F}_{fs}=\frac{1}{2}\left(1+\text{erf}\left(2\log\left(5\psi_{fs}\right)\right)\right)\times\\\frac{\bm{n}_{b}}{\left\lvert\bm{n}_{b}\right\rvert}\left\{\left\lvert\bm{u}_{rel}\cdot\hat{\bm{n}}_{b}\right\rvert\left(1+C_{vm}\beta\right)\frac{V_{b}}{\max\left(t_{m}-t,\delta{t}\right)} - \left[\bm{F}_{b}+\bm{F}_{d}+\bm{F}_{l}+\beta{C}_{vm}V_{b}\frac{d\bm{u}_{l}}{dt}\right]\cdot\frac{\bm{n}_{b}}{\left\lvert\bm{n}_{b}\right\rvert}\right\}.\end{multline}
The first term $\frac{1}{2}\left(1+\text{erf}\left(2\log\left(5\psi_{fs}\right)\right)\right)$ varies smoothly from $0$ when $\psi_{fs}$ is small, to $1$ for large $\psi_{fs}$, and is approximately $1$ for $\psi_{fs}>0.4$. This term ensures the free-surface interaction force is switched on smoothly as a bubble approaches the free surface. The free-surface interaction force decelerates the bubble in the direction normal to the free surface, such that the bubble velocity approaches the free-surface velocity over the timescale $T_{c}$ (or $\delta{t}$, whichever is greater). The momentum exchange is modified to include $\bm{F}_{fs}$ as
\begin{equation}\bm{M}=\left(-\bm{F}_{d}-\bm{F}_{l}-\bm{F}_{vm}-\bm{F}_{fs}\right)/\beta.\end{equation}
We additionally impose a special treatment for bubbles interacting with spray. Where an individual SPH particle $i$ becomes separated from the bulk of the fluid (such that $\mathcal{P}_{i}$ contains only $i$), it is subjected only to a gravitational body force, and all other terms in~\eqref{eq:mom_le} are set to zero. If a bubble has fewer than $20$ (in three dimensions) SPH particle neighbours, \emph{and} all of those SPH particle neighbours are identified as free-surface particles, then the bubble is discounted, and assumed to have burst. This provides increased stability in regions of violent spray, such as around breaking waves, and prevents bubbles from becoming completely isolated from the SPH simulation.

\begin{figure}
\centerline{\includegraphics[width=0.49\textwidth]{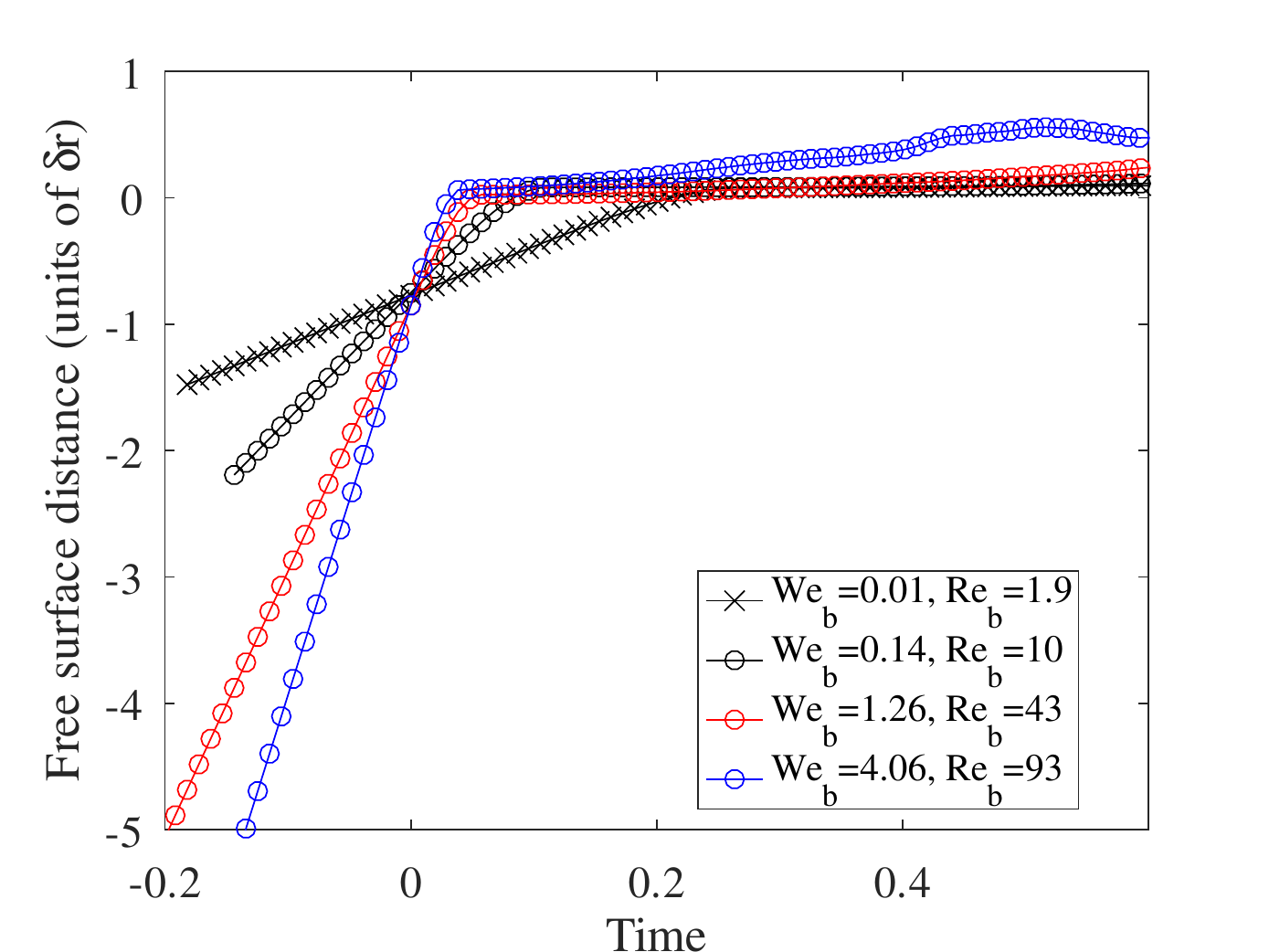}
\includegraphics[width=0.49\textwidth]{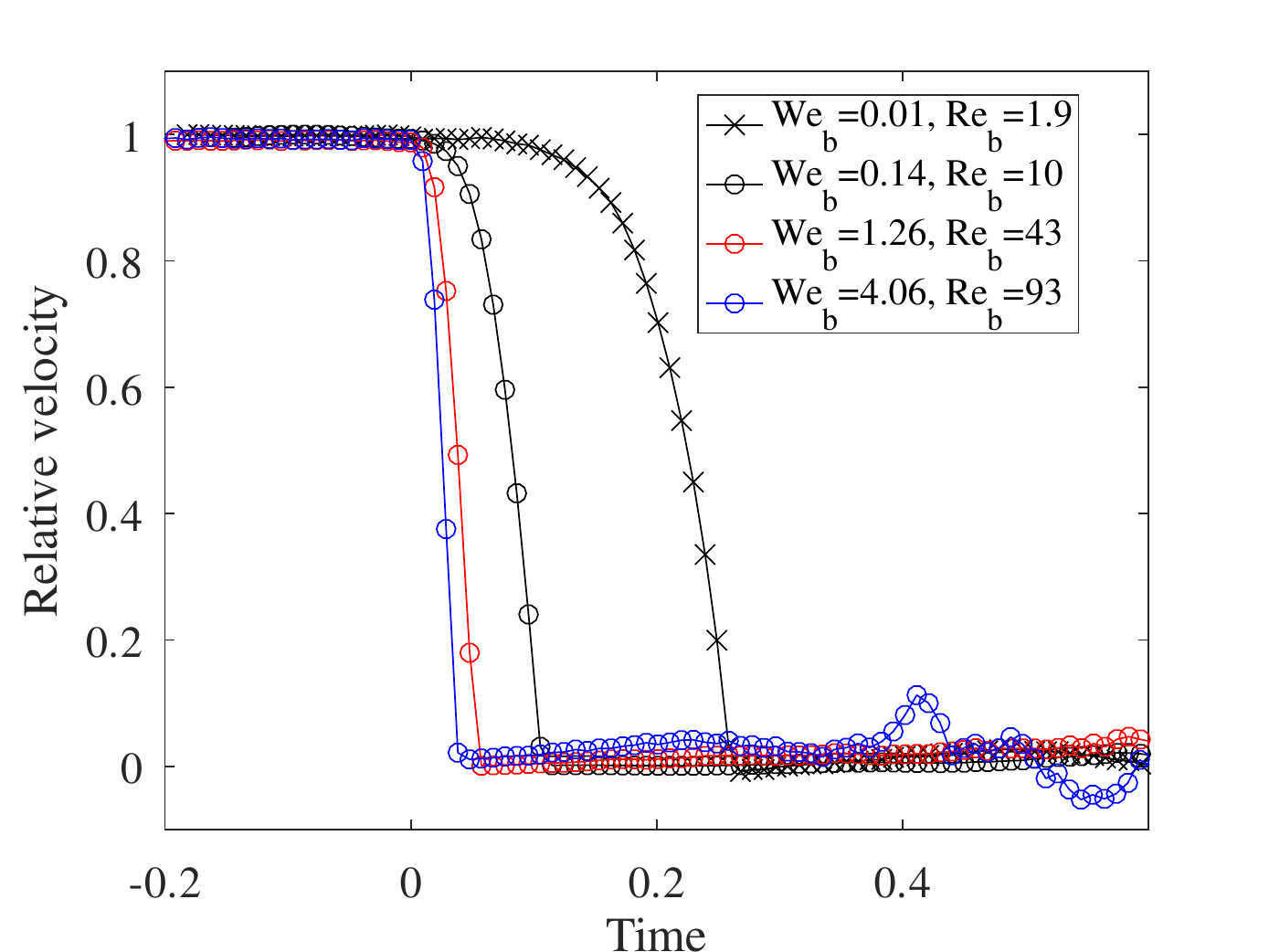}}
\caption{Trajectories of individual bubbles approaching a free surface, for a range of bubble Weber numbers $We_{b}$. Left panel: variation of distance to free surface, scaled with particle spacing $\delta{r}$, with dimensionless time. Right panel: variation of relative normal velocity $\left\lvert\bm{u}_{rel}\cdot\hat{\bm{n}}_{b}\right\rvert$ with dimensionless time, scaled with the terminal velocity of the rising bubble. In both panels, a time shift has been applied such that $t=0$ corresponds to the time at which the free-surface interaction begins.\label{fig:rise}}
\end{figure}

The effect of the free-surface interaction model is demonstrated in Figure~\ref{fig:rise}. A tank of still water is simulated, with an individual bubble rising due to buoyancy. We vary $a_{b}$, and taking the characteristic velocity scale as the terminal velocity of the bubble $u_{t}$, the bubble Reynolds number varies as $Re=10^{5}u_{t}a_{b}$, and the bubble-scale Weber number is $We_{b}=1.4\times{10}^{4}a_{b}u_{t}^{2}$. Figure~\ref{fig:rise} shows the bubble trajectory (left) and rise velocity (right) for several values of $We_{b}$. In both plots, a time shift has been applied so that the coordinate on the abscissa corresponds to the time since the bubble-surface interaction begins. We see that for larger $We_{b}$, the bubble is decelerated more quickly. For all $We_{b}\in\left[0.01,4.06\right]$, the bubble comes to rest on the free surface. For larger bubble Weber numbers $We_{b}$ (corresponding to larger $a_{b}$), the system becomes less stable after the bubble reaches the surface, and this is due to the settling over a finite time of the SPH particles to accommodate the bubble volume. In all cases, though, the final position of the bubble remains within a distance $\delta{r}$ of the free-surface location. 

The final feature of the free-surface interaction model is the persistence time. Once a bubble has reached the free surface, it remains subject to~\eqref{eq:mom_b_fs} for a persistence time $T_{p}$. If the motion of the liquid phase is such that the bubble (now moving with the liquid) moves away from the free surface (i.e., if $\psi_{fs}<0.1$), then the bubble is assumed to no longer interact with the free surface, and its motion is again governed by~\eqref{eq:mom_b}. In reality, when bubbles reach a free surface a thin lubrication layer forms, and the liquid in this layer drains until the layer ruptures (see e.g.~\citet{modini_2013} for a description of the mechanism). This process is complex, and is strongly influenced by the local geometry and the salinity~\citep{scott_1975} (or other contaminants and surfactants) and gradients thereof. We cannot seek to accurately capture this process in our model, and instead we seek an order of magnitude estimate for $T_{p}$ which has a physical basis. We use the (here made non-dimensional) expression of~\citet{poulain_2018}, where
\begin{equation}T_{p}=\frac{We^{3/4}}{Re}Fr^{1/2}Sc{a}_{b}^{1/2},\label{eq:tp}\end{equation}
in which $Sc$ is the Schmidt number, taken as $Sc=700$ for sea water at $20^{\circ}C$~\citep{debruyn_1997}. Equation~\eqref{eq:tp} is based on a mechanistic model for film drainage, acounting for molecular diffusion and curvature-pressure induced flow, and provides a rough scaling of the persistence time, with a level of approximation which is consistent with the present framework: a more detailed physical model is beyond the scope of this work. Bubbles which remain in proximity to the free surface beyond $t_{m}$ are assumed to burst (and are removed from the simulation) at $t=t_{m}+T_{p}$.

\section{Bubble plumes}\label{sec:plume}

We first use our model to simulate a buoyant bubble plume in a tank. Our simulation is configured to match the experimental set up of~\citet{fraga_2016} as follows. The domain is a unit cube of liquid, with no-slip conditions at the lateral and lower boundaries, and a free-surface boundary at the upper surface of the liquid. The origin of our coordinate system is at the centre of the base of the tank, with the unit vector in the $z$ direction pointing upwards. A bubble sparger is simulated, centred at $\left(x,y,z\right)=\left(0,0,0.09\right)$, and radius $r_{0}=0.02$. The governing parameters are $\beta=1000/1.2$, $Re=10^{6}$, $We=1.4\times{10}^{4}$, and $Fr=1/\sqrt{9.81}$. The (dimensionless) volumetric flow rate of the bubble sparger is $\dot{Q}=1.67\times{10}^{-5}$. With this dimensionless scaling, the Hinze scale is estimated at $a_{H}=We^{-3/5}/2=1.62\times{10}^{-3}$. 

\begin{figure}
\centerline{\includegraphics[width=0.9\textwidth]{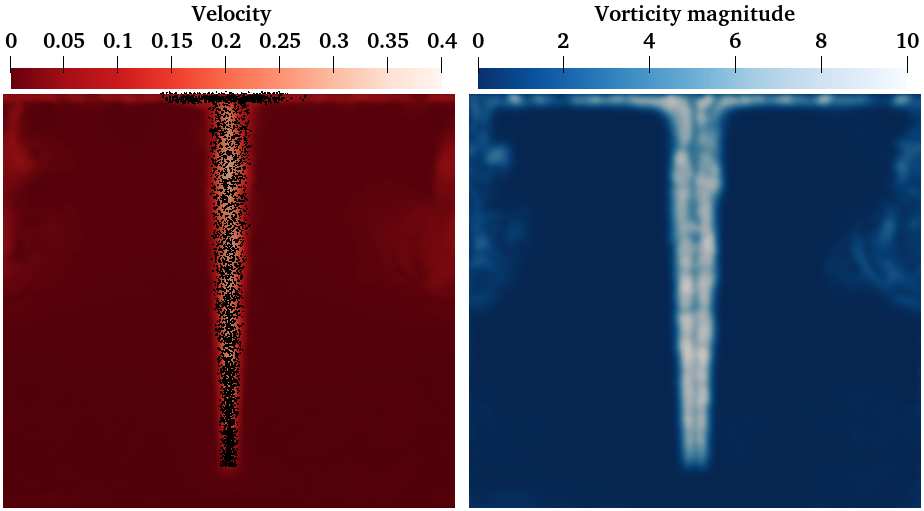}}
\caption{Velocity (left) and vorticity (right) magnitude of the flow around the bubble plume at $t=12$. Bubbles are shown as black circles (not to scale) in the left panel. The images show a slice through the plane $y=0$.\label{fig:plume_vis}}
\end{figure}

\begin{figure}
\centerline{\includegraphics[width=0.49\textwidth]{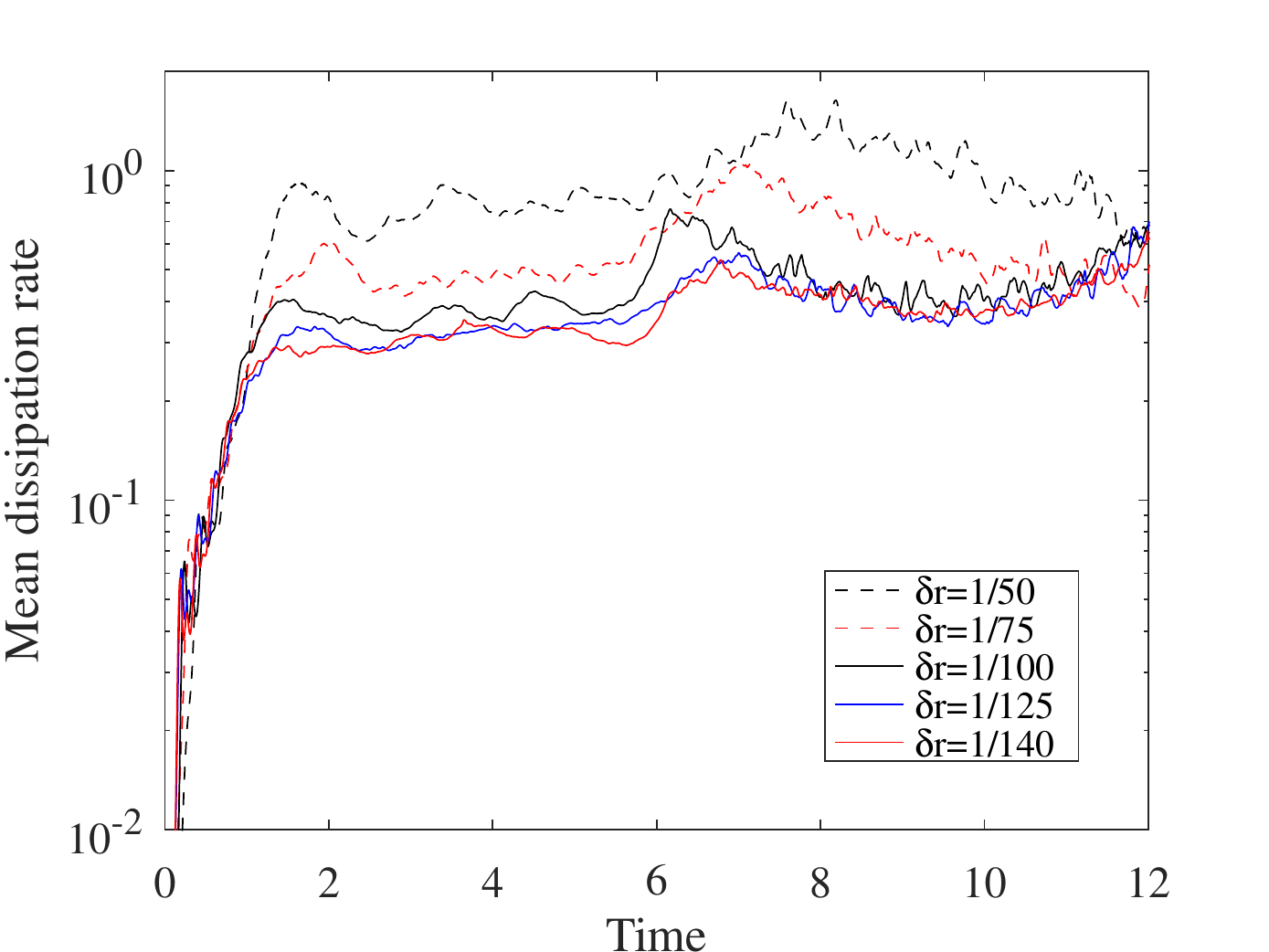}
\includegraphics[width=0.49\textwidth]{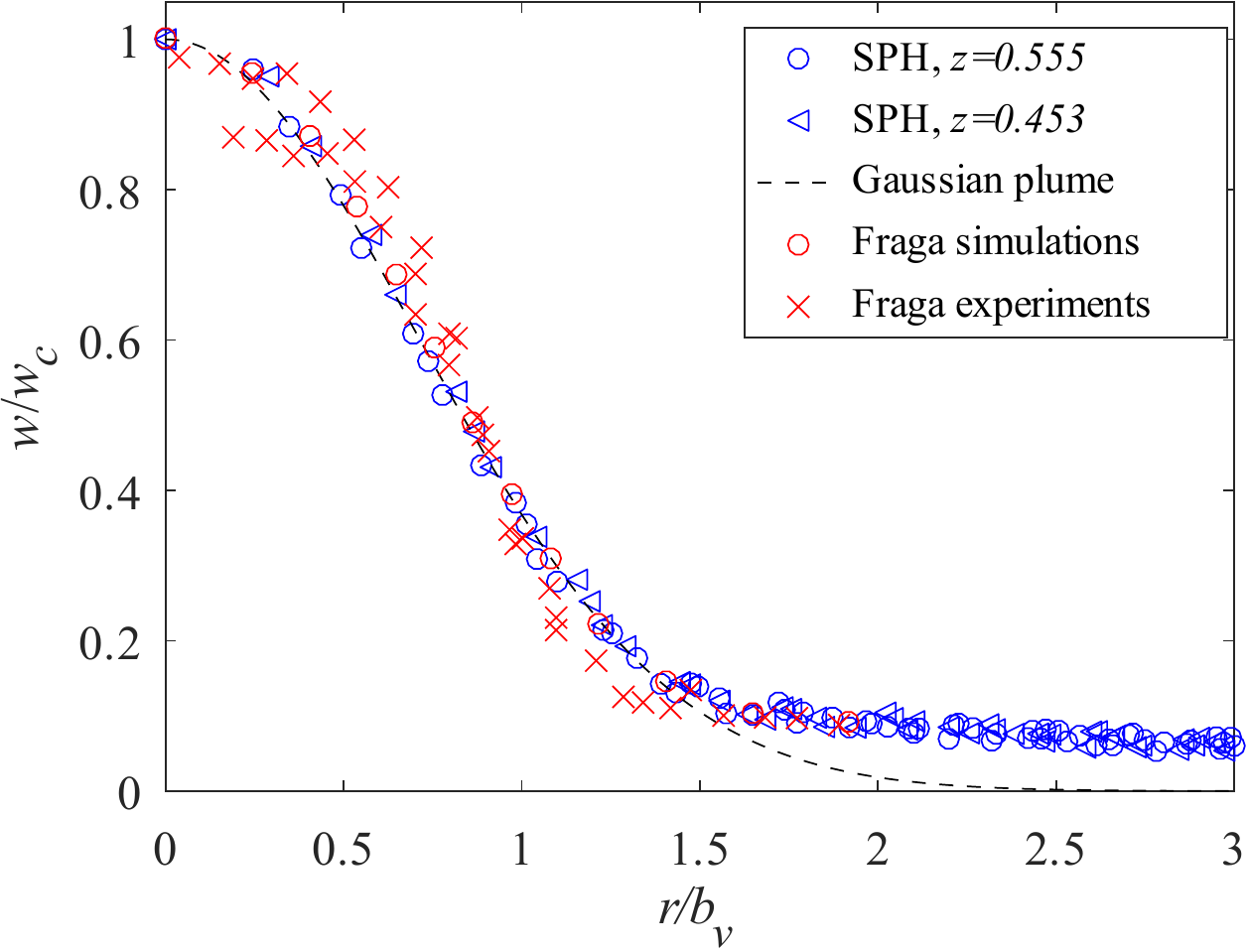}}
\caption{Left panel: Evolution of the mean dissipation rate in the liquid for several resolutions. Right panel: Variation of dimensionless mean vertical velocity in liquid with dimensionless radial position. Blue symbols indicate the results of our simulations at depths $z=0.555$ (circles) and $z=0.453$ (triangles), which are compared with the self-similar Gaussian solution (dashed black line), and the experimental (red crosses) and numerical (red circles) results of~\citet{fraga_2016}.\label{fig:fraga_profile}}
\end{figure}

Initially, we simulate the release of uniform bubbles with $a_{b}=10^{-3}$, as in~\citet{fraga_2016}. \rrb{Figure~\mbox{\ref{fig:plume_vis}} shows the velocity and vorticity magnitude fields in the $y=0$ plane at $t=12$.} The left panel of Figure~\ref{fig:fraga_profile} shows the evolution of the mean dissipation rate in the liquid for several resolutions. For all resolutions there are differences in the dissipation rate, although for $\delta{r}<1/100$, the dissipation rates approximately converge for $t>6$. We note that for the isotropic turbulence with $Re=10^{6}$ investigated in Appendix~\ref{les}, a resolution of $\delta{r}=1/100$ is sufficient to yield accurate dissipation rates compared with the high-order reference data in~\citet{antuono_2021}. For the remainder of this section we set $\delta{r}=1/100$. \rra{The increase in high-frequency noise in the mean dissipation rate at approximately $t=6$ visible in the left panel of Figure~\mbox{\ref{fig:fraga_profile}} is associated with the bubble plume reaching the free surface. Whilst previous authors (e.g.~\mbox{\citet{fraga_2016}}) have used finite volume methods with a fixed free surface, our SPH simulation captures the motion of the free surface, and includes a model for the interaction of the bubbles with the free surface. The high-frequency noise is a numerical artefact of the SPH scheme: as the liquid upwells in the centre of the tank and moves radially outwards, SPH particles join and leave the free surface, resulting in a small redistribution of the SPH particle arrangement.}

The right panel of Figure~\ref{fig:fraga_profile} shows the variation of the mean vertical velocity in the liquid with radial position, at different depths. Here the vertical velocity is scaled by the (local) velocity at the plume centre $w_{c}\left(z\right)$, and the radial position is scaled by the local plume width $b_{v}$, taken as $w\left(r=b_{v}\left(z\right),z\right)=e^{-1}w_{c}\left(z\right)$. The blue symbols indicate the results of our SPH framework just above ($z=0.555$, circles) and below ($z=0.453$, triangles) the centre of the tank. The numerical (red circles) and experimental (red crosses) data from~\citet{fraga_2016} are also shown, as is a Gaussian distribution (dashed black line). We see agreement at both vertical locations with both the Gaussian profile and the results of~\citet{fraga_2016}. For larger $r/b_{v}>1.5$ our results deviate from the Gaussian profile, but match those of~\citet{fraga_2016}, tending towards a non-zero $w/w_{c}$. As discussed in~\citet{fraga_2016}, this deviation from the Gaussian is due to the confined plume in our simulations.

\begin{figure}
\centerline{\includegraphics[width=0.49\textwidth]{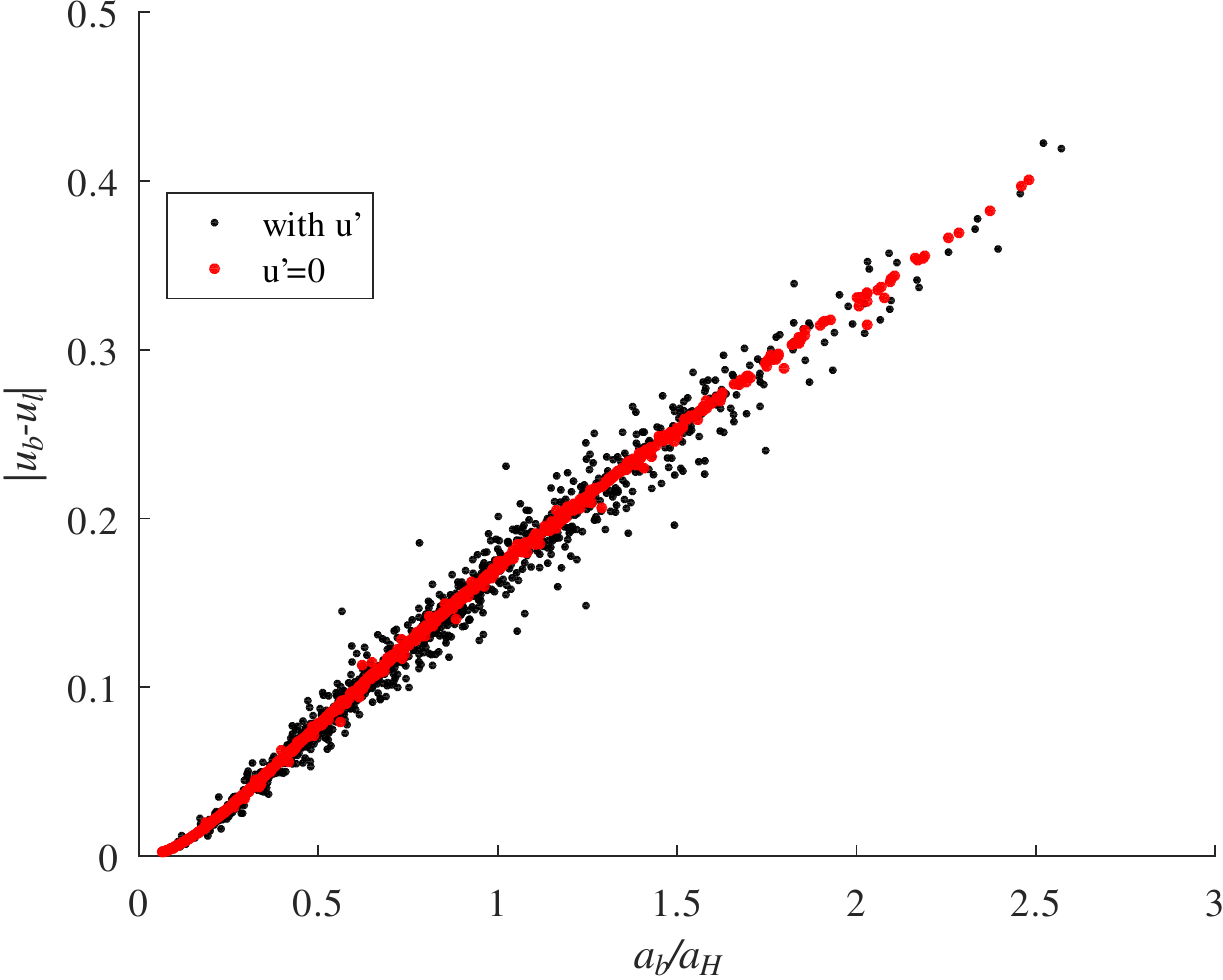}}
\caption{Variation in the relative velocity magnitude (excluding fluctuations) $\lvert\bm{u}_{b}-\tilde{\bm{u}}_{l}\rvert$ with bubble size at a depth of $z=0.555$, for the polydisperse bubble plume, both with (black symbols) and without (red symbols) the Langevin model for sub-resolution velocity fluctuations. \label{fig:plume_a_v_urel}}
\end{figure}

\begin{figure}
\centerline{\includegraphics[width=0.49\textwidth]{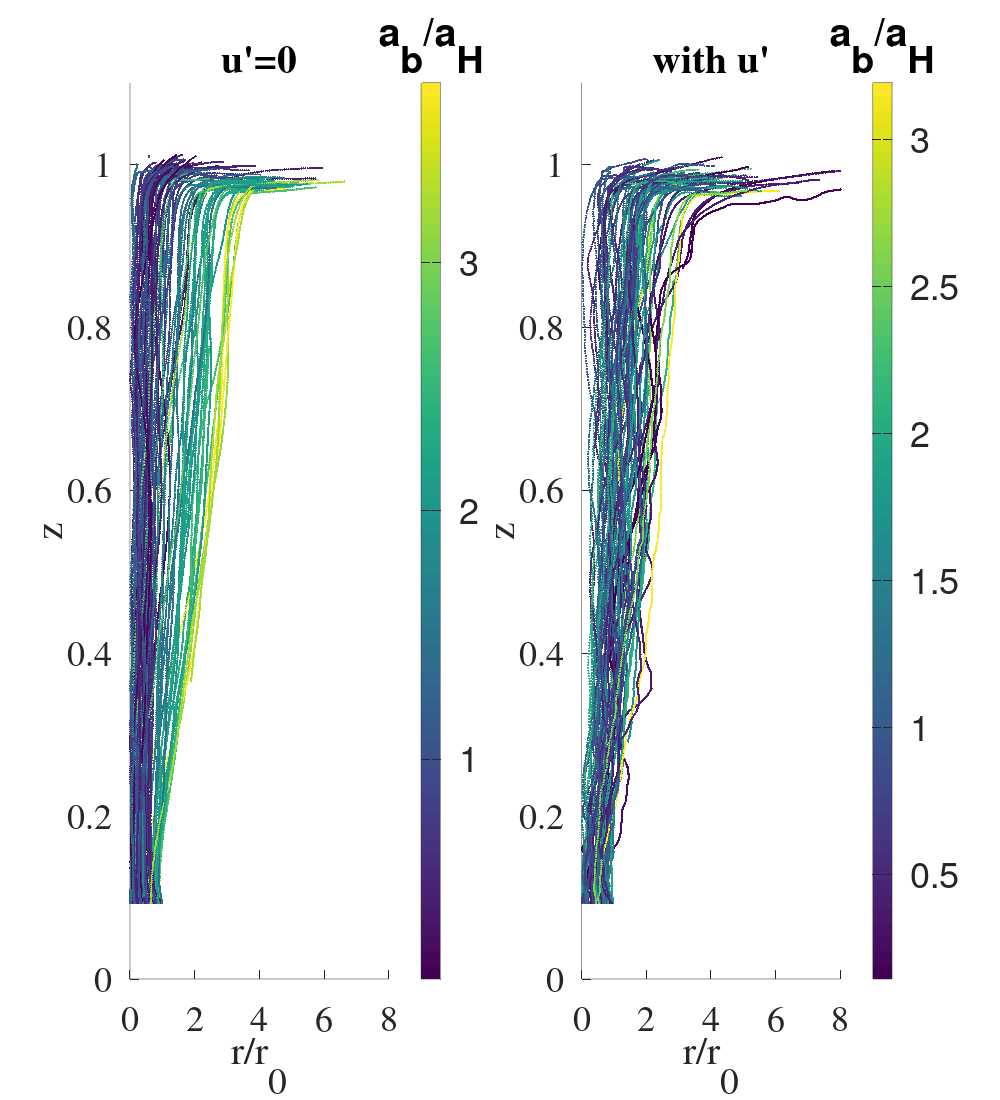}}
\caption{Bubble traces for the polydisperse plume, without (left panel) and with (right panel) the Langevin model for sub-resolution velocity fluctuations. Bubble positions are projected onto polar coordinates $\left(r/r_{0},z\right)$, where $r_{0}=0.02$ is the radius of the plume source.  Traces are coloured by bubble radius $a_{b}/a_{H}$.\label{fig:plume_traces}}
\end{figure}

\begin{figure}
\centerline{\includegraphics[width=0.99\textwidth]{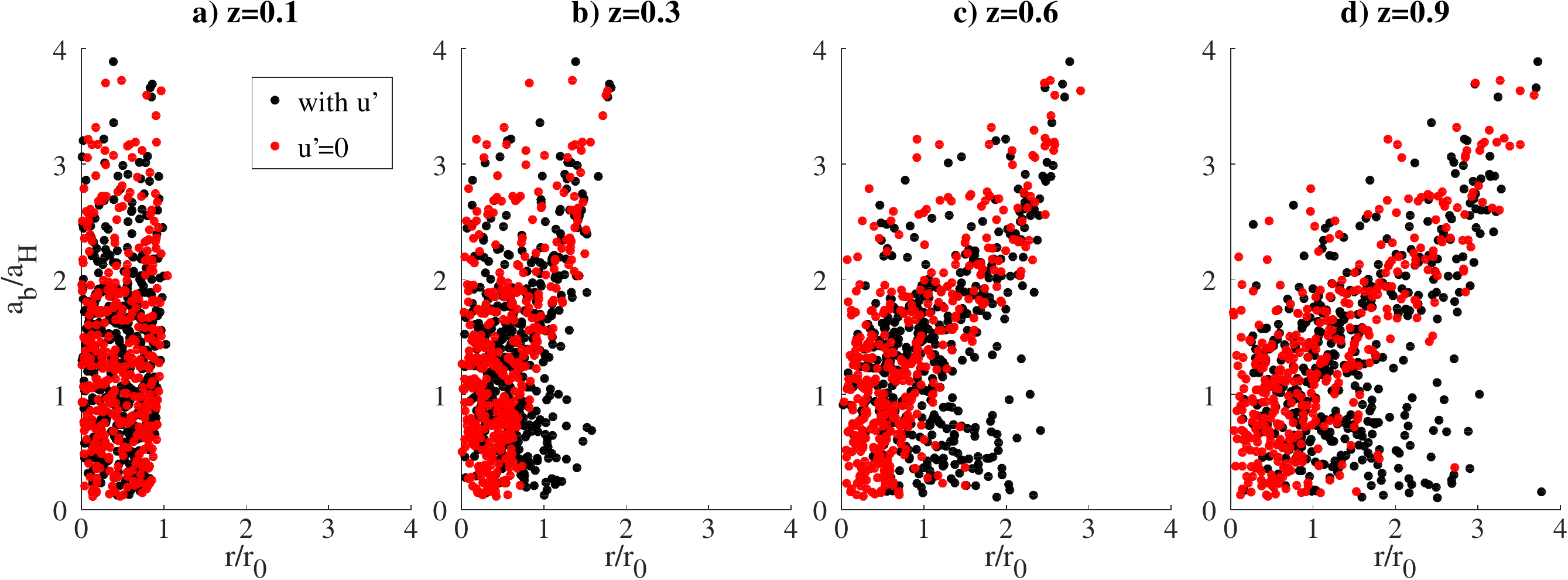}}
\caption{Plots of bubble radius $a_{b}/a_{H}$ against normalised radial position $r/r_{0}$ at several depths, for simulations with (black symbols) and without (red symbols) the Langevin model for sub-resolution velocity fluctuations. \label{fig:plume_r_v_a}}
\end{figure}

We now simulate a plume in the same configuration, but with a polydisperse bubble size distribution. The initial bubble size distribution is a truncated Gaussian with mean $a_{b}=10^{-3}$, standard deviation $10^{-3}$, and a minimum radius cut-off at $a_{b}=10^{-4}$. Figure~\ref{fig:plume_a_v_urel} plots the magnitude of the relative velocity $\lvert\bm{u}_{b}-\tilde{\bm{u}}_{l}\rvert$ against the bubble size near the centre of the tank, at $z=0.555$. Each data point corresponds to a separate bubble, and the data are collected for $t\in\left[10,20\right]$. We see a clear, almost linear trend, with larger bubbles rising faster relative to the liquid than smaller ones. There is also an obvious discrepancy between the results with (black symbols) and without (red symbols) the Langevin model for sub-resolution fluctuations. The Langevin model increases the variation of the relative velocity (without changing the mean). This observation is consistent with Figure~\ref{fig:plume_traces}, which shows the traces of a random sample of bubbles as they rise through the plume. The positions of bubbles are projected onto polar coordinates $\left(r/r_{0},z\right)$, and are coloured by the bubble radius. The left panel shows bubble traces where we set $\bm{u}_{l}^{\prime}=\bm{0}$, whilst for the right panel, we calculate $\bm{u}_{l}^{\prime}$ using the Langevin model. Without the Langevin model, there is a clear trend for larger bubbles to migrate away from the centre of the plume, and smaller bubbles to rise more vertically. This self-organising phenomenon has been previously observed in numerical simulations~\citep{fraga_2016b}, and experimentally~\citep{ye_2022}. The phenomenon can be explained by the linear dependence of the lift force on the relative velocity: as bubbles rise through the plume, the relative velocity is close to orthogonal to the gradient of the velocity in the liquid, and as larger bubbles have a larger relative velocity, they experience an increased lift force, resulting in greater lateral migration. When the Langevin model is included, the bubbles still self-organise, but there is increased lateral migration of small bubbles due to the sub-resolution fluctuations. This self-organisation is clearly depicted in Figure~\ref{fig:plume_r_v_a}, which plots the normalised bubble radius $a_{b}/a_{H}$ against radial position $r/r_{0}$ at several depths. At $z=0.1$ (frame a)), just above the base of the plume, the bubble sizes are evenly distributed across the radius of the plume. Further up the plume, the bubbles migrate radially outwards, and the plume gets wider. There is a clear trend, both with (black symbols) and without (red symbols) the Langevin sub-resolution model, for larger bubbles to migrate further outwards, as can be seen in frames b) to d) of Figure~\ref{fig:plume_r_v_a}. When the Langevin model is included, this trend remains, although as observed in~\ref{fig:plume_traces}, the lateral migration of smaller bubbles is increased - the greater spread in radial position of small bubbles is visible in frames b) to d). This behaviour is expected, as small scale fluctuations have a greater effect on small bubbles than larger bubbles through the increased relative importance of the drag force.

Returning to Figure~\ref{fig:plume_traces}, both with and without the Langevin model, the effect of the free-surface interaction model is clear: as bubbles approach the free surface, they decelerate and move with the liquid, which takes them radially outwards. This behaviour is in qualitative agreement with numerical simulations~\citep{cloete_2009,pan_2021}, although in both cases, these works relied on resolution of the motion of the gas flow above the free surface to capture the bubble-free-surface interactions.

\section{Breaking waves}\label{sec:bw}

We now use our numerical framework to simulate bubble entrainment in a breaking wave. We consider a periodic $3^{rd}$-order Stokes wave, which has been extensively studied in the literature with multi-phase mesh-based schemes~\citep{chen_1999wave,iafrati_2009,deike_2015,deike_2016,mostert_2022,chan_2020,chan_2021b}. We take the wavelength as the integral length-scale, and the Froude speed as the characterstic velocity scale. The domain is a unit cube, periodic in the lateral directions, with a free-slip condition at the lower boundary. We set $Re=4\times{10}^{4}$, $Fr=1$, $We=1.98\times{10}^{4}$ and $\beta=1000/1.2$. These parameters correspond to the configuration of~\citet{mostert_2022} (although we note here that with our scaling, wave overturning occurs shortly after $t=1$, as in~\citet{chen_1999wave} and~\citet{chan_2021b}), and gives a Bond number of $Bo=We\left(\beta-1\right)/\left(4\pi^{2}\beta\right)=500$. \rra{Furthermore, we highlight that the Weber number in our numerical framework only influences the bubbles, as we do not include a surface tension model for the resolved liquid free surface.} The location of the initial free surface is given by
\begin{equation}\eta\left(x\right)=\frac{1}{2\pi}\left\{\chi\cos\left(2\pi{x}\right)+\frac{1}{2}\chi^{2}\cos\left(4\pi{x}\right)+\frac{3}{8}\chi^{3}\cos\left(6\pi{x}\right)\right\},\end{equation}
where $\chi$ is the initial wave steepness. The velocity within the liquid is given by 
\begin{subequations}\begin{align}
u\left(x,y,t=0\right)&=\frac{1}{2\pi}\chi\sqrt{1+\chi^{2}}\cos\left(2\pi{x}\right)\exp\left(2\pi{y}\right)\\
v\left(x,y,t=0\right)&=\frac{1}{2\pi}\chi\sqrt{1+\chi^{2}}\sin\left(2\pi{x}\right)\exp\left(2\pi{y}\right),
\end{align}\end{subequations}
and $w\left(t=0\right)=0$. In many of the results that follow, we refer to the time since impact, $t-t_{im}$, where $t_{im}$ is the time at which the wave first breaks, indicated by a sharp increase in the (spatial) mean value of $\varepsilon$, and confirmed visually. An estimate of the Hinze scale is $a_{H}=We^{-3/5}/2\approx1.36\times{10}^{-3}$, and throughout the following, we report bubble sizes as relative to the Hinze scale.

\begin{figure}
\includegraphics[width=0.49\textwidth]{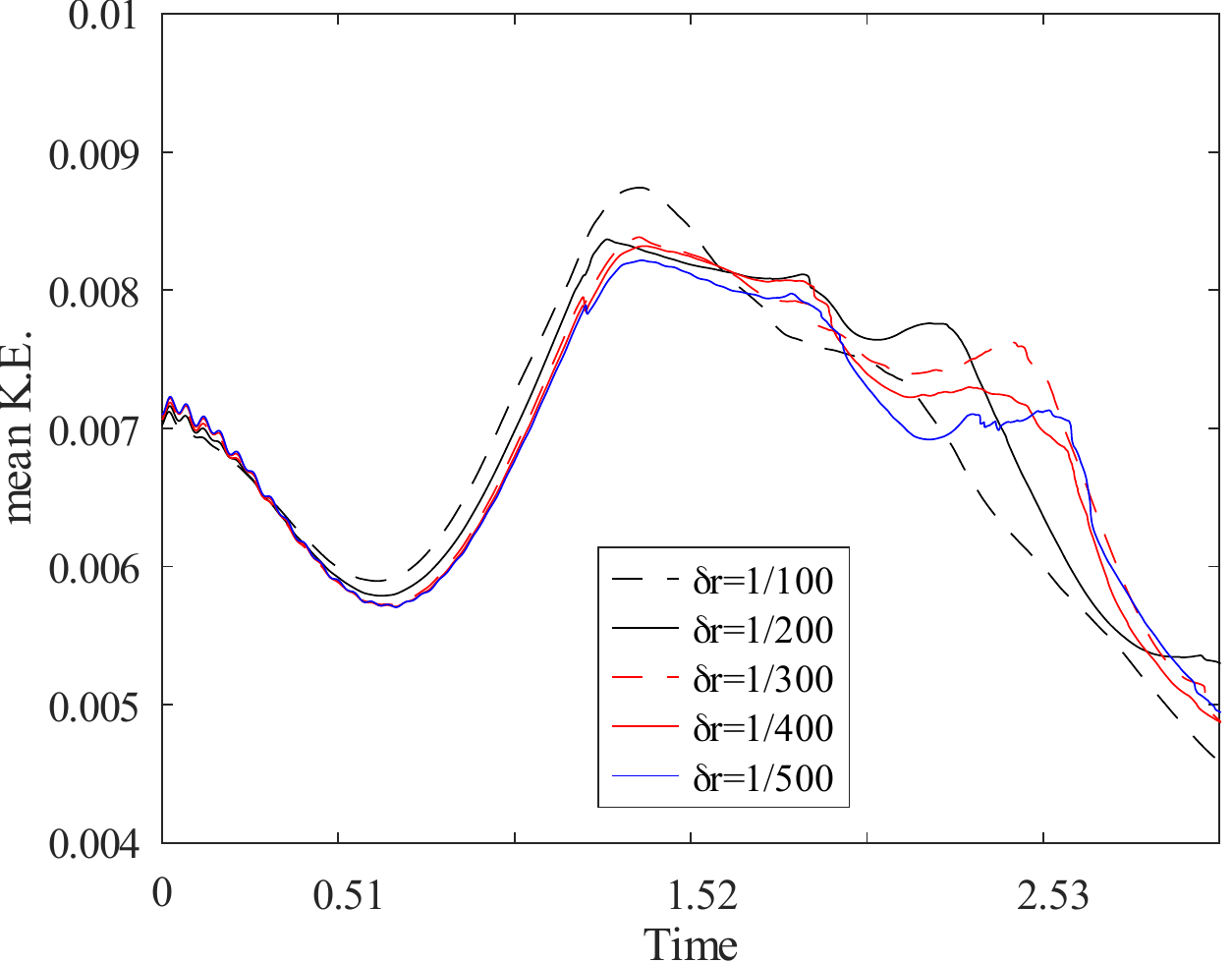}
\includegraphics[width=0.49\textwidth]{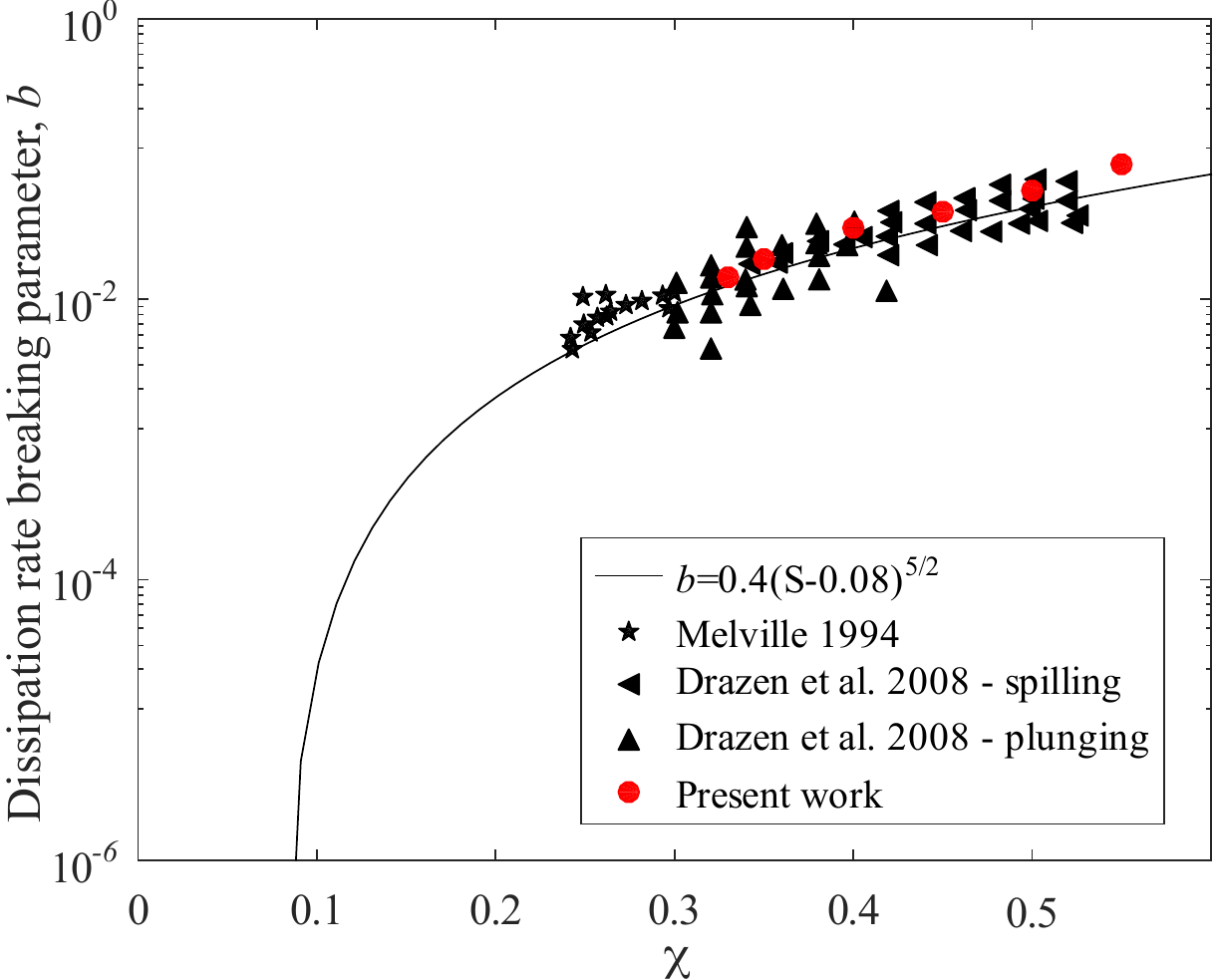}
\caption{\rrhc{Left panel: }Variation of mean kinetic energy (over the entire liquid domain) with time, for a single-phase wave breaking event in the absence of bubbles, for various resolutions, with initial wave steepness $\chi=0.55$.\rrhc{ Right panel: variation of dissipation rate-based breaking parameter $b$ with initial steepness $\chi$ for our simulations, and experimental data of~\mbox{\cite{drazen_2008,melville_1994}}}\label{fig:wave_conv}}
\end{figure}

First, we perform simulations with $\chi=0.55$ in the absence of bubbles, to ensure our SPH model provides a converged solution. \rrhc{The left panel of }Figure~\ref{fig:wave_conv} shows variation of the mean kinetic energy for the single-phase wave with time. We see that the evolution of the energy is approximately converged for resolutions of $\delta{r}\le1/300$. \rra{For the purposes of the present study, this degree of convergence in the kinetic energy is sufficient, and except where otherwise specified, we set $\delta{r}=1/300$ in the following.} \rra{Note that in o}ur single phase SPH model\rra{ surface tension is neglected, although surface tension effects are included in the closure models governing bubble dynamics. This assumption is reasonable for the present case, as the integral scale Weber number is large}. Accordingly, there is no physical limit imposed on the minimum droplet sizes expected during the breaking process. However, we note that the numerics of the SPH algorithm - specifically the particle shifting technique - introduces a surface tension-like effect into the simulation. Although we are unable to accurately quantify this effect, we note that droplets consisting of individual SPH particles are formed during the simulation at all resolutions, \rra{implying that} the \rra{effective numerical surface tension} of the single phase SPH simulation is governed by resolution, for the range of resolutions explored. 

\rrhc{We further validate the numerical framework by performing simulations for a range of initial wave steepness $\chi$, and evaluating the breaking parameter $b$, related to the dissipation rate by}
\begin{equation}b=\frac{\varepsilon_{l}g}{\rho{c}^{5}},\label{eq:bparam}\end{equation}
\rrhc{where $c$ is the phase velocity given by $c=\sqrt{g\lambda}$ as defined for experiments, with $g$ the acceleration due to gravity and $\lambda$ the wavelength. The quantity $\varepsilon_{l}$ is the dissipation rate per unit length of breaking crest, which can be evaluated following~\mbox{\cite{deike_2015}} as the product of the initial wave energy and the decay rate during breaking, calculated by numerical integration over the simulation domain. The right panel of Figure~\mbox{\ref{fig:wave_conv}} shows the variation of the breaking parameter with initial wave steepness for our simulations, alongside numerical data from~\mbox{\cite{melville_1994}} and~\mbox{\cite{drazen_2008}}. The semi-empirical fit of $b=0.4\left(\chi-0.08\right)^{5/2}$~\mbox{\citep{romero_2012}} is shown by the solid line. For all the breaking cases studied ($\chi\in\left[0.33,0.55\right]$) our simulations show a close match with the experimental data and empirical fit.}

\subsection{Bubble size distributions and the effect of resolution}

\begin{figure}
\centerline{\includegraphics[width=0.45\textwidth]{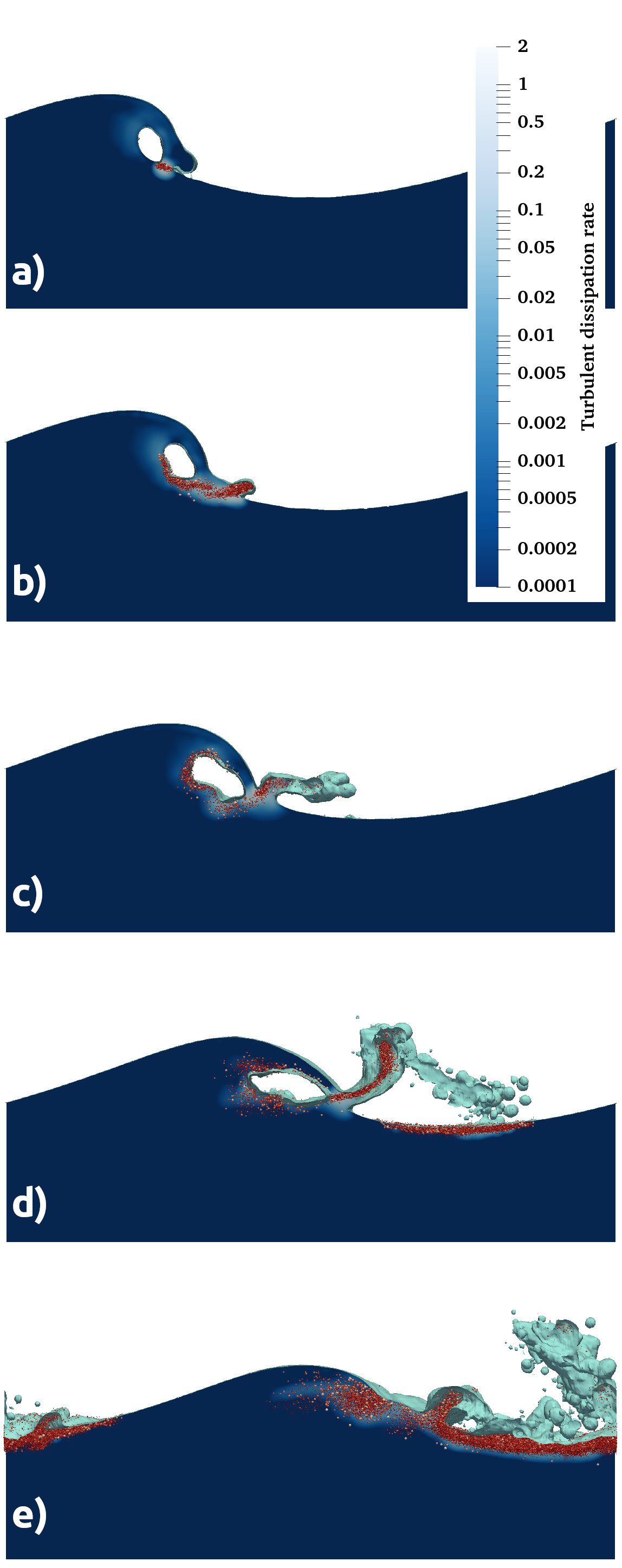}
\includegraphics[width=0.45\textwidth]{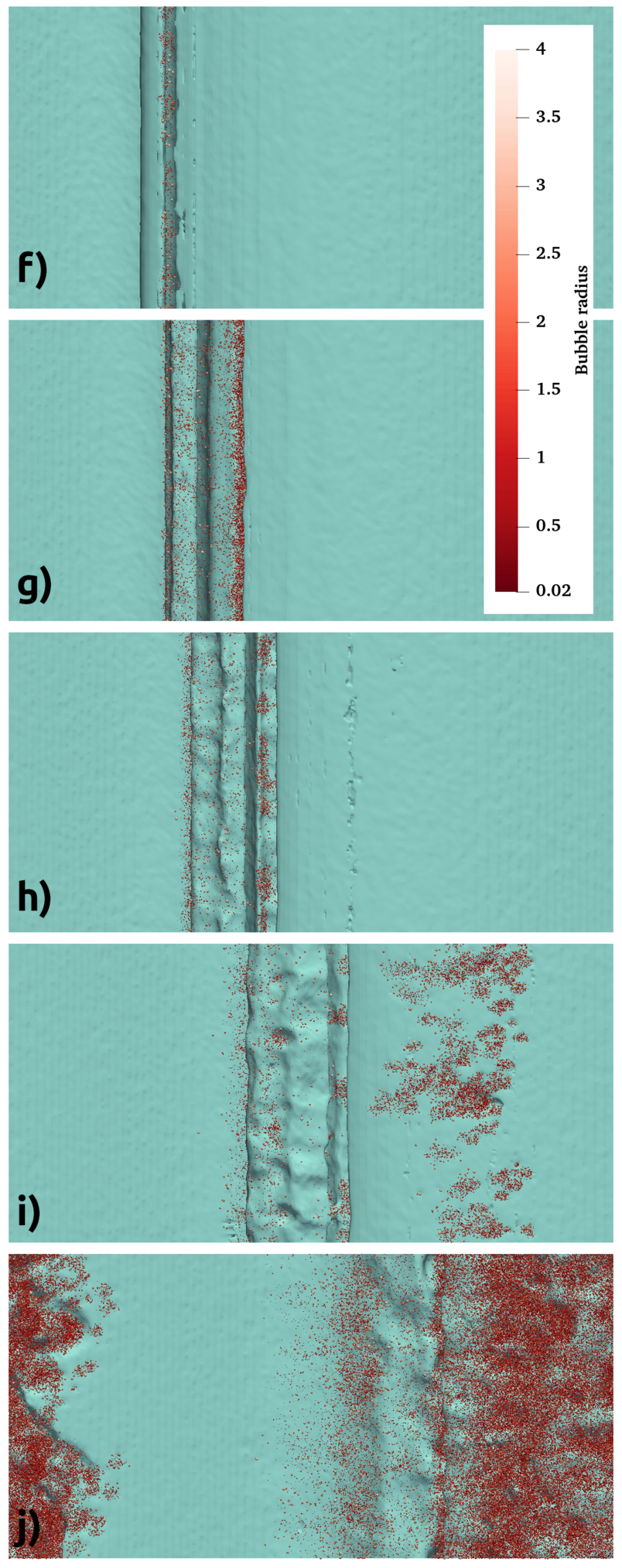}}
\caption{Visualisation of the wave at several instants in time after breaking. The left column (a-e) shows the wave from the side. The right column (f-j) is a view from beneath the wave. The free surface is shown in turquoise. In the left column, the turbulent dissipation rate $\varepsilon$ is shown in a blue colourscale. In all panels, the bubbles are coloured and scaled by $a_{b}/a_{H}$. The corresponding times are a,f) $t-t_{im}=0.015$, b,g) $t-t_{im}=0.09$, c,h) $t-t_{im}=0.215$, d,i) $t-t_{im}=0.465$, e,j) $t-t_{im}=0.840$. SPH particles have been interpolated to a coarse regular mesh for visualisation of the free surface, and the vertical striation visible in the right column is simply a numerical artefact of this grid.\label{fig:wave_img1}}
\end{figure}

We keep $\chi=0.55$ and now include bubbles in our simulation. Figure~\ref{fig:wave_img1} shows the wave at several instants during the first (dimensionless) time unit after breaking, from the side (left) and beneath (right). An animation showing the same views of the wave during the complete breaking process is available in the supplementary material. As the plunging breaker impacts on the surface below, bubbles are entrained in the region of impact (panels a) and f)). At the early times after impact (panels a-c) and f-h)) the flow is largely two-dimensional, with little variation in the transverse direction. Later, as the (resolved) topologically-induced gyre continues to roll, three-dimensional structure forms under the wave (panels d,e) and i,j)). In our simulation, air entrainment occurs through several mechanisms. Firstly, bubbles are entrained on impact of the plunging breaker. Secondly, bubbles are entrained at the surface ahead of the wave due to the impact of spray forward from the plunging breaker; this effect is particularly clear in panel i) of Figure~\ref{fig:wave_img1}. At approximately $t-t_{im}=0.8$ (panels e) and j) in Figure~\ref{fig:wave_img1}), the topologically-induced gyre collapses, and results in further entrainment. We note here a limitation of our approach. As our model for the continuum is single phase (we do not resolve the gas above the free surface), although we predict the entrainment of a large topologically-induced gyre, the mass of this entrained air is partially lost when this bubble collapses. This limitation is common to all single-phase SPH models, and also to the model of~\citet{kirby_2014}. Our model does predict bubble entrainment as the topologically-induced gyre collapses, due to the large values of $\varepsilon$ as the free surfaces come together. Combining the present work with a multi-phase SPH scheme \rra{would overcome this limitation, allowing} large bubbles \rra{to be} resolved, \rra{whilst} bubbles with $a_{b}<\delta{r}$ are modelled\rra{. This }is an active area of research for us, but we reserve such a model for a future work.

\begin{figure}
\centerline{\includegraphics[width=0.49\textwidth]{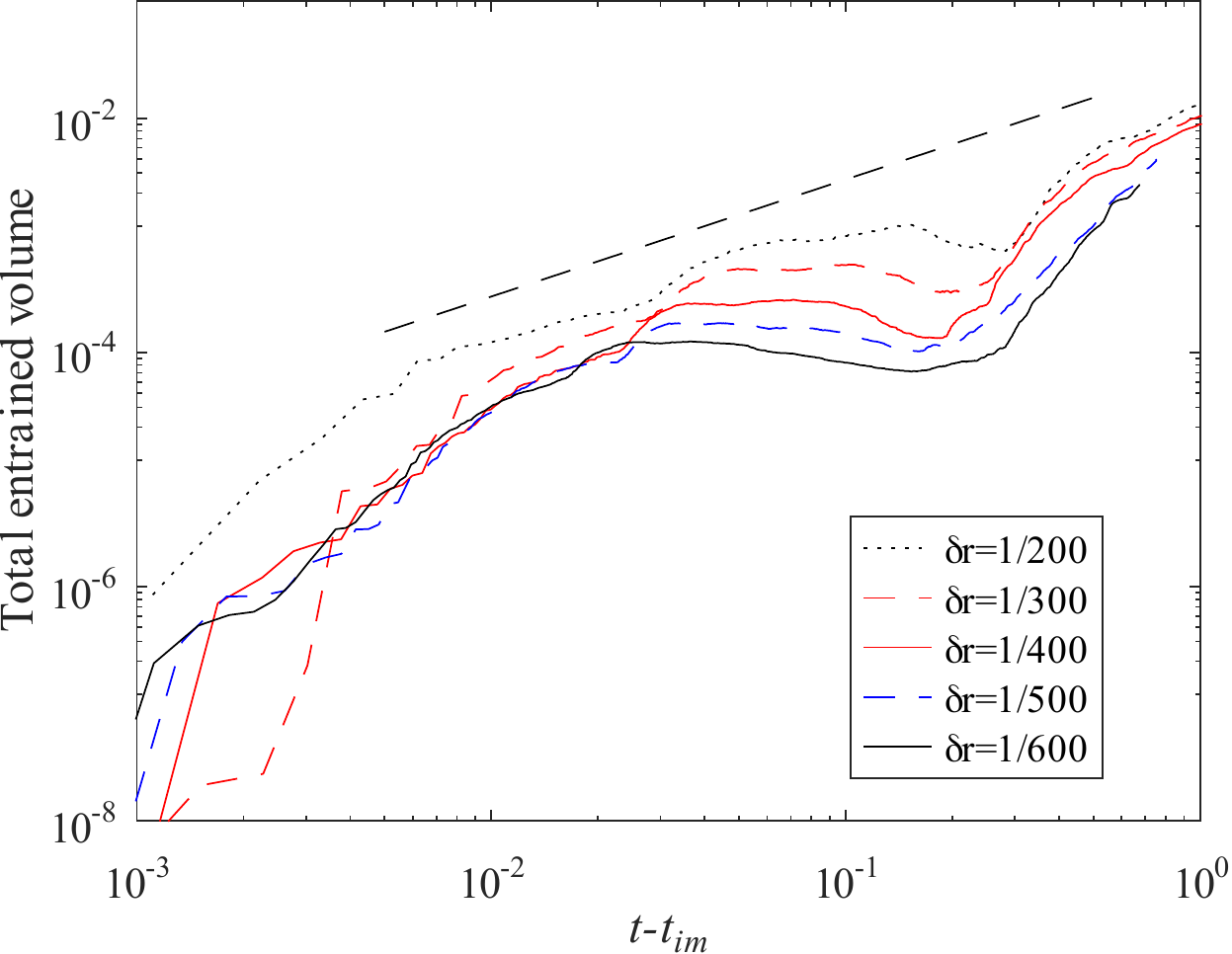}
\includegraphics[width=0.49\textwidth]{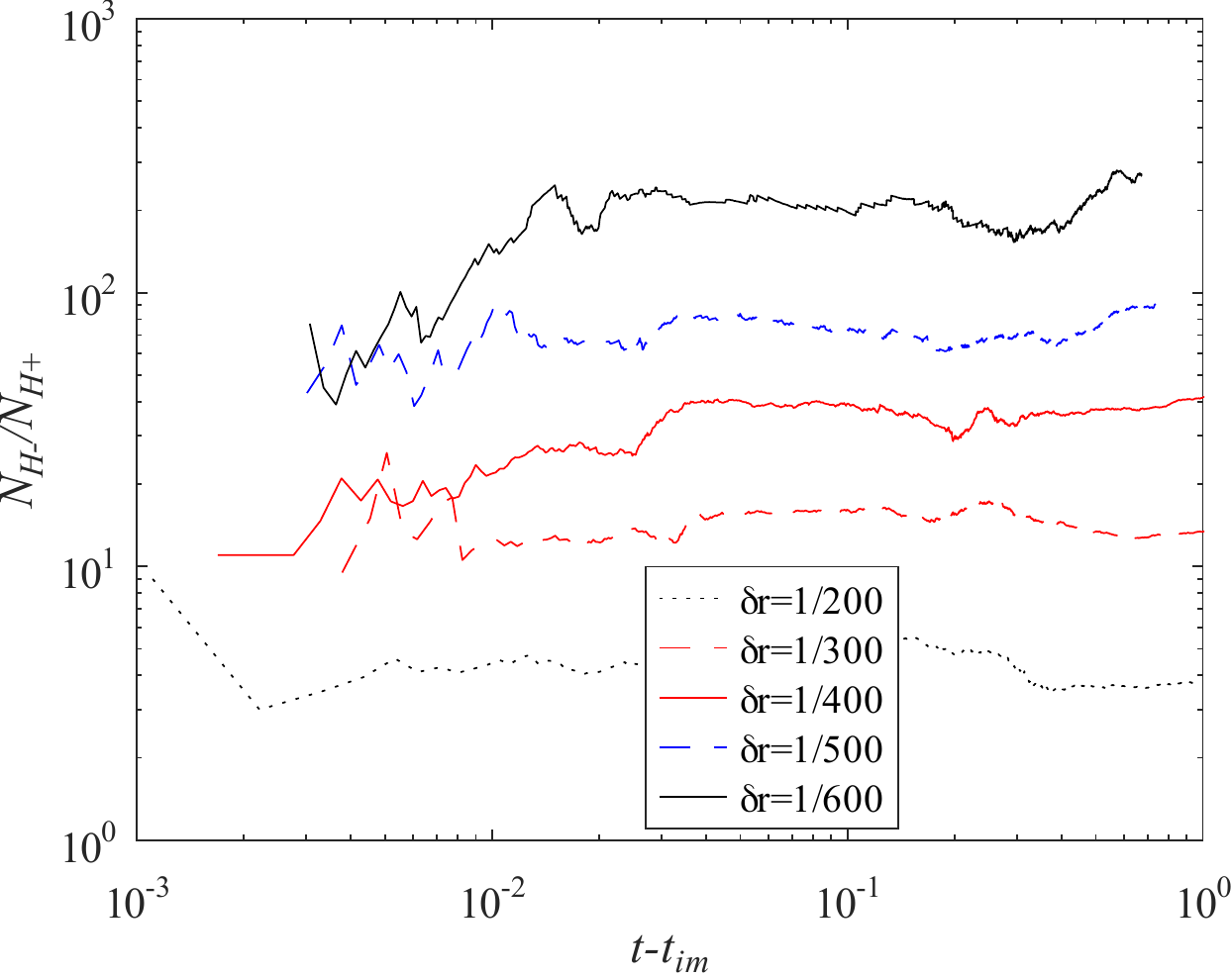}}
\caption{\rrha{Evolution of the bubble population. Left panel: variation of total entrained volume with time after impact $t-t_{im}$, for several resolutions. Right panel: time variation of the number ratio of sub- to super-Hinze scale bubbles, for several resolutions. For both panels, $\chi=0.55$.}\label{fig:wave_pop1}}
\end{figure}

The left panel of Figure~\ref{fig:wave_pop1} shows the evolution of the total volume of entrained bubbles (normalised by the total liquid volume), for several resolutions. There is approximately linear growth volume of air entrainment (the dashed black trendline has slope of $1$), as found in the adaptive-mesh simulations of~\citet{mostert_2022}. \rrha{We see a close agreement in the total entrained volume for $\delta{r}\le1/400$, suggesting the method is converged in terms of the total entrained volume with respect to the SPH resolution. This agreement is particularly close at early times, prior to the collapse of the topologically-induced gyre. At later times there is moderate divergence in the total entrained volume, which we believe is due to the inability of our method to account for different modes of bubble entrainment (discussed further below), and the inevitable dependence of the bubble size distribution on the resolution in a model of this type.}

The right panel of Figure~\ref{fig:wave_pop1} shows the evolution of the ratio $N_{H-}/N_{H+}$ for several resolutions, where $N_{H-}$ is the total number of sub-Hinze scale bubbles, and $N_{H+}$ is the number of bubbles larger than the Hinze scale. For each resolution $N_{H-}/N_{H+}$ is approximately constant, though it varies between resolutions. Our framework only admits bubbles smaller than the SPH resolution, and so for smaller $\delta{r}$, the maximum bubble size is smaller. This reduction in the available range of super-Hinze scale bubble sizes reduces the number of super-Hinze scale bubbles. Given our entrainment model is based on considerations of energy and volume balance (which are roughly invariant with $\delta{r}$), the total volume entrained is approximately the same for all \rrha{six} values of $\delta{r}$ in the right panel of Figure~\ref{fig:wave_pop1}. Hence, we see a corresponding increase in sub-Hinze scale bubbles as $\delta{r}$ is decreased. The reduction in entrained volume at approximately $t-t_{im}=0.2$ for all resolutions (visible in the left panel of Figure~\ref{fig:wave_pop1}) corresponds to the period shown in panels b-c) and g-h) of Figure~\ref{fig:wave_img1}, when a portion of the bubbles entrained during the initial impact are ejected in the spray. A similar pattern is visible in the total bubble population evolution in the results of~\citet{mostert_2022}.

\begin{figure}
\centerline{\includegraphics[width=0.49\textwidth]{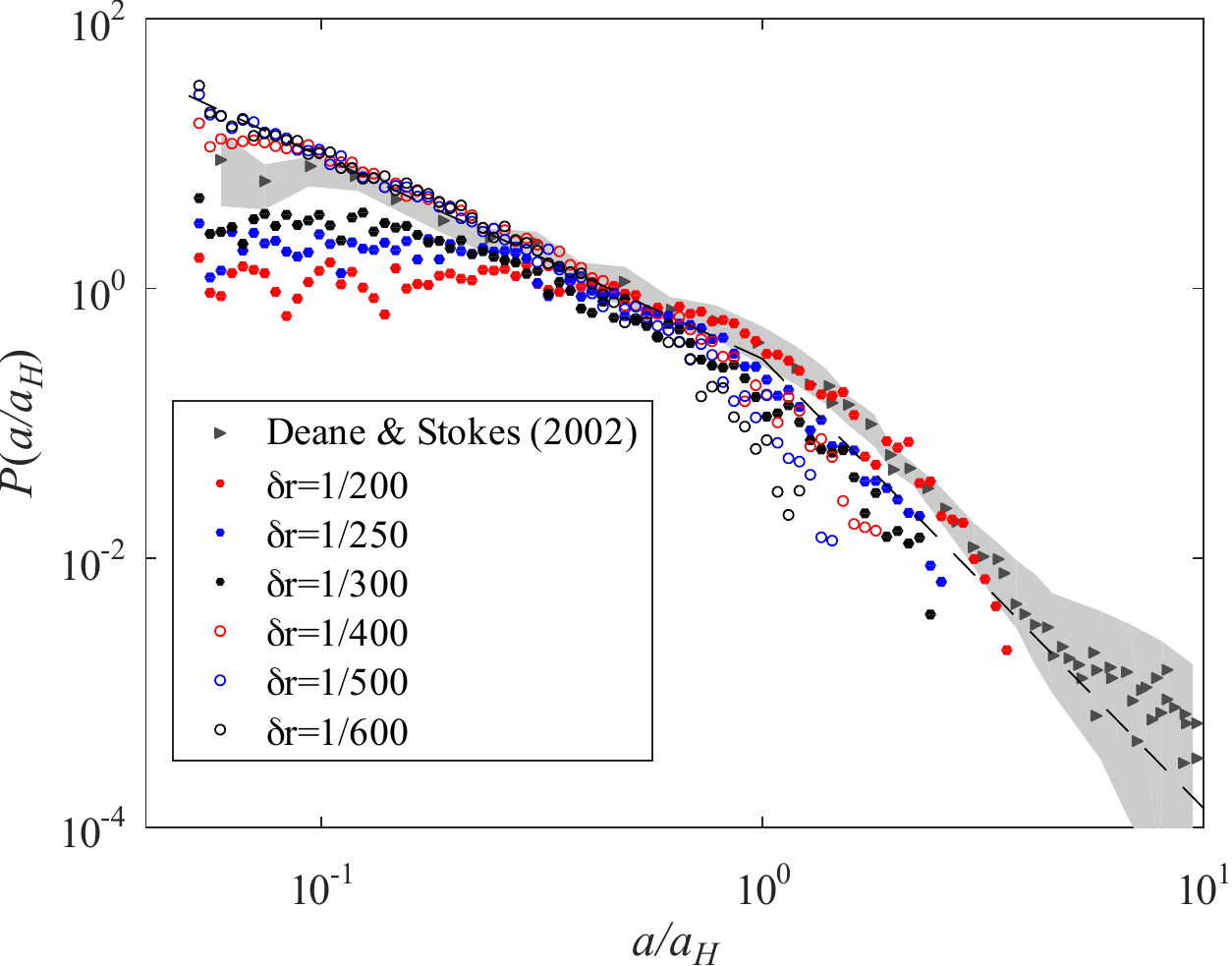}
\includegraphics[width=0.49\textwidth]{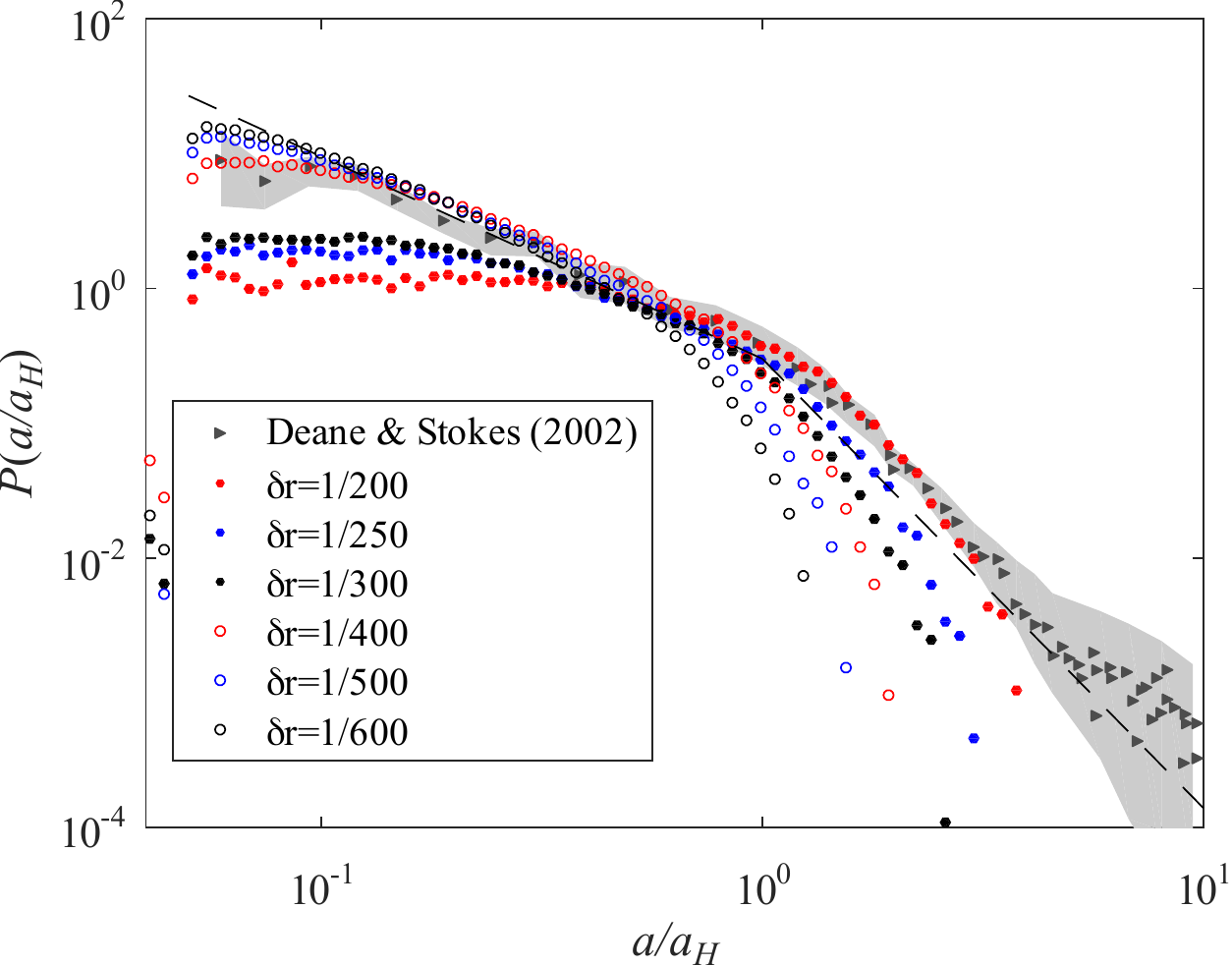}}
\caption{\rrha{Bubble size distribution for $\chi=0.55$ at several resolutions (circles). Left panel: shortly after impact, at $t-t_{im}=0.09$. Right panel: $t-t_{im}=0.840$.}\rrhc{ The grey triangles correspond to the experimental data of~\mbox{\cite{deane_2002}}, and the light grey shaded area shows $\pm$ one standard deviation of the experimental measurements. Note that the magnitude of the experimental data has been scaled to match the non-dimensionalisation of our numerical results.}\label{fig:wave_bsd1}}
\end{figure}

Figure~\ref{fig:wave_bsd1} shows the bubble size distribution $P\left(a_{b}/a_{H}\right)$ for various resolutions at two times after impact. The left panel corresponds to $t-t_{im}=0.09$, and panels b) and g) in Figure~\ref{fig:wave_img1}, whilst the right panel corresponds to $t-t_{im}=0.84$, and panels e) and j) in Figure~\ref{fig:wave_img1}. In both panels of Figure~\ref{fig:wave_bsd1}, the Deane and Stokes spectrum~\citep{deane_2002} is plotted (dashed black lines), with slope $-3/2$ for sub-Hinze bubbles, and $-10/3$ for super-Hinze bubbles.\rrhc{ The experimental data from~\mbox{\cite{deane_2002}} is also plotted (grey triangles), along with the variation of the experimental data (grey shaded region corresponds to $\pm$ one standard deviation).} Our model yields bubble size distributions which \rrhc{closely match the data of~\mbox{\cite{deane_2002}}, including the expected super- and sub-Hinze scale slopes,} over more than an order of magnitude of bubble radii. This match includes accurately predicting the change in slope of the bubble size distribution about the Hinze scale. This result is significant, given the simplicity of our entrainment and breakup models, which are based on simple energy and volume balance arguments. The shift in bubble size distribution as $\delta{r}$ is reduced is clear in Figure~\ref{fig:wave_bsd1}. For smaller $\delta{r}$, $P\left(a_{b}/a_{H}\right)$ drops below the $-10/3$ slope and decays to zero at a smaller $\delta{r}$, whilst the extent to which the sub-Hinze $-3/2$ slope is predicted increases. The match between the expected and simulated bubble size distribution is consistent across both time instants, although $P\left(a_{b}/a_{H}\right)$ shows more noise at early times, as the bubble population provides a smaller sample then. \rrha{Again we mention that this result shows that the energetics of our simulations are converged with respect to the resolution. We also highlight the significant result that for all six resolutions tested, the model predicts the slopes of the bubble size distribution, and where the Hinze scale falls within the range of possible bubbles sizes, the model predicts the Hinze scale and both the sub- and super-Hinze scale bubble size distribution slopes. Figure~\mbox{\ref{fig:wave_bsd1}} also highlights the limitation of this approach - that the bubble size distribution, particularly at the largest bubble scales, depends on the resolution of the SPH. This further supports our plans as discussed above to extend the SPH model to a multiphase one, to allow the full range of bubble scales to be captured.}

We again mention a limitation of our framework, and of models of this type. Our entrainment model is relatively simple, and as such the local (in time and space) entrainment bubble size spectrum depends only on $\varepsilon$ and $\alpha$, and not on the mechanism of bubble entrainment. The three separate entrainment processes discussed earlier - by plunging breaker, spray impact, and topologically-induced gyre collapse - in reality involve quite different mechanisms. Our model, however, cannot differentiate between them. This similarly applies to the entrainment models used in Eulerian-Eulerian work ~\cite{ma_2011,kirby_2014}. It is this limitation which drives us to propose the coupling of a multi-phase SPH scheme, with the gas above the liquid surface resolved, to the discrete bubble model. This would enable the entrainment of large scale bubbles to be captured more accurately, with the computationally cheaper discrete bubble model used once large bubbles have broken to $a_{b}\approx\delta{r}$, and for breakup, entrainment and tracking of smaller bubbles. \rrb{Additionally, in the present model, the equation of motion for the bubbles becomes increasingly stiff for small bubbles, due to the drag model. Meanwhile, if no limit is imposed on the minimum bubble size, the number of bubbles increases significantly. The value of the time step used to integrate the equation of motion for the bubbles is specific to each bubble, and for small bubbles can be more than an order of magnitude smaller than the SPH time step. Consequently in the present model, a large population of small bubbles becomes computationally expensive to simulate. A valuable extension to the present model would be to account for the smallest bubbles with a continuum-continuum approach (as in~\mbox{\citet{kirby_2014}}), with a simplified drag model. Thus we would have a scheme which resolves large bubbles, treats intermediate sized bubbles discretely (as in this work), and treats small bubbles as a continuum, allowing the complete range of bubble scales to be modelled.}

\rrc{In the present work we have focussed on unidirectional periodic waves. An important future development of this work is to include a piston-type wavemaker functionality. There is an extensive body of experimental work on air entrainment in wavemaker-generated waves (e.g.\mbox{\citet{lamarre_1991,drazen_2008}}), which may be used to provide further validation of our numerical framework. This would allow us to model a broader range of representative sea states, increasing the value of the model.}

\begin{figure}
\centerline{\includegraphics[width=0.95\textwidth]{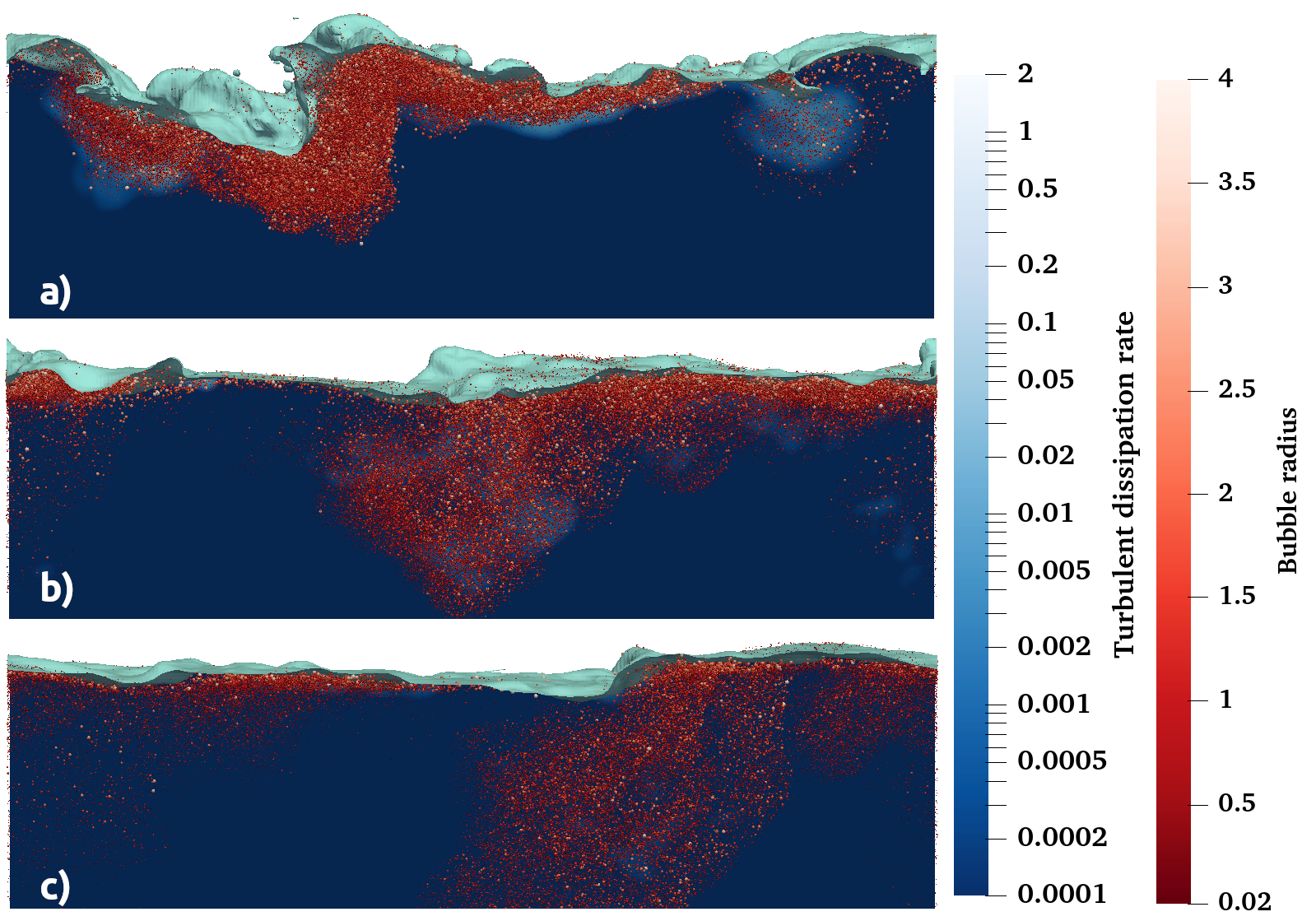}}
\caption{Visualisation of the wave at several time units after breaking. The free surface is shown in turquoise, the turbulent dissipation rate $\varepsilon$ is shown in a blue colourscale, and the bubbles are shown coloured and scaled by $a_{b}/a_{H}$. The corresponding times are a) $t-t_{im}=1.96$, b) $t-t_{im}=3.84$, c) $t-t_{im}=5.71$.\label{fig:wave_late}}
\end{figure}

Figure~\ref{fig:wave_late} shows the wave at several times after $t-t_{im}=1$. At $t-t_{im}\approx{2}$, a large cloud of bubbles is entrained through a reverse breaking wave, visible in panel a) of Figure~\ref{fig:wave_late}. Our simulations predict the obliquely descending eddies observed by~\citet{nadaoka_1989}\rrb{and \mbox{\citet{bonmarin_1989}}}, and modelled \rrb{using a single-phase SPH scheme} by~\citet{dalrymple_2006}\rrb{,~\mbox{\citet{landrini_2007}}} and~\citet{farahani_2014}. The obliquely descending eddies drag a cloud of bubbles downwards away from the surface, which persists for several time units, and is clearly visible in panels b) and c). The observations of bubble distribution between breaking waves predicted by our model, both at early (Figure~\ref{fig:wave_img1}) and late (Figure~\ref{fig:wave_late}) times, is in good qualitative agreement with the Eulerian-Eulerian model of~\citet{kirby_2014}, the detailed simulations of~\citet{chan_2021b} and the experimental observations of~\citet{rapp_1990}.

\subsection{Influence of wave steepness $\chi$}

\begin{figure}
\centerline{\includegraphics[width=0.99\textwidth]{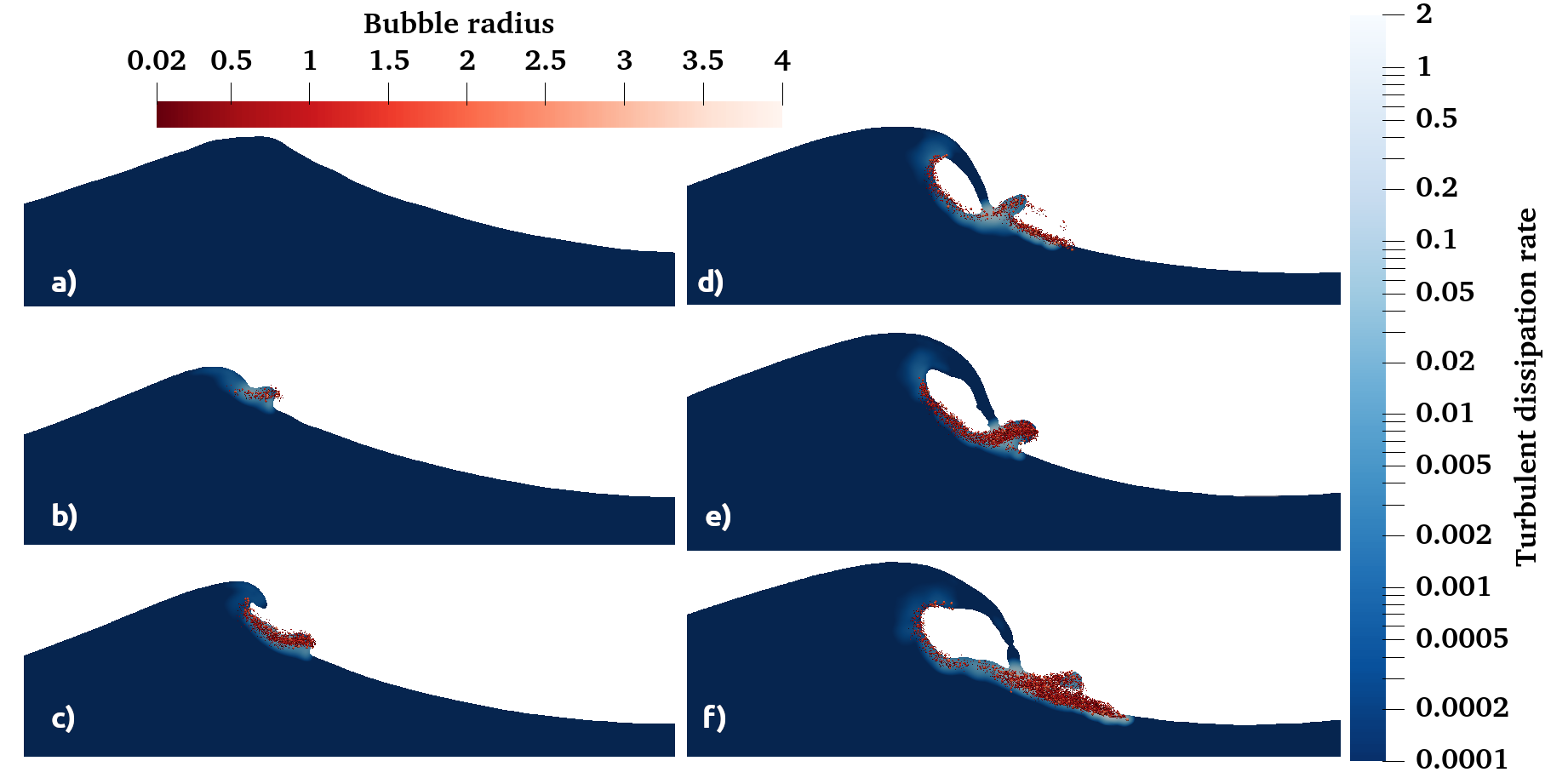}}
\caption{Wave profiles showing dissipation rate $\varepsilon$ (blue colourscale) and bubbles (coloured by $a_{b}/a_{H}$, for various wave steepnesses $\chi$. a) $\chi=0.3$, b) $\chi=0.33$, c) $\chi=0.35$, d) $\chi=0.4$, e) $\chi=0.45$, f) $\chi=0.5$. For panels b-f), the wave is shown at time $t-t_{im}=0.2$. For panel a), in which the wave doesn't break, $t=2.5$.\label{fig:slopeimg}}
\end{figure}

\begin{figure}
\centerline{\includegraphics[width=0.49\textwidth]{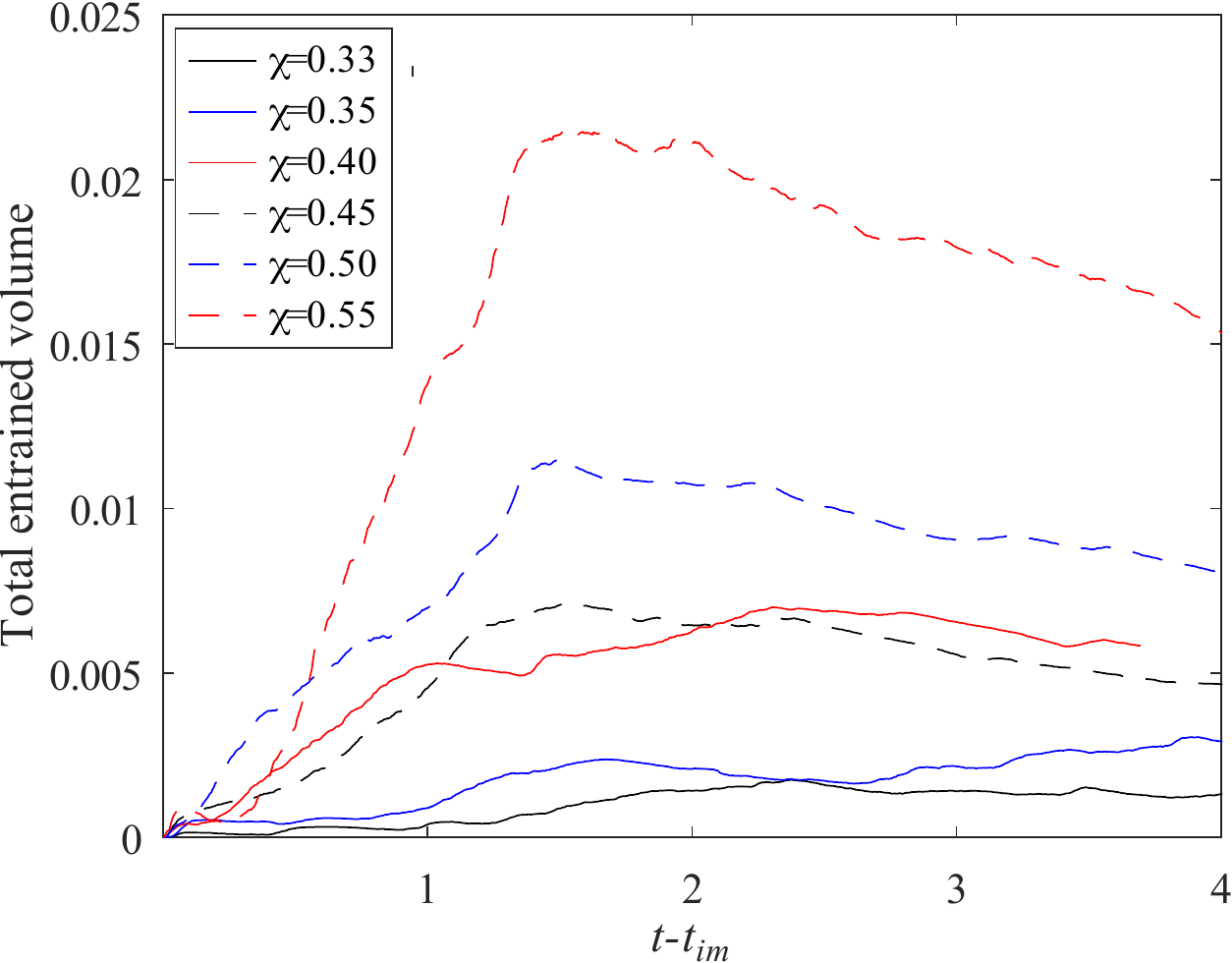}}
\caption{Evolution of total entrained volume with time for various wave steepnesses $\chi$. In all cases, the SPH resolution is $\delta{r}=1/300$.\label{fig:wave_slopevol}}
\end{figure}

Finally we change the wave steepness $\chi$, to cover both breaking and non-breaking waves. Figure~\ref{fig:slopeimg} shows the wave profiles (coloured by dissipation rate), and bubbles (coloured by $a_{b}/a_{H}$), for several wave steepnesses $\chi\in\left[0.3,0.5\right]$. The wave with $\chi=0.3$ (panel a)) doesn't break. For $\chi=0.33$ and $\chi=0.35$ the wave breaks by overspilling, whilst as $\chi$ is increased further, there is a transition to the plunging breaker studied in the previous section. As a limiting test, we observe that for the case where the wave does not break, no bubbles are entrained, and that for all cases where the wave (visibly) breaks, bubbles are entrained. Figure~\ref{fig:wave_slopevol} shows the time evolution of the total entrained volume (normalised by the total liquid volume), for various wave steepnesses $\chi\in\left[0.33,0.55\right]$. For $\chi=0.33$ and $\chi=0.35$, the total entrained volume remains small, although the wave breaking and entrainment process has a long duration. As $\chi$ increases and the plunging breaker regime is approached, the total volume of air entrained increases significantly, whilst the peak moves earlier, to approximately $1.5$ time units after breaking. As in Figure~\ref{fig:wave_pop1}, the linear growth of the entrained volume is clearly visible in the plunging breaker regime.

\rrhc{We note here the work of~\mbox{\cite{lamarre_1991}} and~\mbox{\cite{melville_1994}} who performed measurements on controlled deep-water breaking waves to obtain a picture of the void fraction distribution and the time evolution of integral properties of the bubble plume. Whilst the results of our simulations are not comparable with these experiments (due to the different types and scales of waves considered), we again highlight that the inclusion of a wavemaker in our numerical framework is an area of ongoing development, and will enable direct comparison with these experimental data, along with simulation of a broader range of sea states.}

\section{Conclusions}\label{sec:conc}

In recent years the potential of Smoothed Particle Hydrodynamics (SPH) for simulations of free-surface flows has been widely demonstrated, but limitations in adaptivity remain in comparison to adaptive-mesh methods, preventing the use of SPH high-fidelity simulations of bubbly flows. Approaches which resolve a continuous liquid phase, and include bubbles as discrete Lagrangian particles are established in the mesh-based community, but have not previously been developed in a mesh-free framework, despite the benefits of mesh-free approaches for free-surface flows.

In this work we have presented a numerical framework for Large Eddy Simulations (LES) of bubbly, free-surface flows. The framework employs a continuum-discrete structure, where we use SPH for the LES of the liquid phase, whilst treating bubbles as discrete Lagrangian particles, which interact with the liquid via exchanges of volume and momentum. We introduce a Langevin model for the sub-resolution velocity fluctuations, and several closure models for bubble break-up, entrainment, and free-surface interaction. The modification of a projection method to admit a small degree of compressibility provides a significant reduction in computational costs, whilst preserving smooth pressure fields obtained with incompressible SPH. Exchanges between bubbles and liquid are implemented by SPH interpolation, and the additional construction of neighbour lists required to enable this is straightforward in an SPH framework. Individual bubbles are able to be tracked over a lifetime, including motion due to turbulent structures below the resolution of the SPH scheme. Hence, models for bubble breakup (or in future, deformation and oscillation), which happen over a finite time, may be constructed in integral formulations, rather than as probabilistic events based on an instantaneous flow state.

We have demonstrated the ability of our method to simulate bubble plumes and breaking waves, with quantitative agreement with previous numerical and experimental data in terms of mean flow statistics, bubble size distributions, and bubble population evolution. Despite the inclusion of bubble entrainment and breakup through simplified energy-balance models, the numerical scheme is capable of accurately predicting the Hinze scale, and the multi-slope bubble size distribution present in breaking waves, alongside the bubble population growth rate. The bubble distributions between breaking waves generated by our model compare qualitatively well with experimental data, including the generation of bubble clouds dragged downwards by obliquely descending eddies.

Our investigations have highlighted a limitation of models of this type, which is that bubble entrainment models based on turbulence dissipation rates alone cannot account for the different physical mechanisms of entrainment in different flow types. Further developments which are planned include the extension of the SPH model to multi-phase flows, to yield a framework in which large-scale bubbles are resolved, with small-scale bubbles modelled as in the present scheme. As bubbles influence the forces and loads on structures due to wave impacts, and given the strengths of SPH in computationally affordable free-surface flow simulations, the work herein offers the potential for improvements in the accuracy of predictive modelling for wave-structure interactions.

\section*{Acknowledgements}
We are grateful for financial support from the Leverhulme Trust, via Research Project Grant RPG-2019-206. JK would like to acknowledge funding from the Royal Society via a University Research Fellowship (URF\textbackslash R1\textbackslash 221290). We are grateful to several anonymous reviewers whose insightful comments have helped improve the work.

\textbf{Declaration of Interests:} we report no conflict of interest.

\appendix
\section{Choice and verification of LES closure model}\label{les}

\begin{figure}
\centerline{\includegraphics[width=0.49\textwidth]{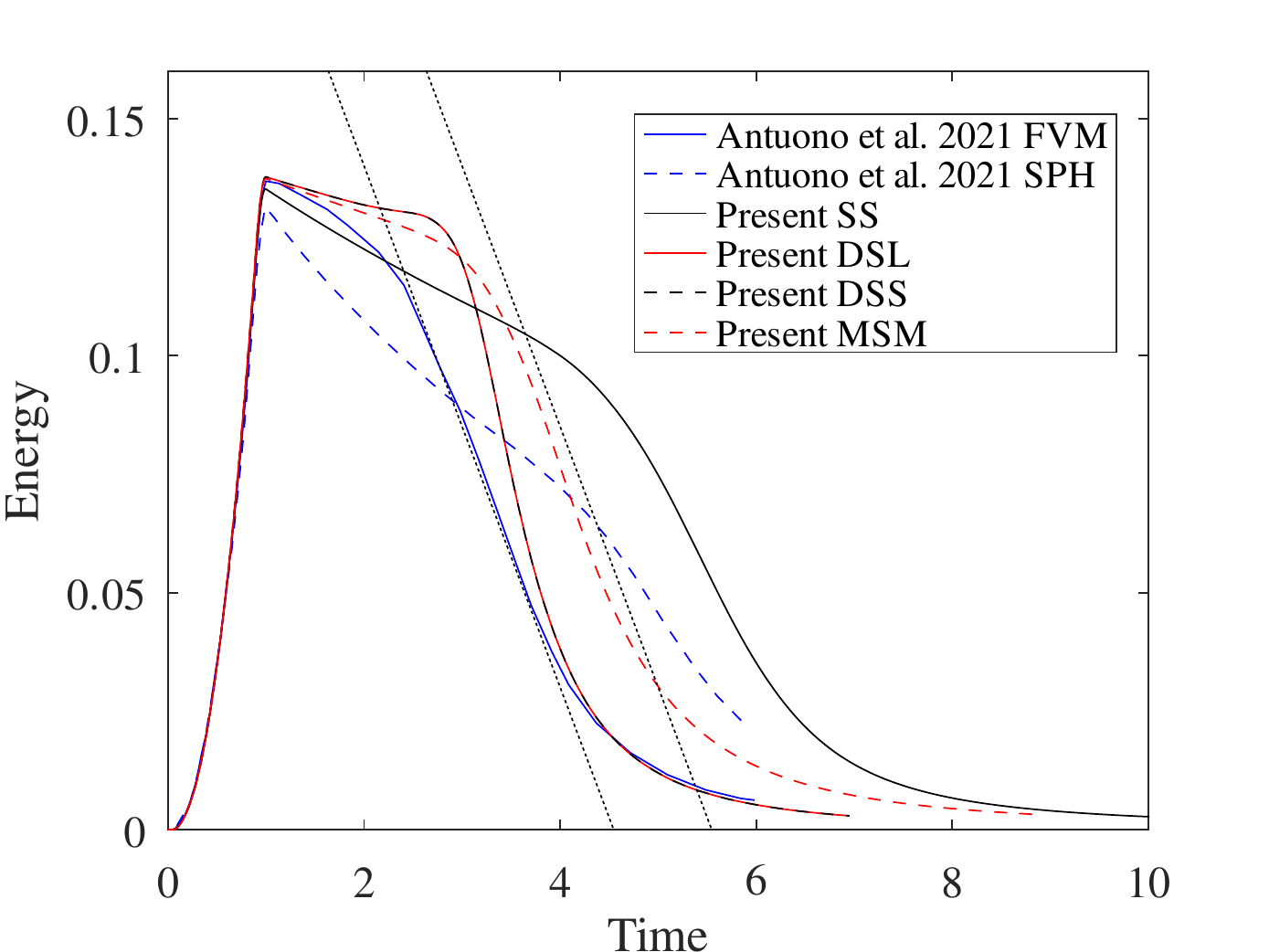}
\includegraphics[width=0.49\textwidth]{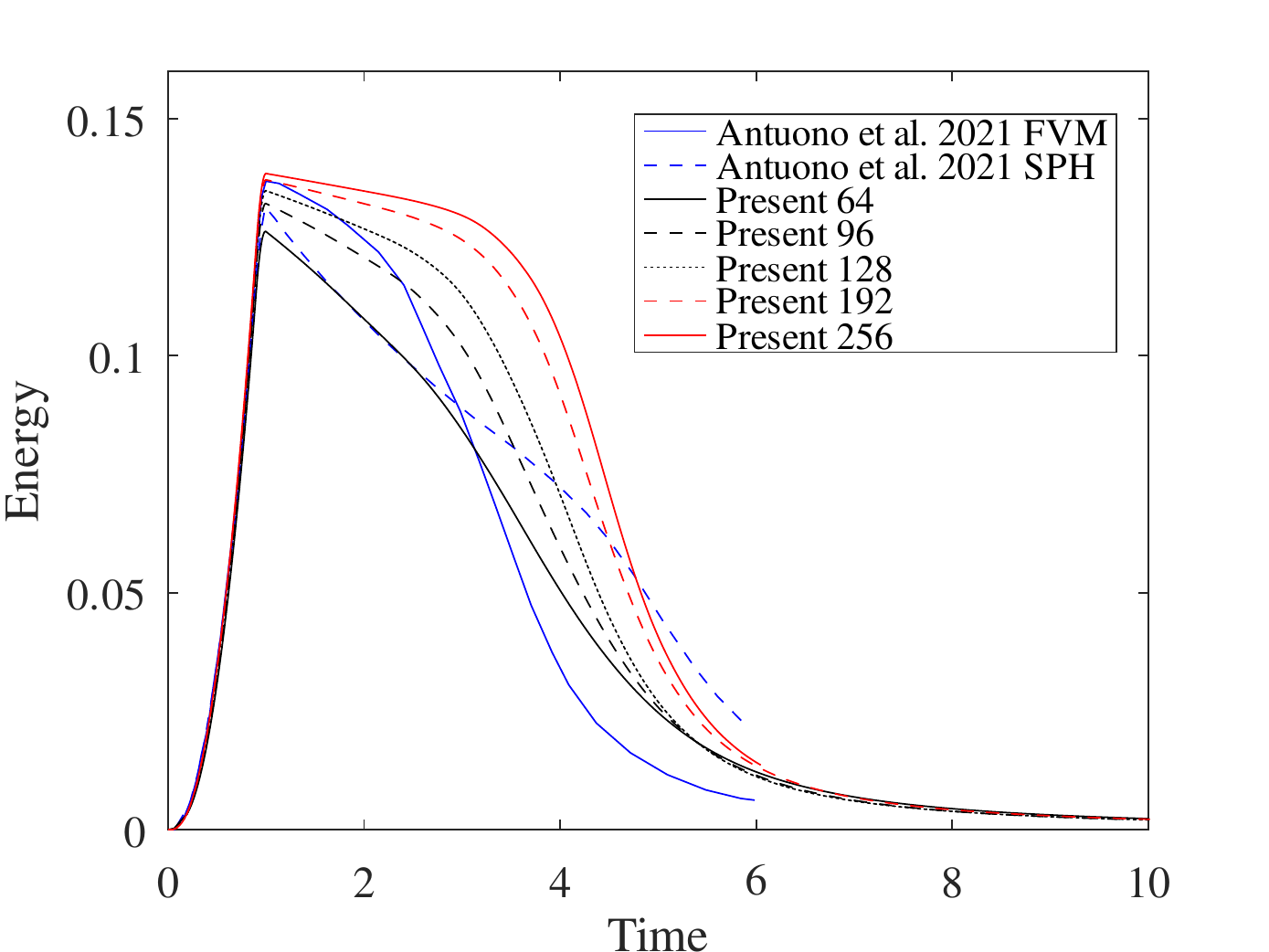}}
\caption{Left panel: Comparison of variation of kinetic energy with time for our approach and the results of~\citet{antuono_2021} (blue lines) for various LES closure models: Standard Smagorinsky (SS) - solid black; Dynamic (Germano) model with Lagrangian averaging (DSL) - solid red; Germano model with Shephard averaging (DSS) - dashed black; and mixed-scale model (MSM) - dashed red. All results were obtained with $\delta{r}=1/128$. Right panel: Comparison of kinetic energy variation with time for our approach at various resolutions, and the results of~\citet{antuono_2021} (blue lines).
\label{fig:isoT_closure}}
\end{figure}

In the development of our numerical framework we have investigated the use of several LES closure models in our SPH scheme. In addition to the mixed-scale model of~\citet{lubin_2006}, we tested a standard Smagorinsky model (referred to here as SS), and the dynamic model of~\citet{germano_1991,lilly_1992}. To avoid locally large fluctuations in $\nu_{srs}$ from the dynamic model, we have implemented both local averaging using a Shepard filter as in~\eqref{eq:shep} (referred to herein as DSS), and Lagrangian averaging following~\citet{meneveau_1996} (referred to as DSL). \rrb{The SS and DSS models have been previously used for LES studies of breaking waves in an finite volume framework by~\mbox{\citet{christensen_2001,christensen_2006}}.} Temporarily neglecting the dispersed bubble phase, we test our single-phase LES SPH scheme on the problem introduced in~\citet{antuono_2021} for weakly-compressible SPH. The problem consists of a triply periodic domain with unit side length. Setting $Re=10^{6}$, a ramped body force is applied over the first time unit, with forcing proportional to the generalised Beltrami flow described in~\citet{antuono_2020}. After this, the flow evolves, with a transition to turbulence occuring within the first few time units. With $Re=10^{6}$, our maximum resolution (of $256^{3}$) is significantly larger than the Kolmogorov scale, and so this is a challenging test for the LES closure model. We compare the results of our code with both the SPH results, and a reference $5^{th}$ order finite volume scheme, presented in~\citet{antuono_2021}. The left panel of Figure~\ref{fig:isoT_closure} shows the evolution of the kinetic energy in the domain with time, for the various LES closure models. All results were obtained with $\delta{r}=1/128$ to match the resolution in~\citet{antuono_2021}. The standard Smagorinsky model (SS, solid black line) is overly dissipative immediately after $t=1$, resulting in significantly delayed transition (identified by the time at which the gradient steepens) compared to the other models. This observation is anticipated: SS models are known to perform poorly, being overly dissipative, in laminar and transitional flows~\citep{kirby_2014}. Compared with the SPH simulations of~\citet{antuono_2021} (dashed blue line), who used a static model, both the dynamic models give an earlier transition, and a post-transition dissipation rate which more closely matches the reference finite volume simulation (solid blue line). Interestingly, there is negligible difference between the DSL and DSS models; for the present case, the choice of averaging procedure has no effect. The mixed-scale-model provides the best match with the reference solution: the post-transition decay rates (the slope of the lines) are a close match, as highlighted by the dotted black lines, which are parallel.

The right panel of Figure~\ref{fig:isoT_closure} shows the variation of kinetic energy with time as the resolution is varied, from $64^{3}$ particles up to $256^{3}$, when using the Mixed Scale Model. In the early stages of decay, the flow is laminar, and the theoretical decay rate has characteristic time $Re/48\pi^{2}\approx2111$. Although this decay rate is never achieved due to the numerical dissipation in all the schemes, we see that as we increase the resolution, the decay rate does reduce (flatter lines immediately after $t=1$). For finer resolutions, there is an increased delay to transition, and the convergence of the post-transition decay rate is apparent. We note that our results converge to a slightly larger post-transition decay rate (steeper lines) than the FV scheme used in~\citet{antuono_2021}, likely due to their use of a resolution of $1/128$. In any case, the post-transition decay rates yielded by our scheme with the MSM are notably closer to the high-order reference solution of~\citet{antuono_2021} than the SPH results of~\citet{antuono_2021} for all resolutions $\delta{r}\le1/96$.

\rrb{We note that the choice of filter length scale $\widetilde{\Delta}$ in SPH is not as clear as for FV schemes (where the implicit filter scale is the same as the cell size). In the present work the implicit filter scale is taken as the smoothing length: $\widetilde{\Delta}=h$, but this choice is by no means unique (indeed,~\mbox{\cite{antuono_2021}} set $\widetilde{\Delta}$ equal to the SPH kernel standard deviation), and may contribute to discrepancies between the LES SPH and the FV reference data. The appropriate specification of $\widetilde{\Delta}$ in SPH is an open problem.}

\begin{figure}
\centerline{\includegraphics[width=0.49\textwidth]{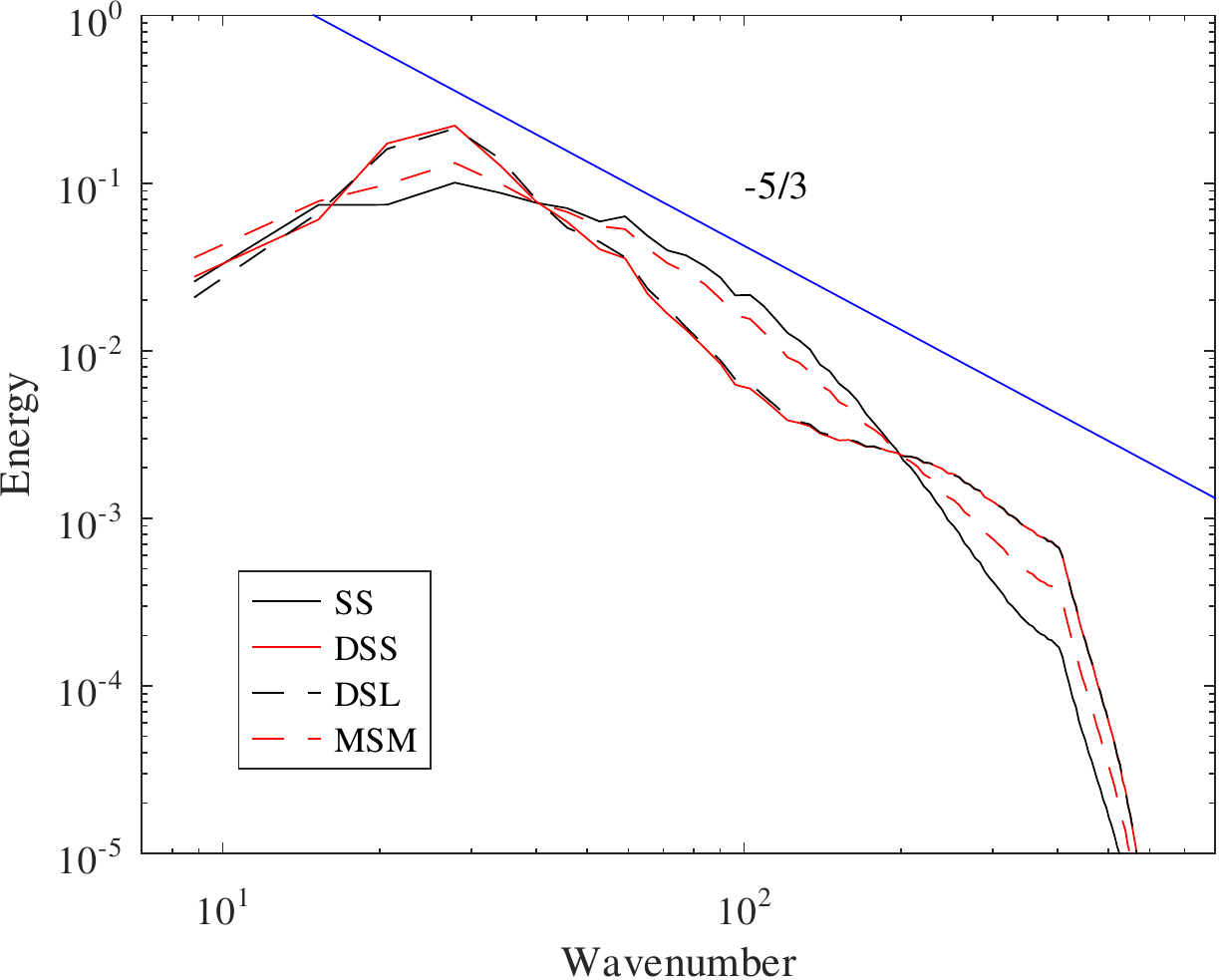}
\includegraphics[width=0.49\textwidth]{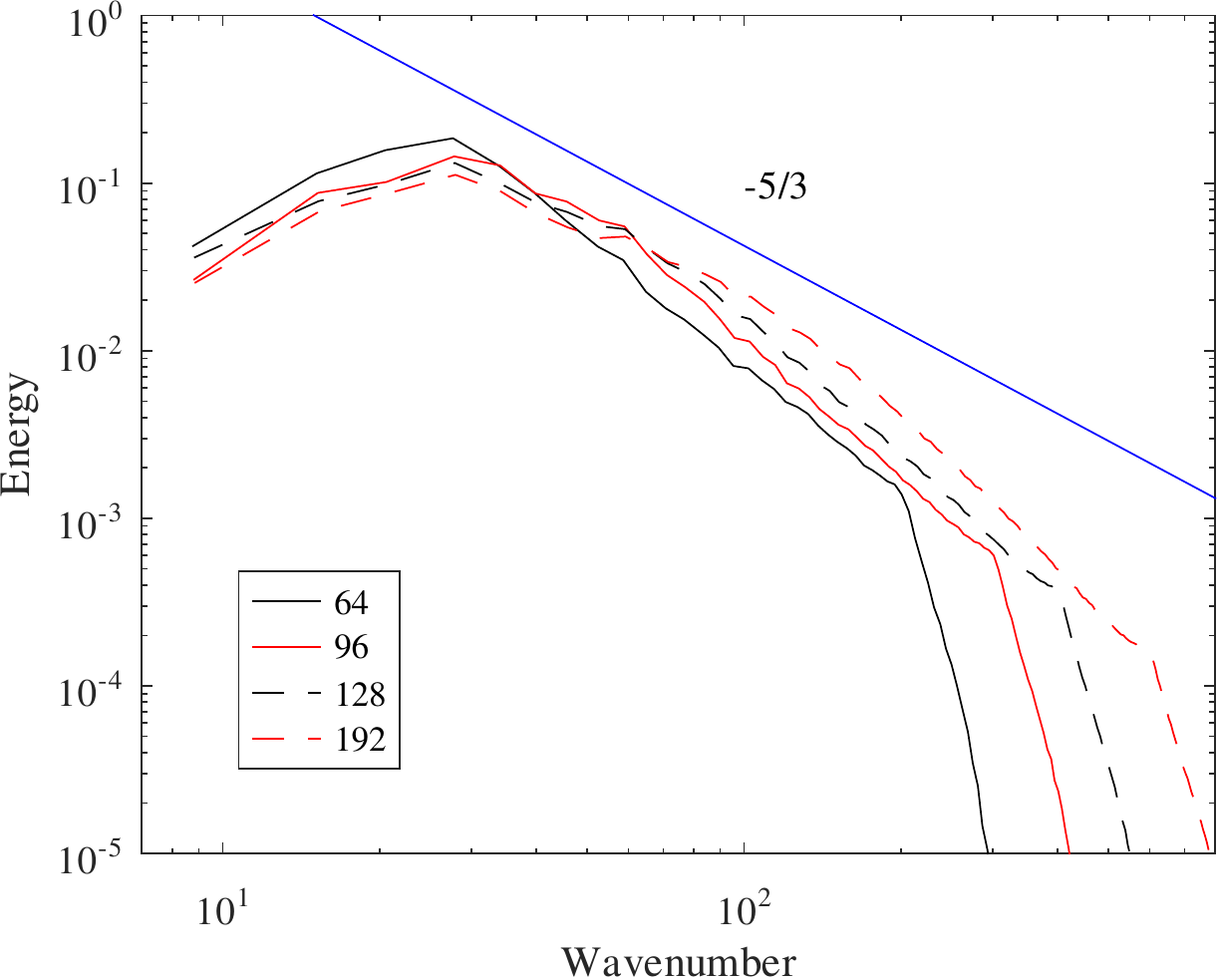}}
\caption{Left panel: Energy spectra for the different closure models: Standard Smagorinsky (SS) - solid black; Dynamic (Germano) model with Lagrangian averaging (DSL) - solid red; Germano model with Shephard averaging (DSS) - dashed black; and mixed-scale model (MSM) - dashed red. Right panel: energy spectra for the MSM closure model at different resolutions. In both panels, a slope of $-5/3$ is indicated by a solid blue line.
\label{fig:isoT_spectra}}
\end{figure}

\rrb{The energy spectra were evalued for the above simulations by interpolating the velocity field from SPH particle positions to a Cartesian grid using a variant of the Local Anisotropic Basis Function Method~\mbox{\citep{king_2020}}, providing $4^{th}$ order consistency. The left panel of Figure~\mbox{\ref{fig:isoT_spectra}} shows the energy spectra (normalised by the instantaneous total energy) for the different closure models. In all cases, the spectra are evaluated when the total energy decays to $0.02$, as in~\mbox{\citet{antuono_2021}}. The spectra for models SS and MSM appear relatively similar, although MSM yields a slope closer to $-5/3$ (solid blue line) for a slightly greater range of wavenumbers. These similarities might be expected - the MSM is the geometric mean of the SS model and a model based on filtering the turbulent kinetic energy. The spectra for the dynamic models (DSS and DSL) are almost identical to each other, and markedly different from the SS and MSM. They exhibit a more pronounced peak at the forcing wavenumber (despite the forcing term being switched off several dimensionless time units prior to evaluation of the spectra), and a distinctly steeper slope than $-5/3$. Furthermore, they have a pronounced peak at high wavenumbers. This is consistent with our observations when using the DSL and DSS models for simulations of breaking waves, that the dissipation rate appeared noisy. Specifically focussing on the Mixed Scale Model, the right panel of Figure~\mbox{\ref{fig:isoT_spectra}} shows the energy spectra for different resolutions. All resolutions have a peak in the energy spectrum at the forcing wavenumber, as expected, followed by a decay. With increasing resolution, the range of wavenumbers for which a $-5/3$ slope (blue line) is observed increases.} 

\rrb{Finally, whilst the tests here provide some quantitative information on the relative performance of different LES closure models in our SPH framework, we note that the performance of the models may change for different flows. For simulations of breaking waves, the turbulence is patchy and transitional, non-isotropic, and multiphase. Quantitative analysis of the performance of LES models under these conditions is not trivial, and remains an open problem.}


\begin{table}\label{los}
\caption{List of notation used herein.Except where explicitly stated, \emph{all} properties are dimensionless.\label{los}}
\begin{tabular}{|l|l|}
\textbf{Symbol} & \textbf{Description} \\
\hline
$\alpha$ & Liquid volume fraction \\
$\beta$ & Density ratio $\rho_{l}/\rho_{b}$ \\
$\gamma$ & Dimensional surface tension \\
$\delta{t}$ & Time increment \\
$\delta{r}$ & Initial SPH particle spacing \\
$\widetilde{\Delta}$ & Implicit LES filter scale \\
$\varepsilon$ & Turbulent dissipation rate \\
$\bm{\zeta}$, $\bm{\zeta}_{r}$  & Random vectors with normally and uniformly (resp.) distributed components \\
$\eta$ & Initial free surface profile for Stokes waves \\
$\kappa$ & An arbitrary property used for exposition of SPH operators \\
$\Lambda$ & Critical bubble radius for breakup \\
$\mu_{l}$ & Dimensional dynamic liquid viscosity \\
$\nu_{srs}$ & Sub-resolution viscosity \\
$\nu_{S}$, $\nu_{B}$ & Shear- and bubble-induced eddy viscosity \\
$\xi$ & Geometric weighting parameter in Mixed-scale-model \\
$\rho_{l}$, $\rho_{b}$ & Dimensional density of liquid and bubble contents \\
$\rho_{N}$ & SPH particle number density \\
$\sigma_{srs}$ & Standard deviation of sub-resolution fluctuating velocity \\
$\tau_{\parallel}$, $\tau_{\perp}$ & Parallel and perpendicular timescales in the Langevin model \\
$\tau_{srs}$ & Sub-resolution velocity fluctuation timescale \\
$\phi$ & An arbitrary property used for exposition of SPH operators \\
$\chi$ & Wave steepness \\
$\psi_{fs}$ & Free-surface-bubble interaction parameter \\
\hline
$a_{b}$, $a_{H}$ & Bubble radius and Hinze-scale radius\\
$b_{v}$ & Bubble plume width \\
$\bm{\mathsf{B}}$ & Diffusion matrix in Langevin model \\
$\mathcal{B}$, $\mathcal{B}_{i}$ & The set of all bubbles, and the set which are neighbours of SPH particle $i$ \\
$Bo$ & Bond number \\
$c$ & Dimensional sound speed \\
$C$ & Timescale constant for Langevin model \\
$c_{def}$ & Bubble deformation coefficient \\
$C_{d}$, $C_{l}$, $C_{vm}$ & Drag, lift and virtual mass coefficients (resp.) \\
$C_{M}$, $C_{\nu,B}$ & Coefficients in mixed-scale model and bubble induced viscosity model \\
$d=3$ & Number of spatial dimensions \\
$\bm{dW}$ & Weiner process \\
$\bm{e_{g}}$ & Unit vector aligned with gravitational acceleration \\
$E_{bc}$, $E_{se}$ & Energy available for bubble creation, and surface energy of a bubble \\
$\bm{\mathsf{E}}$, $E_{\parallel}$, $E_{\perp}$ & Exponential of drift matrix, and components thereof \\
$\bm{F}_{d}$, $\bm{F}_{l}$, $\bm{F}_{vm}$, $\bm{F}_{g}$ & Drag, lift, virtual mass and buoyancy forces on bubbles (resp.) \\
$\bm{F}_{fs}$ & Free-surface interaction force on bubble \\
$Fr$ & Froude number \\
$g$ & Dimensional gravitational acceleration \\
$\bm{\mathsf{G}}$ & Drift matrix in Langevin model \\
$h$ & SPH smoothing length \\
$\bm{\mathsf{I}}$ & Identity matrix \\
$k_{srs}$ & Sub-resolution turbulent kinetic energy \\
$L_{def}$ & Bubble deformation distance \\
\end{tabular}
\end{table}
\begin{table}
\begin{tabular}{|l|l|}
\textbf{Symbol} & \textbf{Description} \\
\hline
$\bm{M}_{b}$, $\bm{M}$ & Interphase momentum exchange on bubble, and evaluated in liquid \\
$Ma$ & Mach number \\
$n_{npb}$, $n_{nb}$ & Number of potential new bubbles, and number of new bubbles \\
$\bm{n}$, $\bm{n}_{0}$ & Surface normal vector, and surface normal vector at plane surface \\
$N$, $N_{b}$ & Total number of SPH particles and bubbles (resp.) \\
$N_{H+}$, $N_{H-}$ & Number of super- and sub-Hinze scale bubbles \\
$\tilde{p}$ & Filtered pressure \\
$\mathcal{P}$, $\mathcal{P}_{i}$ & Set of all SPH particles, and set which are neighbours of particle $i$ \\
$\mathcal{P}_{I}$, $\mathcal{P}_{FS}$ & Set of internal and free-surface (resp.) SPH particles \\
$q_{c}$ & Kinetic energy at test-filter scale \\
$\dot{Q}$ & Volumetric flow rate of bubble source \\
$r_{s}$ & SPH kernel support radius \\
$r_{0}$ & Bubble source radius \\
$\bm{r}$, $\bm{r}_{b}$ & Position vector of SPH particle and bubble (resp.) \\
$\bm{\mathsf{R}}$ & Relative velocity conformation matrix, used in Langevin model \\
$Re$, $Re_{b}$ & Integral- and bubble-scale Reynolds number \\
$\widetilde{\mathcal{S}}$ & Resolved strain rate tensor \\
$Sc$ & Schmidt number \\
$t$ & Time \\
$t_{bc}$, $t_{im}$, $t_{m}$ & Time: of bubble creation, of wave impact, and of bubble-free-surface merging \\
$T_{bu}$, $T_{c}$, $T_{p}$ & Characteristic timescales: for breakup, free-surface interaction, and persistence \\
$u_{def}$ & Bubble deformation velocity \\
$\bm{u}_{b}$ & Bubble velocity \\
$\bm{u}_{l}$, $\tilde{\bm{u}}_{l}$, $\bm{u}_{l}^{\prime}$ & Total, filtered, and fluctuating velocity in liquid \\
$\bm{u}_{ps}$ & Transport velocity \\
$\bm{u}_{rel}$ & Relative velocity between liquid and bubble \\
$V$, $V_{\max}$, $V_{\min}$ & Bubble parent-child volume ratio and limits thereof \\
$V_{b}$ $V_{bc}$ & Bubble volume, and volume available for bubble creation \\
$V_{l,i}$, $V_{i}$ & Volume of liquid, and total volume (resp.) held by SPH particle $i$ \\
$V_{lb}$ & SPH volume interpolated to bubble locations \\
$w$, $w_{c}$ & Vertical velocity component generally, and at plume centreline \\
$W$, $W_{ij}$ & SPH kernel function \\
$\bm{\mathsf{W}}$, $W_{\parallel}$, $W_{\perp}$ & Square root of velocity fluctuation covariance matrix, and components thereof \\
$We$, $We_{b}$ & Integral- and bubble-scale Weber numbers \\
$x,y,z$ & Cartesian coordinate system \\
\end{tabular}
\end{table}

\bibliographystyle{jfm}
\bibliography{jrckbib}

\end{document}